\documentclass[twocolumn]{aastex631}

\usepackage{float}

\shorttitle{Winds and Galaxy-Wide Star Formation at $z\sim 1$}
\shortauthors{Weichen Wang et al.}

\usepackage{amsmath,xcolor,placeins,afterpage, makecell}

\begin{document}

\title{The Importance of Galaxy-Wide Star Formation in Driving Winds at $z\sim 1$}

\correspondingauthor{Weichen Wang}
\email{weichen.wang@unimib.it}

\author[0000-0002-9593-8274]{Weichen Wang}

\altaffiliation{Current address: Department of Physics, University of Milano-Bicocca, Piazza della Scienza, 3, Milano MI 20126, Italy}

\affiliation{Department of Physics \& Astronomy, Johns Hopkins University, 3400 N. Charles Street, Baltimore, MD 21218, USA}

\author[0000-0002-3838-8093]{Susan A. Kassin}

\affiliation{Space Telescope Science Institute, 3700 San Martin Drive, Baltimore, MD 21218, USA}

\affiliation{Department of Physics \& Astronomy, Johns Hopkins University, 3400 N. Charles Street, Baltimore, MD 21218, USA}

\author[0000-0003-4996-214X]{S. M. Faber}

\affiliation{University of California Observatories and Department of Astronomy \& Astrophysics, University of California, Santa Cruz, 1156 High Street, Santa Cruz, CA 95064, USA}

\author[0000-0003-3385-6799]{David C. Koo}

\affiliation{University of California Observatories and Department of Astronomy \& Astrophysics, University of California, Santa Cruz, 1156 High Street, Santa Cruz, CA 95064, USA}

\author[0000-0002-4176-9145]{Timothy M. Heckman}

\affiliation{Department of Physics \& Astronomy, Johns Hopkins University, 3400 N. Charles Street, Baltimore, MD 21218, USA}

\author[0000-0002-9217-7051]{Xinfeng Xu}

\affiliation{Department of Physics and Astronomy, Northwestern University, 2145 Sheridan Road, Evanston, IL 60208, USA}

\affiliation{Center for Interdisciplinary Exploration and Research in Astrophysics (CIERA), Northwestern University, 1800 Sherman Avenue, Evanston, IL 60201, USA}

\author[0000-0002-4176-9145]{Hassen M. Yesuf}

\affiliation{Shanghai Astronomical Observatory, Chinese Academy of Sciences, 80 Nandan Road, Shanghai 200030, People’s Republic of China}

\affiliation{Kavli Institute for Astronomy and Astrophysics, Peking University, Beijing 100871, People’s Republic of China}

\affiliation{Kavli Institute for the Physics and Mathematics of the Universe (WPI), UTIAS, University of Tokyo, Kashiwa, Chiba 277-8583, Japan}

\affiliation{University of California Observatories and Department of Astronomy \& Astrophysics, University of California, Santa Cruz, 1156 High Street, Santa Cruz, CA 95064, USA}

\author[0000-0002-6219-5558]{Alexander de la Vega}

\affiliation{Department of Physics \& Astronomy, University of California, 900 University Ave, Riverside, CA 92521 USA}

\author[0000-0002-6993-0826]{Emily C. Cunningham}
\affiliation{Department of Astronomy, Columbia University, 550 West 120th Street, New York, NY, 10027, USA}
\affiliation{Department of Astronomy, Boston University, 725 Commonwealth Avenue, Boston, MA 02215, USA}

\author[0000-0001-8867-4234]{Puragra Guhathakurta}

\affiliation{University of California Observatories and Department of Astronomy \& Astrophysics, University of California, Santa Cruz, 1156 High Street, Santa Cruz, CA 95064, USA}

\author[0000-0003-2775-2002]{Yicheng Guo}

\affiliation{Department of Physics and Astronomy, University of Missouri, Columbia, MO 65211, USA}

\author[0000-0003-4196-0617]{Camilla Pacifici}

\affiliation{Space Telescope Science Institute, 3700 San Martin Drive, Baltimore, MD 21218, USA}

\author[0000-0003-2249-2539]{John Pharo}

\affiliation{Leibniz-Institut für Astrophysik Potsdam (AIP), An der Sternwarte 16, 14482, Potsdam, Germany}

\author[0009-0002-7553-4747]{Ying Qin}

\affiliation{Department of Physics \& Astronomy, Johns Hopkins University, 3400 N. Charles Street, Baltimore, MD 21218, USA}

\begin{abstract}

In this work, we study winds for a representative sample of 86 star-forming galaxies (SFGs) at $z\sim1$ with $M_\star=10^{9.0}-10^{11.5}\,M_\sun$, by measuring the Mg II line profiles in deep Keck spectra. A total of 50 (58\%) are found to have winds. Unlike local starburst galaxies, the wind detection rate does not exhibit a threshold in star-formation rate (SFR) density $\Sigma_\mathrm{SFR}$ at 0.1$\,M_\odot$/yr/kpc$^2$, but shows a gradual decline around this value.
We find correlations between wind velocity $v_\mathrm{wind}$ and SFR, $\Sigma_\mathrm{SFR}$, and stellar mass, as per previous studies. 
Intriguingly, the $z \sim 1$ SFGs appear to follow the same $v_\mathrm{wind}$--SFR relation as local starbursts.  
A combined fit gives: $\log v_\mathrm{wind} = 0.16\cdot\log\mathrm{SFR} + 2.4$ (3-$\sigma$ significance).  This unified relation spans over 4 dex in SFR and agrees with Illustris-TNG. No unified relation is found between $v_\mathrm{wind}$ and stellar mass, sSFR, or $\Sigma_\mathrm{SFR}$. This suggests winds might be most closely associated with SFR. We examine whether winds in $z\sim1$ SFGs are driven by their most compact star-forming regions.  To do so, we consider whether the relation between $v_\mathrm{wind}$ and the $\Sigma_\mathrm{SFR}$ measured from only these regions is stronger than that for the galaxy-wide $\Sigma_\mathrm{SFR}$.  We do not find a stronger correlation, suggesting that winds are most related to $\Sigma_\mathrm{SFR}$ of the entire galaxy. 
Collectively, these findings suggest a picture in which galaxy-wide star formation plays an important role in driving winds at $z\sim1$.  Wind bubbles from all star-forming regions could combine momentum and help lift their entrained gas out of the galaxy.

\end{abstract}


\section{Introduction} \label{sec:intro}

Galactic outflows are driven by energy and momentum inputs from massive stars, supernovae, and/or active galactic nuclei (AGN). They play a key role in galaxy evolution as a major channel of feedback (e.g., \citealt{Chevalier1985,Heckman1990,Somerville2015,Thompson2024}).
The impact of winds on galaxy evolution is expected to be particularly important at redshift $z\sim 1$, when winds are ubiquitous among normal star-forming galaxies (SFGs; e.g., \citealt{Weiner2009,Rubin2010, Bordoloi2011, Erb2012}).  This is likely related to the high star-formation rates at this epoch (e.g., \citealt{Noeske2007,Noeske2007a,Madau2014,Whitaker2014}). 

Previous studies at $z\sim1$ measured relations between the velocity of the cool wind ($v_\mathrm{wind}$) and galaxy star formation properties, such as the star formation rate (SFR) and SFR surface density ($\Sigma_\mathrm{SFR}$).  Positive correlations with marginal or weak statistical significance (i.e., $\lesssim 3 \sigma$) have been found. 
We present an overview of these $z\sim1$ studies below.

\cite{Kornei2012} and \cite{Prusinski2021} found a marginal correlation between $v_\mathrm{wind}$ and SFR and about a 3-$\sigma$ correlation between $v_\mathrm{wind}$ and $\Sigma_\mathrm{SFR}$. They used samples of 72 and 22 galaxies, respectively. \cite{Rubin2014} used a larger sample of 104 galaxies and found no significant correlation between $v_\mathrm{wind}$ and either SFR or $\Sigma_\mathrm{SFR}$.  
This difference might be caused by the method used to measure $\Sigma_\mathrm{SFR}$ (see, e.g., \citealt{Kornei2012}). In addition to these galaxy-by-galaxy studies, stacking analyses report similar results, i.e., marginally significant or no correlations, at $z \sim 1$ (e.g., \citealt{Rubin2010,Erb2012,Bordoloi2014,sugahara_evolution_2017,Calabro2022}) or higher redshifts (e.g., \citealt{Kehoe2025,Lyu2025}).

Two main questions remain regarding the winds of $z\sim 1$ SFGs. It remains unclear which star formation property of galaxies (SFR, $\Sigma_\mathrm{SFR}$, or specific star formation rate sSFR) is most closely associated with $v_\mathrm{wind}$. It is also unknown how the relations measured for $z\sim1$ SFGs are related to those measured for galaxies in today's universe, especially local starburst galaxies.  
The local starburst galaxies, whose wind physics has been studied comprehensively, are valuable to compare with,  allowing two interesting topics to be addressed. First, their winds are known to be driven by star formation in small regions with high $\Sigma_\mathrm{SFR}$ (e.g., \citealt{Heckman1990,Heckman2015,Xu2022}) and, interestingly, similarly compact, highly star-forming regions are also found in $z\sim1$ SFGs (e.g., \citealt{Guo2012,Guo2015,Claeyssens2025}). It remains to be known whether these similar regions are the main drivers of the winds of SFGs.  Second, starburst galaxies show a threshold in $\Sigma_\mathrm{SFR}$ below which no galaxies have winds \citep{Heckman2015}.
It remains to be known whether a similar threshold exists for the SFGs. These questions are investigated in this paper.

This paper uses a sample of 86 SFGs at $z\sim 1$ which includes twice as many $z\sim1$ galaxies below $10^{9.5} M_\odot$ as previous studies combined. An outline of the paper is as follows.  The observations and data are described in \S\ref{sec:data}.  Measurements of galaxy properties are described in \S\ref{sec:galprop}. Sample selection is described in \S\ref{sec:sample}. Measurements of wind velocities are presented in \S\ref{sec:analysis}. 
In \S\ref{sec:result}, the wind velocities are studied as a function of the galaxy properties including SFR, stellar mass ($M_\star$), and sSFR. In \S\ref{sec:result2},  the wind velocities are studied as a function of the SFR densities, which are measured not only in a galaxy-wide sense but also in smaller sub-galactic regions.  To investigate the two main questions described above, the relations between $v_\mathrm{wind}$ and star formation properties are quantified with correlation significance and compared with the local starburst relations in \S\ref{sec:result} and \S\ref{sec:result2}.
In \S\ref{sec:summary}, we present conclusions.

The wavelengths of spectral lines quoted in this paper are measured in air at Standard Temperature and Pressure \citep{Peck1972}.  Magnitudes are on the AB magnitude system (\citealt{Oke1983}). A $\Lambda$CDM cosmology with a Hubble constant of 70 km/s and $\Omega_M = 0.3$, $\Omega_\Lambda = 0.7$ is adopted.

\section{Spectroscopic Observations and Reductions} \label{sec:data}

The spectroscopic data used in this work were obtained as part of the ``Halo Assembly in Lambda-CDM: Observations in 7 Dimensions'' (HALO7D) survey (\citealt{Yesuf2017,Cunningham2019a,Cunningham2019b}). Observations for this survey were conducted with the DEep Imaging Multi-Object Spectrograph (DEIMOS; \citealt{Faber2003}). The main targets of this survey are the Milky Way halo stars; distant galaxies were used as ``filler'' targets when a slit could not be placed on a halo star. To enable distant galaxy filler targets, HALO7D was performed in areas surveyed by the Cosmic Assembly Near-infrared Deep Extragalactic Legacy Survey (CANDELS; PIs: S.~Faber and H.~Ferguson;  \citealt{Grogin2011,Koekemoer2011}), namely the GOODS-South, GOODS-North, COSMOS, and EGS fields.

Information on the HALO7D observations is as follows; for more details, please see \citet{Yesuf2017}, \citet{Cunningham2019a,Cunningham2019b}, \citet{Pharo2022}, and \citet{Wang2022}. The spectra were obtained using a 600 line mm$^{-1}$ grating blazed at a central wavelength of 7200\,\AA\ and a GG455 order-blocking filter. Spectroscopic slits were 1\arcsec\ wide and the corresponding spectral resolution is 3.5\,\AA\ in terms of the full width at half maximum (FWHM), or equivalently $R = 2000$ and $\sigma_\mathrm{ins} = 60\,$km/s at 7200\,\AA. 
For the galaxies in this study, the exposure times are 1--25 hours with a median of 7 hours. 

The raw spectra are reduced with the DEIMOS {\sc spec2d} pipeline (\citealt{Cooper2012,Newman2013}). Spectra for a given galaxy are generally observed at multiple position angles and these spectra are co-added.  We apply corrections for heliocentric motion to the individual 1D spectra prior to co-adding. Spectra from all individual exposures for a given galaxy are combined as follows.  First, they are ranked in order of S/N in the continuum around the Mg II lines.  Starting with the spectrum with the highest S/N,  we add to it spectra with successively lower S/N and calculate the combined spectrum as the mean weighted by the inverse flux variance. We continue to add spectra until the S/N of the combined spectrum plateaus or we run out of the spectra to add.

\section{Hubble Images} 
\label{sec:imagingdata}
\label{sec:hstdata}

Galaxies in this work were observed by \emph{HST} in multiple wavebands as part of the CANDELS program \citep{Grogin2011,Koekemoer2011} and the GOODS program \citep{Giavalisco2004}. For the purpose of the  morphological analysis in \S \ref{sec:measurements_thiswork}, we use the F435W images (FWHM = 0.\arcsec10) observed with the Advanced Camera for Surveys (ACS)  and the F160W images (FWHM = 0.\arcsec18) observed with the Wide Field Camera 3 (WFC3).  For the galaxies in the GOODS-South and GOODS-North fields, the ACS F435W images are from CANDELS.  For the galaxies in the COSMOS and EGS fields, ACS F435W images are from  UVCANDELS \citep{Wang2020AAS}.

\section{Properties of the HALO7D Galaxies} \label{sec:galprop}

\subsection{Rest-frame Magnitudes and Stellar Masses from CANDELS}

\label{subsec:galprop_rfcolor}
\label{subsec:galprop_mass}

Rest-frame colors (U-V and V-J) and rest-frame near ultraviolet (NUV) magnitudes of the galaxies are adopted from a catalog released by the CANDELS team (PI: D. Kocevski, S. Wuyts, and G. Barro; \citealt{Barro2019}). These quantities were calculated from multi-band photometry from HST, Spitzer, and ground-based facilities using the {\sc EAZY} photometric redshift code \citep{Brammer2008}. 

Stellar masses are also adopted from a catalog released by the CANDELS team \cite{Santini2015}.  These masses are the medians of different measurements of the stellar masses of galaxies made by 10 teams which performed spectral energy distribution fitting to multi-band photometry. As demonstrated by \cite{Santini2015} and other studies (\citealt{Mobasher2015,Barro2019,Pacifici2023}), the uncertainties are mainly caused by the different SED-fitting assumptions and are typically 0.15 dex.

\subsection{Galaxy Properties Measured in This Paper}
\label{sec:measurements_thiswork}

\subsubsection{Sizes $(r_e)$}
\label{sec:sizemeasurements}

For the galaxies selected for this study, which are described in \S\ref{sec:sample}, we measure their half-light radii $r_e$ from \emph{HST}/ACS F435W images ($\lambda_\mathrm{C} \simeq 4300\,$\AA). 
We adopt the F435W waveband to match the rest-frame wavelength of the filter band used for the local starburst measurements in \cite{Xu2022}, namely the rest-frame NUV which traces recent star formation. Among the 86 sample galaxies in this study, 74 have the F435W coverage and therefore only those have sizes measured. The measurements are done in the following two steps. 

First, their half light radii are measured via aperture photometry using the {\sc photutils} package (\citealt{Bradley2020}) as follows. For each galaxy, a series of concentric elliptical apertures are placed, whose ellipticities and position angles are fixed to the values measured from \emph{HST} images by \cite{VanderWel2012}. Since we want the ellipses to trace the stellar structure in the galaxies, we choose to adopt the ellipticities and position angles measured in the longest available waveband of \emph{HST},  namely WFC3 F160W.  The half light radius is taken as the radius along the major axis of the ellipse enclosing half of the integrated galaxy flux. 

The half light radii measured above are adopted only if the measured values are larger than 0.\arcsec 30, which is three times the HST F435W point spread function (PSF). This is the case for 63 galaxies. For the remaining 11 smaller objects, the impacts of the PSF need to be taken into account. Therefore, we re-measure their sizes in the following second step. 

Considering that these 11 galaxies all have smooth and compact morphologies that can be fit by S\'ersic models, we use the {\sc statmorph} python package to perform S\'ersic fitting \citep{Rodriguez-Gomez2019}, which simultaneously models the PSF. Results are inspected to ensure that there are no substantial impacts from bright neighbors or image artifacts. We obtain the PSF-subtracted half-light radii from the fitting and adopt these values as the half light radii of these galaxies.

\subsubsection{Star Formation Rates (SFRs)}
\label{subsec:galprop_sfr}

SFRs are inferred from the rest-frame NUV luminosities $L(\mathrm{NUV})$, and are corrected for dust attenuation using the equation $\mathrm{SFR} = 3.6\times 10^{-10} L_{\mathrm{NUV}} \times 10^{0.4 A_\mathrm{NUV}}$, following e.g., \cite{Wang2018}.  In this equation, the unit of SFR is $M_\sun \mathrm{yr}^{-1}$ and the unit of $L_\mathrm{NUV}$ is $L_\sun$. The NUV attenuation value, $A_\mathrm{NUV}$, is calculated via spectral energy distribution (SED) fitting by \cite{Pacifici2012,Pacifici2015,Pacifici2016}. The coefficient on the right side of the equation, $3.6\times 10^{-10}$, is from the SFR calibration relation by \cite{Kennicutt2012} in which a \cite{Chabrier2003} initial mass function (IMF) is adopted.  The uncertainties on the SFRs are mainly due to the uncertainty of $A_\mathrm{NUV}$ from assumptions adopted in the SED fitting (see \citealt{Pacifici2023} for a discussion of the errors in SED fitting) and are typically 0.36 dex for the galaxies selected in this work. 
We also calculate the specific star formation rate, sSFR, as the SFR divided by the stellar mass.

We compare in Appendix \ref{appendix:sfrmass_cali} the SFR values calculated as described above with the values inferred from the rest-frame UV and infrared (IR) luminosities and from a different SED fitting tool, the BayEsian Analysis of GaLaxy sEds (BEAGLE; \citealt{Chevallard2016}). In brief, we find that the SFR values calculated in this work are consistent with those inferred using the two other methodologies, with systemic offsets smaller than 0.06 dex and scatters of around 0.26 dex among the different methods.

\subsubsection{SFR Densities $(\Sigma_\mathrm{SFR,\,avg}$ and $\Sigma_\mathrm{SFR,\,max})$}
\label{subsec:galprop_sfrd}

In this work, we use two types of SFR densities, $\Sigma_\mathrm{SFR,\,avg}$ and $\Sigma_\mathrm{SFR,\,max}$, each measured on a different spatial scale.

The quantity $\Sigma_\mathrm{SFR,\,avg}$ is measured by averaging over the half-light radius $r_e$ of a galaxy, which is measured along the major axis, so the spatial scale is equal to  $r_e$. The average half-light radius is about 4 kpc. We measure $\Sigma_\mathrm{SFR,\,avg}$, corrected for galaxy inclination to face-on, by dividing the integrated SFR by $2 \pi r_e^2$, where $r_e$ is the half-light radius along the major axis. 
The unit of $\mathrm{SFR}$ is $M_\sun/\mathrm{yr}$, and that of $r_e$ is kpc.   The unit of $\Sigma_\mathrm{SFR,\,avg}$ is then $M_\sun/\mathrm{yr}/\mathrm{kpc}^2$.

To characterize the SFR density of the most highly star-forming peaks in a galaxy, the quantity $\Sigma_\mathrm{SFR,\,max}$ is also introduced and measured. It is measured from a group of the brightest \emph{HST} F435W image pixels, and the spatial scale is equal to the radius of a circle with an area as large as the summed pixel area.  We denote the spatial scale for $\Sigma_\mathrm{SFR,\,max}$ as $r_\mathrm{max,\,SF}$; it has an average value of around 0.4 kpc. We choose to keep the measurement of $\Sigma_\mathrm{SFR,\,max}$ simple for the ease of interpretation, which is done in the following two ways.

For the first way, we select the top 10 brightest pixels in the F435W image of a galaxy.  The corresponding area is similar to that enclosed by the effective radius of a typical low-redshift starburst (\citealt{Berg2022}). Specifically, each pixel in the F435W image is 0.\arcsec03 on a side, corresponding to 0.21 kpc at $z=0.7$ and 0.25 kpc at $z=1.5$. The locations of the selected pixels for each galaxy are shown in Figures \ref{fig:galaxyroster_windnoagn4}-\ref{fig:galaxyroster_flaggednoagn}. The fluxes of these 10 pixels are summed and converted to a luminosity. A SFR is inferred from this NUV luminosity following the calibration given in \S \ref{subsec:galprop_sfr} and assuming an $A_V$ that is the same as the galaxy-integrated value.  The risks of this are discussed below.  This SFR is then divided by the total area of the 10 pixels  (de-projected to take into account the galaxy's inclination), which is around 0.6 kpc$^2$, to obtain the final SFR density.

For the second way, the SFR is measured from only the pixels in the F435W image  with estimated SFR densities above 3 $M_\odot$/yr/kpc$^2$. 
We then select pixels with SFR densities above 3 $M_\odot$/yr/kpc$^2$, and finally take the average as $\Sigma_\mathrm{SFR,\,max}$.

Assuming the dust correction for each pixel in a galaxy is the same as the galaxy-integrated value ($A_\mathrm{NUV}$ in \S\ref{subsec:galprop_sfr}), this correction may lead to overestimated or underestimated SFR densities because the rest-frame UV bright pixels may have different dust attenuation values compared to the galaxy average. \cite{Wang2017} shows that the SFGs at $z\sim 1$ have dust attenuations which vary by up to 1 mag in $A_V$, or equivalently 2.4 mag in $A_\mathrm{NUV}$, from their centers to their outskirts. Adopting this dust gradient, we estimate that the NUV dust attenuation of the pixels may differ from the galaxy average by up to $\pm1.2$ mag, corresponding to an uncertainty in the calculated SFR density of around $\pm0.5$ dex.

\section{Sample Selection} \label{sec:sample}

\begin{figure*}
	\centering
	\includegraphics[width = 7.1 in]{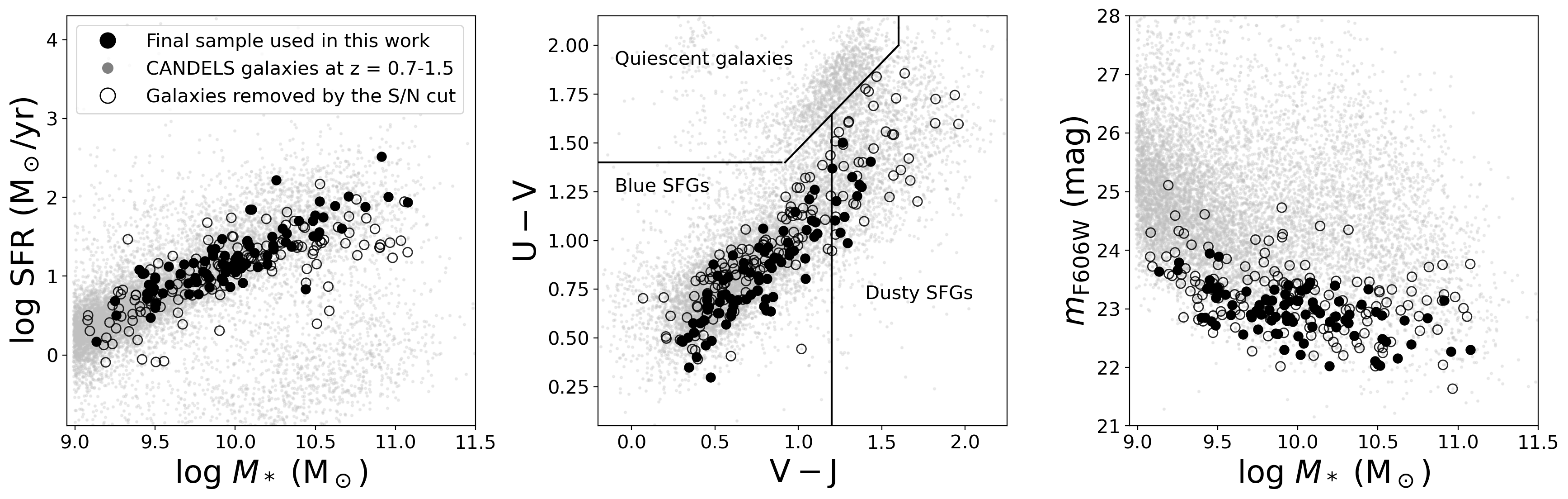}
	\caption{The galaxy sample studied in this paper is representative of star-forming galaxies (SFGs) at $z\sim1$.  We compare our sample (large black points) with galaxies in the CANDELS survey (small gray points) spanning the photometric redshift and stllar mass ranges of our sample: $0.7<z<1.5$ and $M_\star>10^{9}\,M_\sun$.  We also compare  with the galaxies in HALO7D which meet our selection criteria except that they do not pass our S/N $>$ 3 cut (open circles).    From left to right, we show the SFR vs. stellar mass relation, the UVJ diagram, and F606W magnitude vs. stellar mass.  The galaxies in our sample trace the CANDELS SFGs in the left panel. As shown in the middle and right panels, they are slightly bluer in their UVJ colors and brighter in F606W than the galaxies which have S/N $<$ 3. Note that a cut of F160W $<$ 26.5 mag is applied to the CANDELS galaxies to remove false galaxy detections, following common practice (e.g., \citealt{Santini2015}).}\label{fig:halo7dsample}
\end{figure*}

\begin{figure}
\centering
\includegraphics[width = 3.1 in]{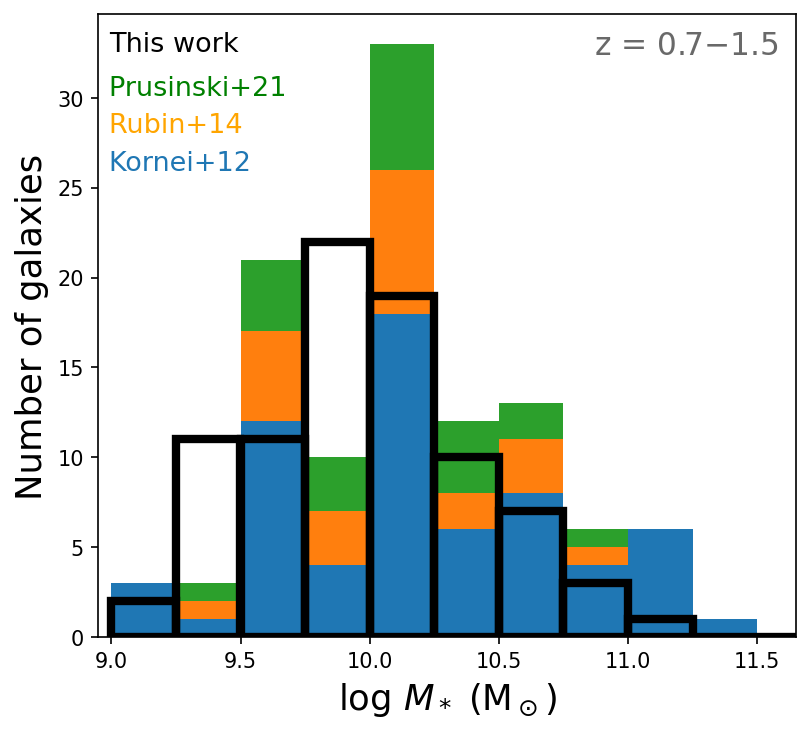}
\caption{Our galaxy sample is similar in mass but includes somewhat more low-mass galaxies than previous studies of galactic winds at $z\sim 1$: \citet{Prusinski2021}, \citet{Rubin2014}, and \cite{Kornei2012}.  Our sample (black) includes 13 SFGs with stellar masses below $10^{9.5}\,M_\odot$, two times as many as the galaxies (6 in total) in the same mass range from three previous works. Regarding the literature studies, only galaxies within the same redshift range of our sample, namely $0.5<z<1.5$, are included. \label{fig:sample_compare_literature}}
\end{figure}

A total of 86 galaxies are selected from the HALO7D survey according to the following criteria, and we refer to these galaxies as the $z\sim1$ SFG sample for the rest of this paper:

\begin{enumerate}
	
    \item Targets are required to be in the CANDELS fields. This ensures the appropriate ancillary data needed for the measurements of galaxy properties. Out of the 2722 targets in HALO7D, a total of 1885 are in CANDELS.

    \item To remove stars, we require that the Source Extractor (\citealt{Bertin1996}) stellarity indicator CLASS\_STAR from the CANDELS catalog is less than 0.8. This removes 315 objects, leaving 1570.
	
    \item  We require that a spectroscopic redshift was able to be measured by \cite{Pharo2022}. \cite{Pharo2022} measured spectroscopic redshifts from the HALO7D spectra using the rest-frame optical spectral lines and found that the galaxies for which the redshifts could not be measured have relatively short exposure times ($\lesssim 4$ hours), F606W magnitudes fainter than 24 mag, and stellar masses lower than $10^{9}\,M_\sun$. This removes 626 galaxies, leaving 944. 
 
    \item The observed wavelengths of the \ion{Mg}{2} $\lambda\lambda$ 2796/2803\,\AA\ lines and the [\ion{O}{2}] $\lambda\lambda$ 3727/3729\,\AA\ lines are required to fall within the wavelength ranges of HALO7D's DEIMOS spectra. In order to secure accurate wavelength solutions, we also require that the observed wavelengths of the \ion{Mg}{2} lines are greater than 4680\,\AA, which is the wavelength of the bluest calibration lamp line. As a result of these two requirements, the sample is confined to a redshift range of $z=0.7-1.5$ and 490 galaxies remain.
	
    \item SFGs are selected according to their rest-frame U-V and V-J colors using the redshift-dependent criteria created by \cite{Williams2009}. Measurements of the colors are described in \S\ref{subsec:galprop_rfcolor}.  A total of 443 galaxies remain after this cut.
	
    \item Galaxies are selected to have stellar masses above $10^9\ M_\sun$. Galaxies with lower masses than this cut-off are not included because of the relatively high incompleteness caused by the spectroscopic redshift cut made in \#3. A total of 343 galaxies remain after this cut.
	
	\item The 2D spectra are visually inspected to be free from artifacts near the Mg II lines and also at least one of the following emission lines: the [\ion{O}{2}] doublet, [\ion{O}{3}] 4959\,\AA, [\ion{O}{3}] 5007\,\AA, or H\,$\beta$.  This means that these lines land on the detector 
    and are not contaminated by sky lines or light from a nearby source.  A total of 70 galaxies are removed due to the requirements on the line locations on the detector and 14 galaxies are removed due to contamination.  A total of 259 galaxies remain after this cut.
	
    \item It is required that the S/N of the continuum of the co-added 1D spectrum of each galaxy be greater than 3.0. The continuum S/N is measured within a rest-frame wavelength range of 50\,\AA\ centered on the \ion{Mg}{2} lines. A total of 108 galaxies remain after this cut. 

    \item Finally, AGNs are removed from the sample. They are identified as follows. First, we look for X-ray sources in the archival \emph{Chandra} data and we take these to be AGNs. We obtain the X-ray source catalogs from the literature \citep{Xue2011,Nandra2015,Civano2016,Xue2016, Luo2017} and match them to our galaxies using a matching radius of 1.\arcsec\ A total of 21 galaxies are identified via this step and removed. Second, we identify infrared AGNs following the criteria by \cite{Donley2012}. One galaxy is identified, and the galaxy also has an X-ray detection. Finally, we visually inspect the \ion{Mg}{2} lines and identify those with very broad emission (FWHM$>$1000 km/s), which signifies the broad line region of an AGN (e.g., \citealt{McLure2002}). A total of 5 broad line AGNs are found, including the object identified to be both an infrared AGN and an X-ray AGN and three other objects identified to be X-ray AGNs.
    A total of 22 galaxies are removed and 86 galaxies remain after this final cut.
	
\end{enumerate}

\subsection{Sample Characteristics}

The final sample contains 86 galaxies which have redshifts between 0.7 and 1.5 with a median at 0.96. In Figure \ref{fig:halo7dsample}, our sample is compared with CANDELS galaxies within the same redshift range and with stellar masses greater than $10^{9}\,M_\sun$ (criterion 6). Our sample is shown as large black points, and the CANDELS galaxies as small gray points.  Galaxies that were removed by the S/N cut (criterion 8) are shown as open circles.  The figure consists of three panels: the ``star-forming main sequence" diagram (SFR vs.\ $M_\star$), a rest-frame color-color diagram ($\mathrm{U-V}$ vs.\ $\mathrm{V-J}$), and a magnitude-mass diagram (F606W magnitude vs.\ $M_*$).  The F606W waveband corresponds to 2400--3600\AA\ in the rest-frame.  From this figure, we conclude that our final galaxy sample traces the star-forming and F606W-bright CANDELS galaxies.  The galaxies removed due to the S/N cut are less star-forming, less massive, more dusty, and fainter in F606W. 

The stellar mass distribution of galaxies in our sample is compared with that of other studies which also measured winds of individual SFGs via UV absorption lines \citep{Kornei2012,Rubin2014,Prusinski2021} in Figure \ref{fig:sample_compare_literature}. 
For this comparison, we consider only galaxies at $z$=0.7--1.5 from these literature studies. From this figure, it is concluded that the galaxy sample in this paper is similar in stellar mass distribution to the literature studies. Additionally, it includes 13 SFGs below $10^{9.5}\,M_\odot$, twice as many as the galaxies (6 in total) in the same mass range from the literature. 

\section{Measuring Wind Velocities for the $\lowercase{z}\sim1$ SFG Sample} \label{sec:analysis}

\begin{figure*}
\centering
\includegraphics[width = 4.5 in]{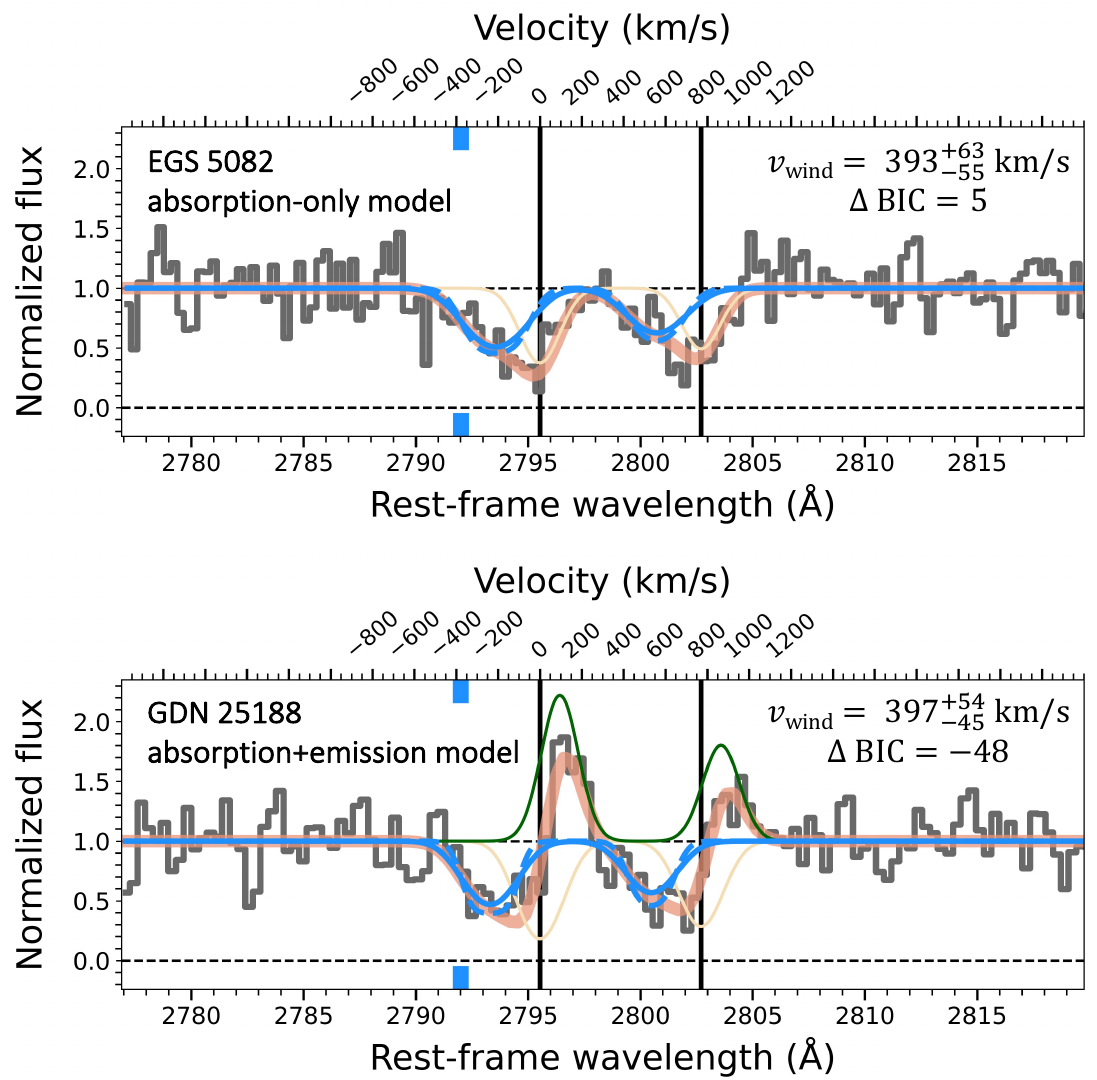}
\caption{Example fits to the Mg II doublet for two galaxies are shown, one which prefers the absorption-only model (top, $\Delta$BIC=5) and one which prefers the absorption+emission model (bottom, $\Delta$BIC=-48). Fits for all the galaxies in the sample are in Figures 10--25. 
Both spectra have S/N values of $\sim5$ which is typical of the sample galaxies.  The observed spectra are plotted as gray lines and the best-fit models are overplotted as thick pale orange lines. 
The systemic absorption component of the models is shown as a thin khaki line.  The blueshifted absorption component of the models is shown as a thin solid blue line. 
Barely visible are the thin dashed blue lines which represent the intrinsic profiles of the wind components prior to convolving with the line spread function. The blue vertical bars at the top and bottom of the plots mark the wavelengths where the intrinsic wind profiles reach half the maximum depth, which are used to infer the wind velocities. The emission components are shown as thin green lines in the bottom panel. The two black vertical lines spanning the height of the plots mark the locations of zero velocity for the \ion{Mg}{2} lines.  
\label{fig:linefittingexamples}}
\end{figure*}

Wind velocities are measured by fitting models to the \ion{Mg}{2} 2796 \AA\  line profiles, as described below. Additional details can be found in \cite{Wang2022}. 

\subsection{Refining the Redshift Measurements}
\label{subsec:analysis_specz}

Spectroscopic redshifts of the HALO7D galaxies were measured by fitting several major spectral lines in the rest-frame UV and optical simultaneously by \cite{Pharo2022}. We refine the redshift measurements by fitting the [\ion{O}{2}] $\lambda\lambda$\,3727/3729\,\AA\ line profiles following \cite{Wang2022}. They differ by less than $6\times10^{-4}$ in terms of $\Delta z / (1+z)$. 
The redshifts measured from the [\ion{O}{2}] lines are taken as the systemic galaxy redshifts, and they are used to convert the galaxy spectra from the observed frame to the rest frame before the wind measurements are conducted.

\subsection{Fitting the \ion{Mg}{2} Line Profiles}
\label{subsec:analysis_fitting}

For each galaxy, two models are used to fit its \ion{Mg}{2} doublet, one which includes only the absorption component (hereafter ``absorption-only model'') and another which includes both absorption and emission components (hereafter ``absorption+emission model''). For both models, the optical depth of the \ion{Mg}{2}\,2803\,Å line is set to 2 times that of the \ion{Mg}{2}\,2796\,Å, as determined by atomic physics (\citealt{Spitzer1978}). 

For the absorption-only model, we fit each line of the \ion{Mg}{2} doublet with two kinematic components: the first component is centered at line-of-sight velocity $v = 0$ km/s and the second at $v<0$ km/s. The former represents absorption by the interstellar medium (ISM) within the galaxy and the latter is caused by outflowing gas clouds in the wind. For each component, the optical depth ($\tau$) is assumed to be a Gaussian function of $v$.  For the wind component, the covering factor ($C_f$) is a free parameter in the fit. It quantifies the fraction of sightlines blocked by the line-absorbing gas. 

The spectral profile of each Mg II line from this model is as follows (same as in \citealt{Wang2022}):
\begin{equation}
F_\mathrm{abs-only}(v) =  [1-C_f+C_f\cdot e^{-\tau_\mathrm{wind}(v)}] \cdot e^{-\tau_\mathrm{\,ISM}(v)}, \label{eqn:fluxprofile_absonly}
\end{equation}
where $F_\mathrm{abs-only}$ is the continuum-normalized flux as a function of $v$, 
$\tau_\mathrm{wind}(v) = \tau_{c,\, \mathrm{wind}}\cdot e^{-(v-v_{c,\,\mathrm{wind}})^2/(2\sigma_\mathrm{wind}^2)},$ 
and 
$\tau_\mathrm{\,ISM}(v) = \tau_\mathrm{c,\ ISM}\cdot e^{-v^2/(2\sigma_\mathrm{\,ISM}^2)}$. 
A total of six free parameters are used in this model: $v_\mathrm{c,\,wind}$, which is the centroid velocity of the wind component, $\sigma_\mathrm{wind}$, which is the velocity dispersion of the wind component, $C_f$, which is the covering fraction of the wind component, $\tau_\mathrm{c,\,wind\,2796}$, which is the peak optical depth of the wind component for \ion{Mg}{2}\,2796\,Å, $\sigma_\mathrm{ISM}$, which is the velocity dispersion of the ISM component, and $\tau_\mathrm{c,\,ISM\,2796}$, which is the peak optical depth of the ISM component for \ion{Mg}{2}\,2796\,Å.

\begin{figure*}
	\centering
		\includegraphics[width = 7.0 in]{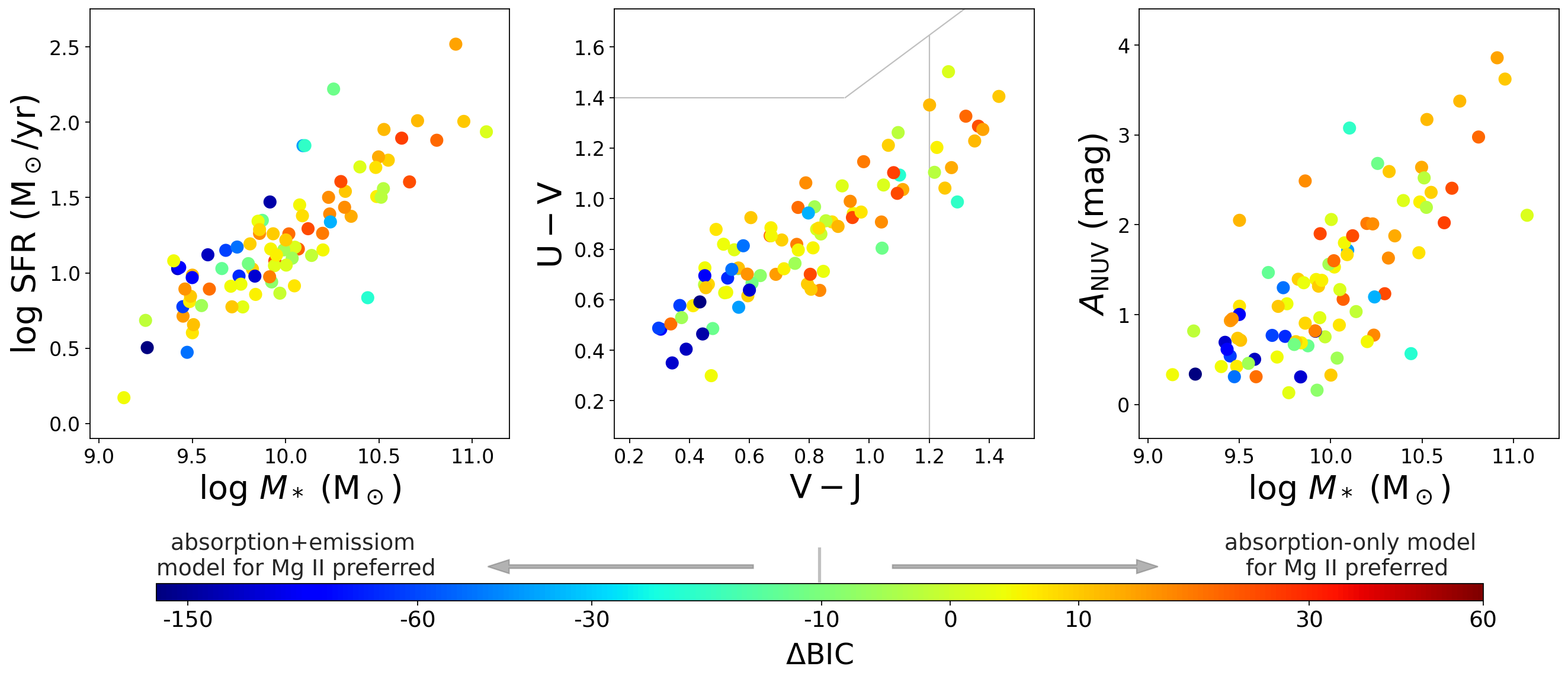}
	\caption{Properties of galaxies with and without \ion{Mg}{2} emission differ. From left to right, the SFR versus $M_\star$ diagram, $\mathrm{U-V}$ versus $\mathrm{V-J}$ diagram, and UV dust attenuation ($A_\mathrm{NUV}$) versus $M_\star$ diagram are shown.  Galaxies with emission, which correspond to those with $\Delta\mathrm{BIC}\geq -10$, generally have lower stellar masses and SFRs (left panel), bluer $\mathrm{U-V}$ and $\mathrm{V-J}$ colors (middle panel), and lower $A_\mathrm{NUV}$ (right panel) than the rest of the sample. The $\Delta$BIC statistic is indicated by the colorbar and shows how strongly the absorption-only model or the absorption+emission model is preferred for fitting the observed \ion{Mg}{2} lines. The lower the $\Delta$BIC value (bluer points), the more the absorption+emission model is preferred; the higher the value (redder points), the more the absorption-only model is preferred. We choose to adopt the absorption-only model if $\Delta\mathrm{BIC}\geq -10$ and the absorption+emission model if $\Delta\mathrm{BIC}<-10$. \label{fig:bic_vs_galprop}}
\end{figure*}

For the absorption+emission model, we simply add emission components to the absorption-only model, one for each line.  These components are defined as Gaussians with centers in the range $-100\ \mathrm{km/s} < v < 100\ \mathrm{km/s}$ \citep{Zhu2015}. The Gaussians are set to be at the same centroid velocity and have the same width, and the ratio of their amplitudes is confined to a range from 1:1, corresponding to an optically thick limit, to 2:1, corresponding to an optically thin limit (see \citealt{Prochaska2011}).  

The spectral profile of each Mg II line from the absorption+emission model is as follows:
\begin{equation}
F_\mathrm{abs+emi}(v) = F_\mathrm{abs-only}(v) + A_\mathrm{\,emi}\cdot e^{-(v-v_\mathrm{c,\,emi})^2/(2 \sigma_\mathrm{c,\,emi}^2)}, \label{eqn:fluxprofile_emiabs}
\end{equation}
where $F_\mathrm{abs-only}$ is defined in Equation \ref{eqn:fluxprofile_absonly} and the second term on the right-hand side is the emission component. 
With the emission components included, a total of ten free parameters are used in this model: six parameters from the absorption-only model, the centroid velocity and velocity dispersion of the emission components ($v_\mathrm{c,\,emi}$, $\sigma_\mathrm{c,\,emi}$), and the amplitudes of the two Gaussian components ($A_\mathrm{\,emi\,2796}$, $A_\mathrm{\,emi\,2803}$).

We conduct the fitting using the Markov Chain Monte Carlo python package, {\sc emcee} \citep{Foreman-Mackey2013,Foreman-Mackey2019} with the same setup as in \cite{Wang2022}. The line profile models are first convolved with the instrumental line spread function and then fit to the observed line profiles. 
We generate a ``best-fit'' \ion{Mg}{2} line profile by adopting the 50th percentile value in the posterior distribution of each free parameter.

Fits for two example galaxies are shown in Figure \ref{fig:linefittingexamples}: one galaxy fit with the absorption-only model (top panel) and one fit with the absorption+emission model (bottom panel). Fits for all the galaxies are shown in Figures 10--25.  In these plots, observed spectra are shown in gray and the best-fit profiles as thick pale-orange lines. The absorption components of the model profiles are shown in blue and khaki, for the blueshifted wind component and the systemic component, respectively, and the emission components in green.

\begin{figure*}
\centering
\includegraphics[width = 7.0 in]{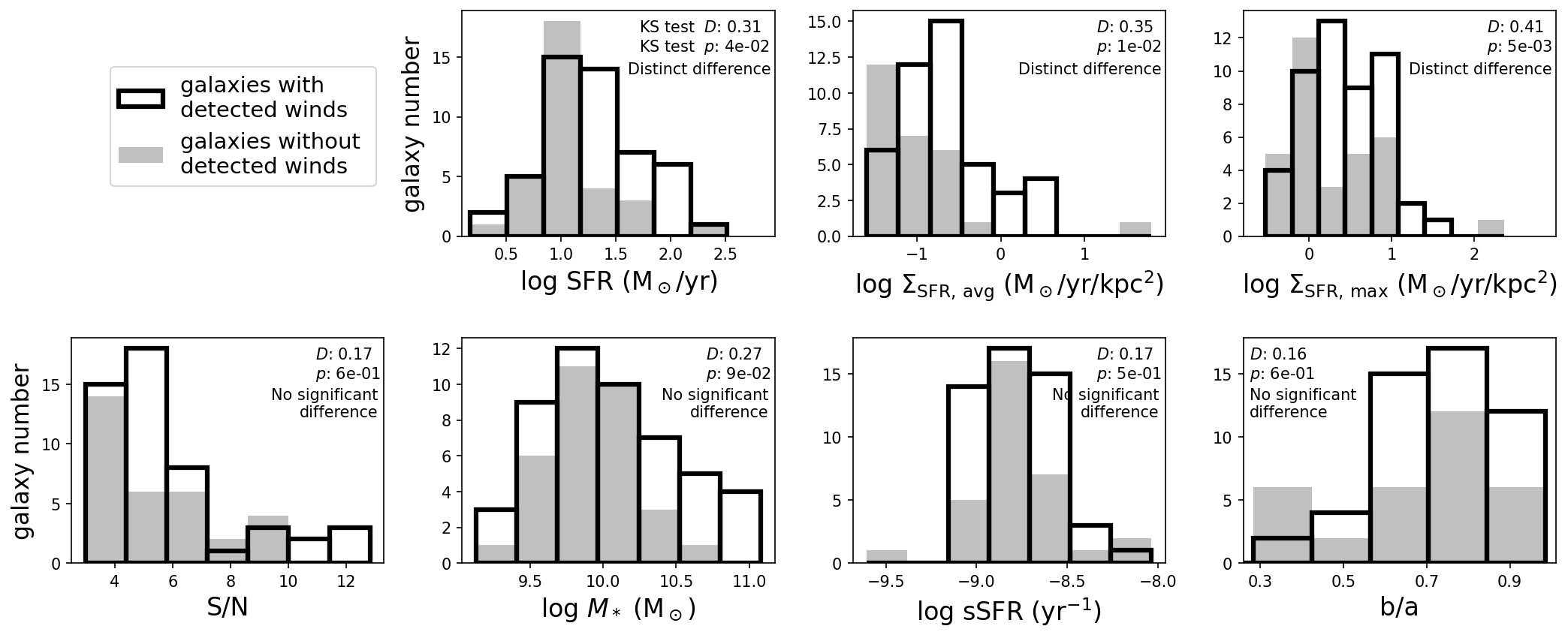}
\caption{The distributions of properties of galaxies with and without winds are shown as clear and filled histograms, respectively. The histograms in each panel are compared via a Kolmogorov–Smirnov test. Values of the resulting $D$ and $p$ statistics are indicated in each panel. Statistically distinct histograms($p<0.05$) are in the top row and those which are indistinguishable ($p\geq0.05$) are in the bottom row.  Whether a galaxy has a detected wind or not depends on SFR densities and SFR but not significantly on $M_\star$, sSFR, or S/N. No significant dependence is found for the axis ratio $b/a$ either, but this is likely because our sample is confined to a narrow range of high $b/a$ values (i.e., nearly face-on galaxies). \label{fig:winddetection_vs_galprop}}
\end{figure*}

\subsection{Identifying Galaxies with Mg II Emission and Selecting Best-fit Mg II Models}

\label{subsec:analysis_emissionvsabsorption}
To determine which model to adopt for a given galaxy, absorption-only or absorption+emission, we adopt the Bayesian Information Criterion (BIC; \citealt{Schwarz1978}). The BIC is defined here as $\mathrm{BIC} = \chi^2 + p\ln(n)$, where $\chi^2$ is the chi-squared of the fit, $p$ is the number of free parameters in a line profile model, and $n$ is the number of independent spectral resolution elements in the observed line profile. For each galaxy, we compute the BIC values of the two models and calculate a quantity $\Delta$BIC which is defined as the BIC of the absorption+emission model minus that of the absorption-only model.  We adopt the absorption+emission model if the $\Delta$BIC is smaller than $-10$, a criterion indicative of strong statistical confidence (\citealt{Liddle2007}) and commonly adopted in the literature for line profile fitting (e.g., \citealt{Swinbank2019,Avery2022,Concas2022}). Otherwise, the absorption-only model is adopted. 
We show the observed \ion{Mg}{2} line profiles of individual galaxies along with the adopted best-fit line profiles for two example galaxies in Figure \ref{fig:linefittingexamples}.
A visual inspection of the line profiles of all the galaxies in the sample in Figures 10--25 confirms that galaxies with smaller $\Delta$BICs have more prominent \ion{Mg}{2} emission. Additionally, for each model, we perform a visual inspection and flag the ones that are likely driven by spurious noise features in the observed spectra. A total of 4 galaxies are flagged, which only account for $<5\%$ of the total sample and are shown in Appendix \ref{appendix:mg2lineprofiles}. These four galaxies are not shown in the figures presented in the next two sections or included when fitting the relations between the wind velocity and galaxy properties. 

In Figure \ref{fig:bic_vs_galprop}, the galaxies for which the absorption+emission models are preferred, i.e., $\Delta\mathrm{BIC}<-10$, generally have lower masses ($<10^{10}\,M_\odot$; left panel), bluer $\mathrm{U-V}$ and $\mathrm{V-J}$ colors ($<0.8$ and $<0.6$, respectively; middle panel), and lower $A_\mathrm{NUV}$ ($<1\,$mag; \S \ref{subsec:galprop_sfr}; right panel) than the galaxies for which the absorption-only model is preferred. This is consistent with previous studies (e.g., \citealt{Weiner2009,Rubin2010,Erb2012,Martin2012,Kornei2013,Zhu2015,Feltre2018}).

\begin{figure*}
\centering
\includegraphics[width = 5.0 in]{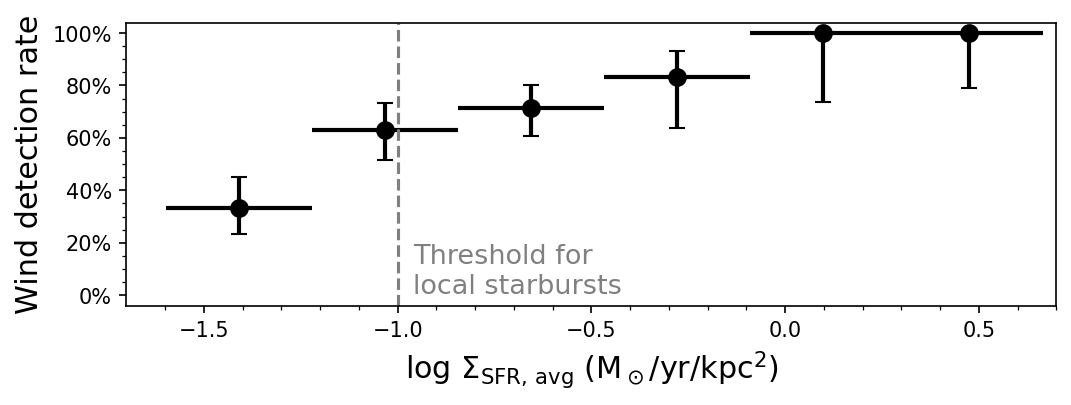}
\caption{The wind detection rate of $z\sim1$ SFGs is shown as a function of $\Sigma_\mathrm{SFR,\,avg}$. The SFGs do not show a distinct threshold in SFR density below which no galaxies have winds, whereas such a threshold exists for local starbursts at 0.1$\,M_\odot$/yr/kpc$^2$ (vertical line; \citealt{Heckman2015}). Instead, for the SFGs, the detection rate shows a gradual decline at this value.  \label{fig:winddetectionrate_vs_sfrd}}
\end{figure*}

\begin{deluxetable*}{lllll}
\tablewidth{0.9\textwidth}
\tablecaption{Best-fit relations between $v_\mathrm{wind}$ and galaxy properties for our $z\sim 1$ SFG sample (first 5 rows) and for the combined sample of $z\sim 1$ SFGs \& $z\sim 0$ starbursts (last row) \label{tab:windfitting}}
\tablehead{
\colhead{Relation} & \colhead{Slope $\alpha$} & \colhead{Intercept $\beta$} & \thead{$p$-value, \\ significance} & \colhead{Measurement range} 
}
\startdata
$\log v_\mathrm{wind} = \alpha\cdot\log\mathrm{SFR}+\,\beta$ & $0.05^{+0.05}_{-0.05}$ & $2.54^{+0.06}_{-0.07}$ & 0.09, 1.7$\sigma$  & $\log\,$SFR = [0.2, 2.5] \\  
$\log v_\mathrm{wind} = \alpha\cdot(\log M_{\star}-9)$+$\,\beta$ & $0.06^{+0.05}_{-0.05}$ & $2.55^{+0.05}_{-0.05}$ & 0.11, 1.6$\sigma$  & $\log M_{\star}$ = [9.1, 11.1] \\ 
$\log v_\mathrm{wind} = \alpha\cdot(\log\mathrm{sSFR}+9)+\beta$ & $-0.07^{+0.12}_{-0.13}$ & $2.63^{+0.03}_{-0.03}$ & 0.68, 0.4$\sigma$  & $\log\,$sSFR = [$-9.1$, $-8.2$] \\ 
$\log v_\mathrm{wind} = \alpha\cdot\log\Sigma_\mathrm{SFR,\,avg}$+$\,\beta$ & $0.13^{+0.04}_{-0.04}$ & $2.69^{+0.03}_{-0.03}$ & 0.003, 3.0$\sigma$  & $\log\Sigma_\mathrm{SFR,\,avg}$ = [$-1.6$, 0.4] \\ 
$\log v_\mathrm{wind} = \alpha\cdot\log\Sigma_\mathrm{SFR,\,max}$+$\,\beta$ & $0.14^{+0.04}_{-0.05}$ & $2.55^{+0.03}_{-0.02}$ & 0.012, 2.5$\sigma$  & $\log\Sigma_\mathrm{SFR,\,max}$ = [$-0.5$, 1.4] \\ %
\hline 
\multicolumn{5}{c}{Fit to the combined sample of $z\sim 1$ SFGs and $z\sim 0$ starbursts}\\
$\log v_\mathrm{wind} = \alpha\cdot\log\mathrm{SFR}$+$\,\beta$  & $0.16^{+0.03}_{-0.03}$ & $2.41^{+0.04}_{-0.04}$ & 0.0004, 3.5 $\sigma$ &  $\log\mathrm{SFR}$ = [$-1.4$, 2.5] \\ %
\enddata
\tablecomments{The $v_\mathrm{wind}$, SFR, $M_\star$, and sSFR are in units of km/s, $M_\sun$/yr, $M_\sun$, and yr$^{-1}$, respectively. The SFR densities have units of $M_\sun$/yr/kpc$^2$. The uncertainties are 1-$\sigma$ ranges. The $p$-values and significance values are calculated from the Spearman rank test.}
\end{deluxetable*}

\subsection{Determining Whether a Galaxy Has a Wind \& Comparing the Properties of Galaxies With versus Without Winds}

\label{subsec:analysis_winddetect}

We detect winds using the best-fit model favored by the BIC criterion. For each galaxy, we identify the blueshifted absorption component of the best-fit \ion{Mg}{2} 2796\,\AA\ line profile. These are shown as blue dashed lines in Figure \ref{fig:linefittingexamples} for two example galaxies, whereas the full sample are shown in Figures 10--25.  We then calculate the maximum depth of the blueshifted component.
To avoid false detections due to noise, a galaxy is determined to have a wind if the maximum depth of its continuum-normalized spectrum is greater than 0.3 in normalized flux units.  We choose this threshold value so that any wind is detected with at least a 2-$\sigma$ significance (i.e., $>95$\%): The S/N of our spectra is at least 3 per pixel so the standard error per resolution element (5 pixels) is at most 0.15 (=1/(S/N)/$\sqrt{5}$) in normalized flux units, and hence the required depth value of 0.3 for detection will be at least two times this standard error value. Among the 86 sample galaxies, 50 are found to have winds, corresponding to a wind detection rate of 58\%. 

After identifying galaxies with and without winds, we examine whether wind detectability depends on galaxy properties and S/N. 
We do so by comparing the distributions of galaxies with and without winds among various galaxy properties and S/N.  A Kolmogorov–Smirnov test is used.
In each panel of Figure \ref{fig:winddetection_vs_galprop}, we show the distributions of galaxies with and without winds in terms of galaxy properties and S/N. Based on $p$-values from the Kolmogorov–Smirnov test, we show that wind detection depends on SFR and SFR density (integrated and those of UV-bright pixels) (top row), but not significantly on the S/N, $M_\star$, sSFR, or axis ratio $b/a$  (bottom row). The relations between winds and galaxy properties will be discussed further in Sections \ref{sec:result} and \ref{sec:result2}. 

The lack of dependence of wind detectability on the axis ratio $b/a$  is probably because the sample is confined to a relatively narrow range of high $b/a$ values, namely $b/a>0.5$ or low inclinations. As expected from bi-conical outflows with wide opening angles and also found by previous observations (e.g., \citealt{Rubin2014}), the wind detection rate drops significantly below $b/a=0.5$, which is poorly traced by our sample. 

In Figure \ref{fig:winddetectionrate_vs_sfrd}, we show the wind detection rate as a function of $\Sigma_\mathrm{SFR\,avg}$ and find that there is no distinct SFR density threshold below which a $z \sim 1$ SFG does not have a wind. Such a threshold was previously found for local starbursts at 0.1$\,M_\odot$/yr/kpc$^2$ (\citealt{Heckman2015}). However, for the SFGs, we do find that the wind detection rate gradually decreases with decreasing SFR density. In other words, the SFGs with detected winds tend to have SFR density values higher than 0.1$\,M_\odot$/yr/kpc$^2$. The wind detection rate is 100\% at $0<\log\,\Sigma_\mathrm{SFR,\,avg}<1$, 74\% at $-1<\log\,\Sigma_\mathrm{SFR,\,avg}<0$, and drops to 41\% at $\log\,\Sigma_\mathrm{SFR,\,avg}<-1$. The wind detection rates are calculated from the galaxy numbers in each bin in the top middle panel of Figure \ref{fig:winddetection_vs_galprop} and their error bars (1-$\sigma$) are calculated following \cite{Cameron2011}.

\begin{figure*}
	\centering 
	\includegraphics[width = 5.3in]{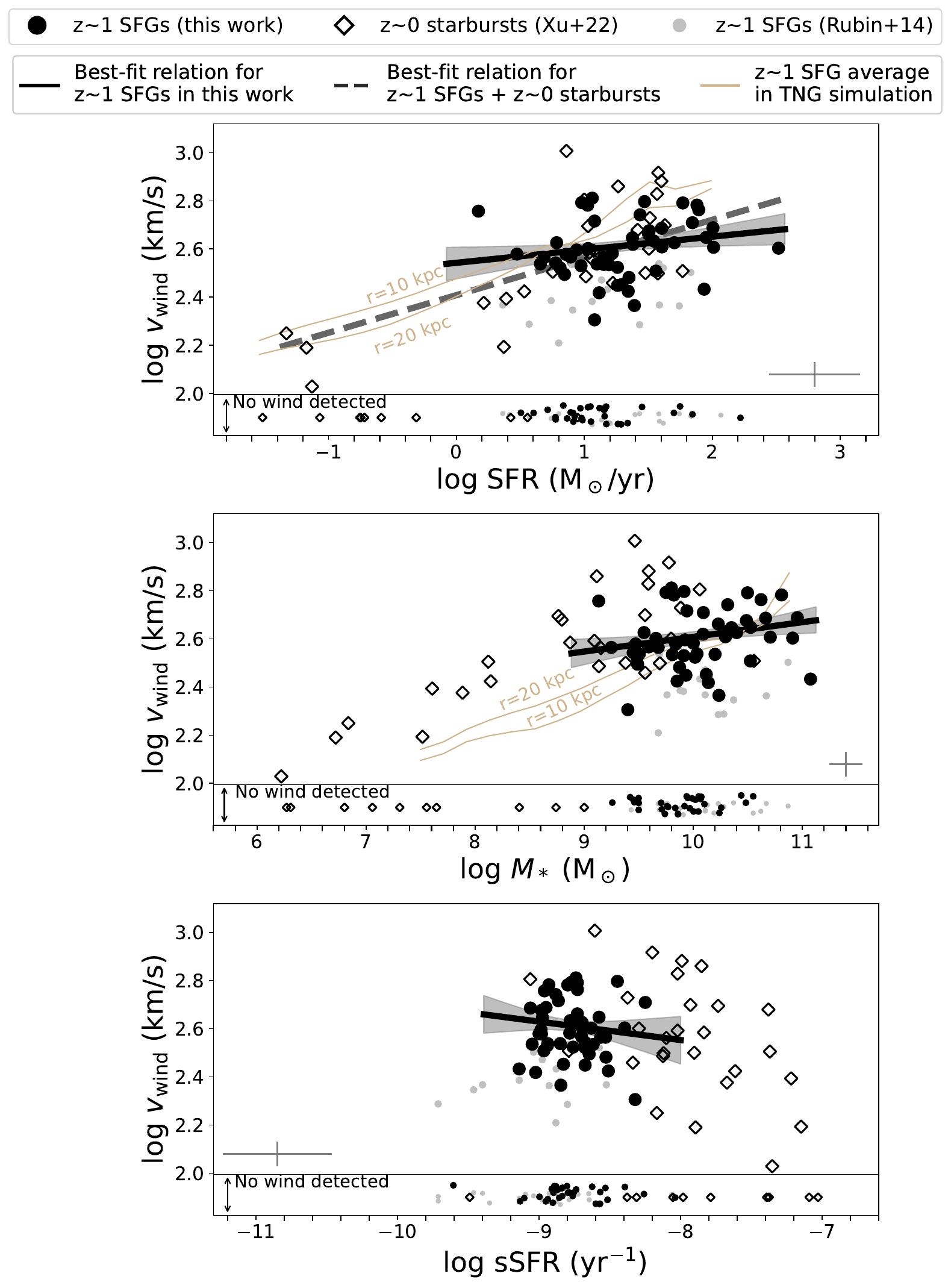}
	\caption{The wind velocities ($v_\mathrm{wind}$) of our $z\sim 1$ SFGs (filled black circles) are shown as a function of SFR (top), stellar mass (middle), and sSFR (bottom). The $v_\mathrm{wind}$ of the $z\sim 1$ galaxies shows weak correlations (1.7-$\sigma$ and 1.6-$\sigma$ significance) with SFR and mass and no correlation with sSFR. The best-fit relations are shown as solid black lines and are given in Table \ref{tab:windfitting}. We also plot the $z\sim 1$ SFGs from \citealt{Rubin2014} (small gray dots) and $ z\sim 0$ starbursts from \cite{Xu2022} The measurements by \cite{Rubin2014} are lower by $\sim$0.1 dex in $v_\mathrm{wind}$ compared to ours, which may be caused by different ways of measuring $v_\mathrm{wind}$ (Appendix  \ref{sec:literaturemeasurements}). Interestingly, the $z \sim 0$ starbursts appear to follow a similar relation to that of our $z\sim 1$ SFGs on the $v_\mathrm{wind}$--SFR diagram (top panel), but not for other relations. A fit to the combined sample of the $z \sim 0$ starbursts and our $z\sim 1$ SFGs results in the dashed line in the top panel (parameters given in Table 1). Remarkably, galaxies at $z\sim 1$ from the Illustris-TNG simulations (khaki lines; \citealt{Nelson2019}) follow this combined relation. Although simulated galaxies lie to lower $v_\mathrm{wind}$ at a given mass, this is not unexpected given how different the wind measurement is conducted (\S \ref{subsec:result_sfr}). 
    Galaxies with no detected winds are shown at the bottom of the panels. Typical measurement errors are shown as gray cross-hairs. \label{fig:vwind_vs_masssfr_compare}}
\end{figure*}

\subsection{Measuring Wind Velocities}

\label{subsec:analysis_vwindmeasure}

Next, for the 50 galaxies that have winds, we measure their wind velocities from the best-fit wind absorption components (blue dashed lines in Figures \ref{fig:linefittingexamples} \& 10--25). We measure the wavelength where the best-fit wind component reaches 50\% of its maximum depth, which we call $\lambda_\mathrm{wind,\,2796}$. This quantity is shown as thick blue vertical bars in Figures \ref{fig:linefittingexamples} \& 10--18. We calculate the wind velocities from it as 
    $v_\mathrm{wind} = c \cdot (1 - \lambda_\mathrm{wind,\,2796}/ \lambda_\mathrm{rest,\,2796}), \label{eqn:vwindandlambda}$
where $c$ is the speed of light and $\lambda_\mathrm{rest,\,2796}$ is the rest-frame wavelength of the \ion{Mg}{2} 2796\,\AA\ line, i.e., 2795.528 \AA. 
To calculate the uncertainty on the wind velocity for each galaxy, we perform a Monte Carlo analysis by repeating the velocity measurement $8\times10^3$ times. Each time, we sample the posterior distributions of the parameters of the fit, use the sampled parameter values to generate a new line profile, and measure a wind velocity from the new profile. The 16th and 84th percentiles of the wind velocity are quoted as its 1-$\sigma$ bounds. Typical 1-$\sigma$ uncertainties of the velocities are $\pm 60$ km/s.

The wind velocities of our sample galaxies range 200--700 km/s with a median of 397 km/s. These values are within the ballpark of the wind velocities of the $z\sim 1$ SFGs measured by other studies. We also note that the way a wind velocity in this work, namely by calculating the velocity shift at 50\% of the maximum absorption line depth, is similar to how the quantity $v_\mathrm{max}$ is measured in the literature (e.g., \citealt{Heckman2016,Xu2022}).

\section{Wind Velocity as a Function of SFR, Stellar Mass, and sSFR} 
\label{sec:result}
\label{subsec:result_z1measurements}

The wind velocities as a function of SFR, $M_\star$, and sSFR for our sample of $z \sim 1$ SFGs are presented in the top, middle, and bottom panels of Figure \ref{fig:vwind_vs_masssfr_compare}, respectively, as filled black circles. A Spearman correlation test and a log-linear fit are conducted for each of the relations, the results of which are given in Table \ref{tab:windfitting}. The best-fit relations are shown as thick solid lines in Figure \ref{fig:vwind_vs_masssfr_compare}, and the 1-$\sigma$ uncertainty is indicated by the gray shading. 
The fits follow the recipes described in \cite{Weiner2006} and \cite{Hogg2010}, in which  measurement uncertainties and the impact of outliers are taken into account. The galaxies without wind detections, shown at the bottom of the figure, are not included in the fits.  The intrinsic scatter of $v_\mathrm{wind}$ at a given SFR, $M_\star$, and sSFR is substantial, around 0.1 dex or equivalently 100 km/s.  Each of these relations is discussed in more detail below. We also compare with a study of compact starbursts at $z\sim0$ (\citealt{Xu2022}) and a study of SFGs at $z\sim1$ (\citealt{Rubin2014}), the measurements of which are described in detail in Appendix \ref{sec:literaturemeasurements}.

\subsection{Wind Velocity as a Function of SFR}
\label{subsec:result_sfr}

\subsubsection{Main Results}

As shown in the top panel of Figure \ref{fig:vwind_vs_masssfr_compare}, a weak positive correlation is found between $v_\mathrm{wind}$ and SFR. It has a Spearman $p$-value of 0.09, corresponding to a 1.7-$\sigma$ significance. The best-fit relation from the log-linear fit is $v_\mathrm{wind} \propto \mathrm{SFR}^{0.05}$ (Table \ref{tab:windfitting}). 

Furthermore, we compare our $z\sim1$ SFGs with the $z\sim0$ starbursts from \cite{Xu2022}, which are shown as open diamonds in the figure. 

Intriguingly, we find that the SFGs and starbursts appear to follow a unified $v_\mathrm{wind}$ versus SFR relation. 
In the SFR range where the SFGs and starbursts overlap (log SFR = [0.5, 1.5]), the $v_\mathrm{wind}$ medians of the two samples differ by less than 0.05 dex or 40 km/s. The physical implications of this finding are discussed in \S\ref{sec:discussion}. 

We fit the SFG sample and starburst sample together to derive a unified relation: $v_\mathrm{wind} \propto \mathrm{SFR}^{0.16}$, which is shown as the dashed line in the figure. The corresponding Spearman correlation $p$-value is 0.004, corresponding to a 3.5-$\sigma$ significance. We validated that this fit and correlation are not driven by three low-SFR objects below $\log\,\mathrm{SFR}\, [M_\sun/\mathrm{yr}] = 0$ in the starburst sample. If excluding them, the best-fit relation becomes $v_\mathrm{wind} \propto \mathrm{SFR}^{0.12}$ and the $p$-value increases slightly to 0.006, corresponding to a 2.7-$\sigma$ significance.

We note that this unified relation exists only if the \emph{total} SFRs are adopted for the $z\sim1$ SFGs. No unified $v_\mathrm{wind}$--SFR relation is found if the SFRs of only their compact star-forming regions (namely, top 10 brightest UV pixels) are adopted instead. 

\subsubsection{Comparing with the $z \sim 1 $ Observations by Rubin et al.\ (2014) and the Illustris-TNG Simulations}
In Figure \ref{fig:vwind_vs_masssfr_compare}, we also compare the measurements of our $z\sim 1$ SFG sample with another observational study in the same redshift range by \cite{Rubin2014} and the Illustris-TNG simulations by \cite{Nelson2019}. 

\cite{Rubin2014} reported no strong correlation between $v_\mathrm{wind}$ and SFR, which is consistent with the fit to our $z \sim 1$ SFGs which has only a 1.7-$\sigma$ correlation significance. Compared to the wind velocities measured from our $z \sim 1$ SFGs, the measurements from \cite{Rubin2014} are also offset to lower wind velocities by $\sim 0.1$ dex in Figure \ref{fig:vwind_vs_masssfr_compare}. This is most likely because the wind velocity was calculated in a different way in \cite{Rubin2014}, tracing slower parts of a wind. This is detailed in Appendix \ref{sec:literaturemeasurements}. 

We also compare our $z \sim 1$ wind measurements with those from the Illustris-TNG simulations (dashed khaki lines in Figure \ref{fig:vwind_vs_masssfr_compare}). As shown in the top and middle panels, the $v_\mathrm{wind}$ of simulated galaxies falls within the ballpark of our observational results. In the top panel, the simulation relations also trace the best-fit relation to the combined sample of $z\sim 1$ and $z\sim0$ observed galaxies, except at $\log~$SFR$>1$ where it deviates to higher wind velocities. Nevertheless, we caution that this is only a qualitative comparison because different methods are used to measure $v_\mathrm{wind}$. In the simulations, $v_\mathrm{wind}$ is the 90th percentile of the velocities of the outflowing gas in all phases, located at 10 or 20 kpc from a galaxy. In the observations, the $v_\mathrm{wind}$ is measured only from the cool phase of gas and regardless of the distance from a galaxy along the line of sight.

\subsection{Wind Velocity as a Function of Stellar Mass}
\label{subsec:result_mass}

As shown in the middle panel of Figure \ref{fig:vwind_vs_masssfr_compare}, a weak positive correlation is found for our $z\sim1$ SFGs between $v_\mathrm{wind}$ and $M_\star$, which has a Spearman $p$-value of 0.11 corresponding to a 1.6-$\sigma$ significance. The best-fit relation from the log-linear fit is $v_\mathrm{wind} \propto \mathrm{SFR}^{0.05}$.

It is found that the $z\sim 1$ SFGs and the $z\sim0$ starbursts do not follow the same $v_\mathrm{wind}$ versus $M_\star$ relation. In the mass range where the two types of galaxies overlap (log $M_\star$ = [9.2, 9.7]), the $z \sim 1$ SFGs have, on average, 0.14 dex or 130 km/s lower wind velocities than the starbursts.

The measurements at $z\sim 1$ by \cite{Rubin2014} found a $>3\sigma$ correlation between $v_\mathrm{wind}$ and $M_\star$, which might be in tension with the weak correlation ($<2\sigma$) found in this work. Although the root cause of this potential difference is not clear, we note that our sample covers a broader mass range, extending to lower masses than their sample, which can be seen in Figure \ref{fig:vwind_vs_masssfr_compare}.

\subsection{Wind Velocity as a Function of sSFR}
\label{subsec:result_ssfr}

As shown in the bottom panel of Figure \ref{fig:vwind_vs_masssfr_compare}, there is no correlation between $v_\mathrm{wind}$ and sSFR: The fit has a $p$-value of 0.7, corresponding to only a 0.4-$\sigma$ significance. 
Additionally, our $z\sim 1$ SFGs and the $z\sim 0$ starbursts are in different regions of the diagram that do not overlap significantly with each other. Interestingly, in the narrow sSFR range where the two types of galaxies do overlap (log sSFR = [-8.5, -8.0]), the $z \sim 1$ SFGs have, on average, 0.08 dex or 80 km/s slower wind velocities than the $z\sim 0$ starburst galaxies.

\begin{figure*}
	\centering
	\includegraphics[width = 5.5in]{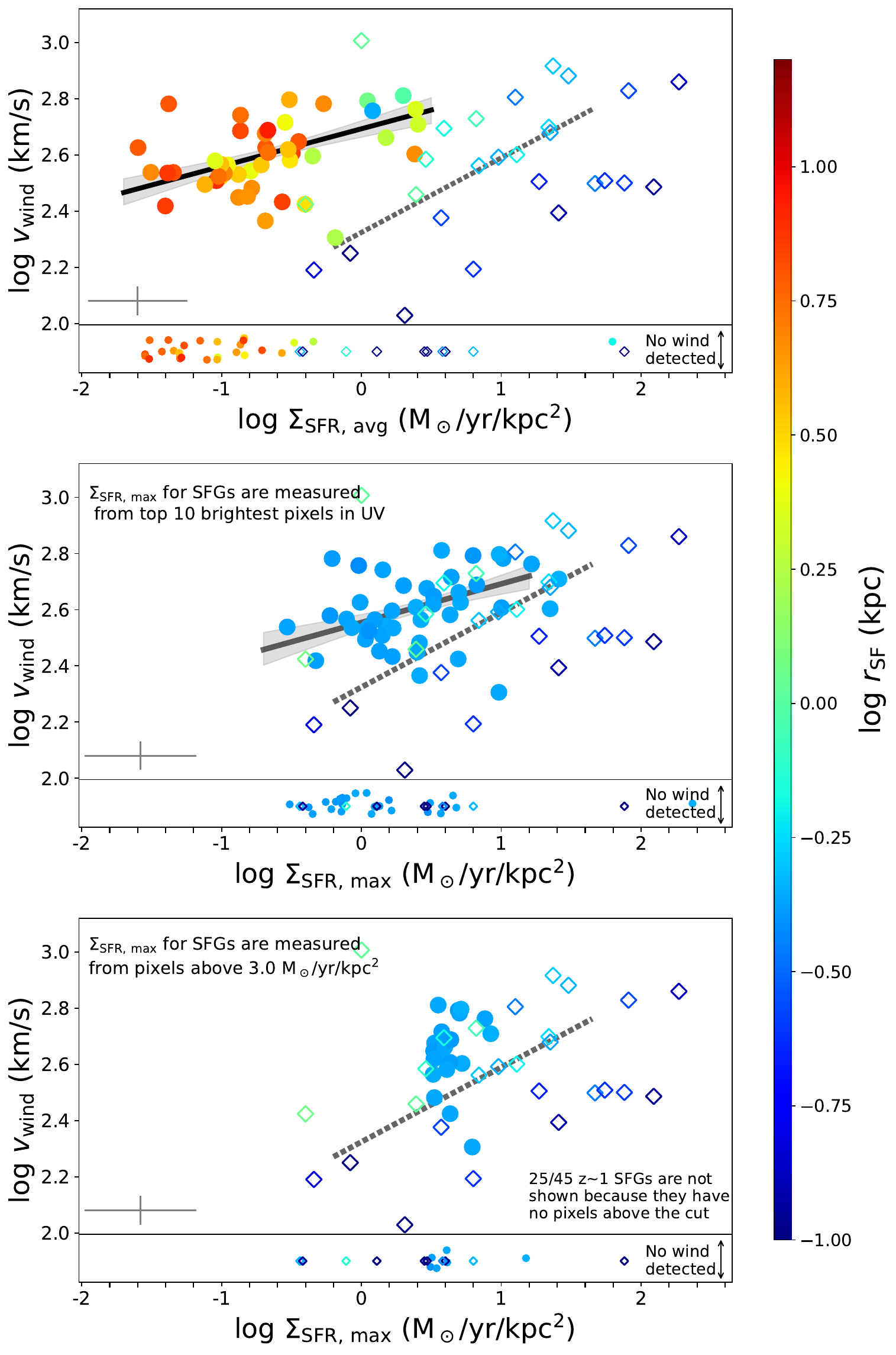}
	\caption{The wind velocities ($v_\mathrm{wind}$) of $z\sim 1$ SFGs (filled circles) are shown as a function of $\Sigma_\mathrm{SFR,\,avg}$ (top panel) and $\Sigma_\mathrm{SFR,\,max}$, the latter of which is measured in two ways (middle and bottom panels; \S \ref{subsec:galprop_sfrd}). The $v_\mathrm{wind}$ of our $z\sim 1$ SFGs show a significant positive correlation with $\Sigma_\mathrm{SFR,\,avg}$ (3.0-$\sigma$ significance), which is stronger than with the correlation with $\Sigma_\mathrm{SFR,\,max}$ in the middle panel (2.5-$\sigma$ significance). The best-fit relations of the $z\sim 1$ SFGs are shown as the solid lines in the two panels. For comparison, the measurements of $z\sim 0$ starbursts are shown as open diamonds and are identical across the three panels. The $z\sim 1$ relations have substantially shallower slopes and are offset to lower SFR densities compared to $z\sim 0$ starbursts, the best-fit relation of which is shown as the dotted lines. The colorbar indicates $r_\mathrm{SF}$, the spatial scale from which a SFR density is measured (\S \ref{subsec:galprop_sfrd}). All other symbols follow the convention of Figure \ref{fig:vwind_vs_masssfr_compare}. 
    \label{fig:vwind_vs_sfrdthreshold_compare} \label{fig:vwind_vs_sfrd_compare}
    }

\end{figure*}

\section{Wind Velocity As a Function of SFR Density} 

\label{sec:result2}
\label{subsec:result2_z1measurements}

The wind velocities as a function of SFR densities for our sample of $z \sim 1$ SFGs are presented in Figure \ref{fig:vwind_vs_sfrd_compare} as filled circles. The top panel shows the relation for the integrated SFR density $\Sigma_\mathrm{SFR,\,avg}$. The middle and bottom panels show the relations with the $\Sigma_\mathrm{SFR,\,max}$ that are measured in two different ways, as described in \S \ref{subsec:galprop_sfrd}: namely measured from the top 10 brightest pixels in the UV (middle panel) and measured from only pixels that have SFR densities above 3.0 M$_\sun$/yr/kpc$^2$ (bottom panel). These two latter relations would be tighter than the first one if the winds come from only highly star-forming regions in the galaxies. The colorbar indicates $r_\mathrm{SF}$, the spatial scale from which a SFR density is measured (\S \ref{subsec:galprop_sfrd}). 

It is found that the $v_\mathrm{wind}$ correlates positively with both $\Sigma_\mathrm{SFR,\,avg}$ and $\Sigma_\mathrm{SFR,\,max}$ at $>2\sigma$-significance, and the correlation with $\Sigma_\mathrm{SFR,\,avg}$ (top panel) shows a higher significance than that with $\Sigma_\mathrm{SFR,\,max}$ (middle panel) (3.0$\sigma$ vs. 2.5$\sigma$). This is justified by the Spearman correlation test and log-linear fit for each of the two relations, the results of which are listed in Table \ref{tab:windfitting}. The best-fit relations are shown as black solid lines in the figure.

In other words, the integrated SFR density $\Sigma_\mathrm{SFR,\,avg}$ relates better to $v_\mathrm{wind}$ than to the other quantities designed to trace only the highest regions of SFR density in the galaxies. 

As for the $\Sigma_\mathrm{SFR,\,max}$ measured in the second way (\S \ref{subsec:galprop_sfrd}) and shown in the bottom panel, no fitting or correlation test is conducted considering the small sample size and narrow range of values spanned in $\Sigma_\mathrm{SFR,\,max}$.

\subsection{Comparing with $z\sim 0$ Starbursts}
\label{subsec:result2_comparez0}

Furthermore, we compare our $z\sim 1$ SFGs with the $z\sim 0$ starbursts. The starbursts are shown as open diamonds in Figure \ref{fig:vwind_vs_sfrd_compare} in the three figure panels. All these galaxies are compact ($\lesssim 1\,$kpc) and their SFR densities are measured in only one way (see Appendix \ref{sec:literaturemeasurements}). 
We also note that the SFR densities presented in Figure \ref{fig:vwind_vs_sfrd_compare} are measured from different spatial scales. The spatial scales, which we denote as $r_\mathrm{SF}$, are indicated by the figure colorbar. For the SFGs, $r_\mathrm{SF}$ is equal to the galaxy size for $\Sigma_\mathrm{SFR,\,avg}$ (top panel) and the $r_\mathrm{max,\,SF}$ for $\Sigma_\mathrm{SFR,\,max}$ (middle and bottom panels; see \S \ref{subsec:galprop_sfrd}). For the starbursts, $r_\mathrm{SF}$ is equal to the starburst size, which is much smaller than the typical $r_\mathrm{SF}$ for the $\Sigma_\mathrm{SFR,\,avg}$ of a SFG but comparable to the $r_\mathrm{SF}$ for the $\Sigma_\mathrm{SFR,\,max}$.

The main finding, which can be seen in the top and middle panels of Figure \ref{fig:vwind_vs_sfrd_compare}, is that the SFGs and starbursts do not follow the same $v_\mathrm{wind}$ versus $\Sigma_\mathrm{SFR}$ relation. Compared to the starbursts, the SFGs have shallower $v_\mathrm{wind}$ versus $\Sigma_\mathrm{SFR}$ relations and are offset to substantially lower SFR density values in both panels.

To demonstrate this quantitatively, we perform a log-linear fit to the $v_\mathrm{wind}$ versus $\Sigma_\mathrm{SFR}$ relation of the starbursts, shown as the dashed line in each panel of Figure \ref{fig:vwind_vs_sfrd_compare}, and find the slope to be 0.27. This slope is substantially steeper than the values measured for the $z\sim 1$ SFGs, which are ${0.13^{+0.04}_{-0.04}}$ for  $\Sigma_\mathrm{SFR,\,avg}$ (top panel) and $0.14^{+0.04}_{-0.05}$ $\Sigma_\mathrm{SFR,\,max}$ (middle panel). 
Additionally, in the $v_\mathrm{wind}$ range where both the SFGs and starbursts can be found ($\log\,v_\mathrm{wind} = $[2.4, 2.9]), the SFGs have substantially lower SFR densities than the starbursts, by around 2 dex for $\Sigma_\mathrm{SFR,\,avg}$ (top panel) and around 1 dex for $\Sigma_\mathrm{SFR,\,max}$ (middle panel). 

The offset between the SFGs and starbursts still persists if adopting the second method to measure $\Sigma_\mathrm{SFR,\,max}$, which is shown in the bottom panel of Figure \ref{fig:vwind_vs_sfrd_compare}. 
Nonetheless, at a given $v_\mathrm{wind}$, the $\Sigma_\mathrm{SFR,\,max}$ values of our $z\sim 1$ SFGs measured in this way are still around 0.7 dex lower than the SFR densities of the $z\sim0$ starbursts.

\section{Conclusions} 
\label{sec:discussion}
\label{sec:summary}

Cool $10^4\,$K winds in a representative sample of star-forming galaxies (SFGs) at $z\sim 1$ are studied using \ion{Mg}{2} $\lambda\lambda$ 2796/2803\,\AA\ absorption lines. From a sample of 86 galaxies, a total of 50 (58\%) are found to have winds, similar to previous studies (e.g., \citealt{Rubin2010,Rubin2014,Kornei2012}).  There are four main findings:

\begin{enumerate}

\item We find that the threshold in SFR density below which local starburst galaxies do not have winds (0.1$\,M_\odot$/yr/kpc$^2$; \citealt{Heckman2015}) does not apply to $z\sim1$ SFGs (Figure \ref{fig:winddetectionrate_vs_sfrd}). Instead, the wind detection rates of the SFGs show a gradual decline at this value by about a factor of 2.

\item We find correlations between $v_\mathrm{wind}$ and SFR, SFR density, and stellar mass (Figures \ref{fig:vwind_vs_masssfr_compare} \& \ref{fig:vwind_vs_sfrd_compare}), as per other studies (e.g., \citealt{Kornei2012,Rubin2014,Prusinski2021}). The strongest correlation is for SFR density, which has a 3-$\sigma$ significance, confirming most other studies (e.g., \citealt{Kornei2012,Prusinski2021}; Figure \ref{fig:vwind_vs_sfrd_compare}, top).

\item Intriguingly, the $z \sim 1$ SFGs appear to follow the same $v_\mathrm{wind}$--SFR relation as local starbursts and galaxies in the Illustris-TNG simulations \citep{Nelson2019} (Figure \ref{fig:vwind_vs_masssfr_compare}, top).  A fit to the combined observations gives: $\log v_\mathrm{wind} = 0.16\cdot\log\mathrm{SFR} + 2.4$, which has a correlation significance of  3-$\sigma$.  This unified relation spans over 4 dex in SFR. 
No unified relation is found between $v_\mathrm{wind}$ and stellar mass (Figure \ref{fig:vwind_vs_masssfr_compare}, middle), sSFR (Figure \ref{fig:vwind_vs_masssfr_compare}, bottom), or SFR density (Figure \ref{fig:vwind_vs_sfrd_compare}, top). If the SFRs are measured from only the compact star-forming regions inside galaxies, no unified relation is found.  These findings suggest that winds might be most closely associated with total SFR.

\item We examine whether winds are driven from only the most compact star-forming regions in galaxies.  To do so, we consider whether the relation between $v_\mathrm{wind}$ and the SFR density measured only from these compact regions is stronger than that for integrated SFR density.  The compact regions are identified as the brightest pixels in the rest-frame ultraviolet images.  We do not find a stronger correlation.  This suggests that winds are more likely driven by the star formation of the entire galaxy, although we caution that resolved spectral maps will be more definitive.

\end{enumerate}

These two latter conclusions support a picture in which the wind of a $z\sim 1$ SFG is closely connected with its total star formation. Although $z\sim 1$ SFGs commonly have compact regions of concentrated star formation, the wind bubbles from all regions of a SFG could combine momentum from supernovae to lift and accelerate their gas out of the galaxy.

\newpage

\begin{acknowledgments}

WW and SAK would like to acknowledge support from NASA's Astrophysics Data Analysis Program (ADAP) grant number 80NSSC20K0760 and an RSAC grant from the Space Telescope Science Institute. DCK acknowledges support from the National Science Foundation grant AST-1615730. HMY was supported by JSPS KAKENHI Grant Number JP22K14072 and the Research Fund for International Young Scientists of NSFC (11950410492). ECC acknowledges support for this work provided by NASA through the NASA Hubble Fellowship Program grant HST-HF2-51502 awarded by the Space Telescope Science Institute, which is operated by the Association of Universities for Research in Astronomy, Inc., for NASA, under contract NAS5-26555. Some of the data presented herein were obtained at the W.~M.~Keck Observatory, which is operated as a scientific partnership among the California Institute of Technology, the University of California and the National Aeronautics and Space Administration. The Observatory was made possible by the generous financial support of the W.~M.~Keck Foundation. The authors wish to recognize and acknowledge the very significant cultural role and reverence that the summit of Maunakea has always had within the indigenous Hawaiian community.  We are most fortunate to have the opportunity to conduct observations from this mountain. This work is based on observations taken by the CANDELS Multi-Cycle Treasury Program with the NASA/ESA HST, which is operated by the Association of Universities for Research in Astronomy, Inc., under NASA contract NAS5-26555. WW would like to thank Dylan Nelson for providing the data tables of wind velocities from the Illustris-TNG simulations, Vicente Rodriguez-Gomez for providing instructions for {\sc{statmorph}}, and Zuyi Chen for providing details about the SED fitting for the CLASSY galaxies. 

This research made use of Astropy,\footnote{http://www.astropy.org} a community-developed core Python package for Astronomy \citep{astropy2013,astropy2018}, including the photutils package \citep{Bradley2020}, an Astropy package for
detection and photometry of astronomical sources, and the statmorph package \citep{Rodriguez-Gomez2019}.

\end{acknowledgments}

\appendix

\section{Comparing SFRs Measured in Different Ways}
\label{appendix:sfrmass_cali}

In this paper we use SFRs inferred from rest-frame NUV luminosities and corrected for dust, as described in \S \ref{subsec:galprop_sfr} and \cite{Wang2018}. In this appendix, we show that they are consistent values measured using other methods. 

\begin{figure}
\centering
\includegraphics[width = 5.3 in]{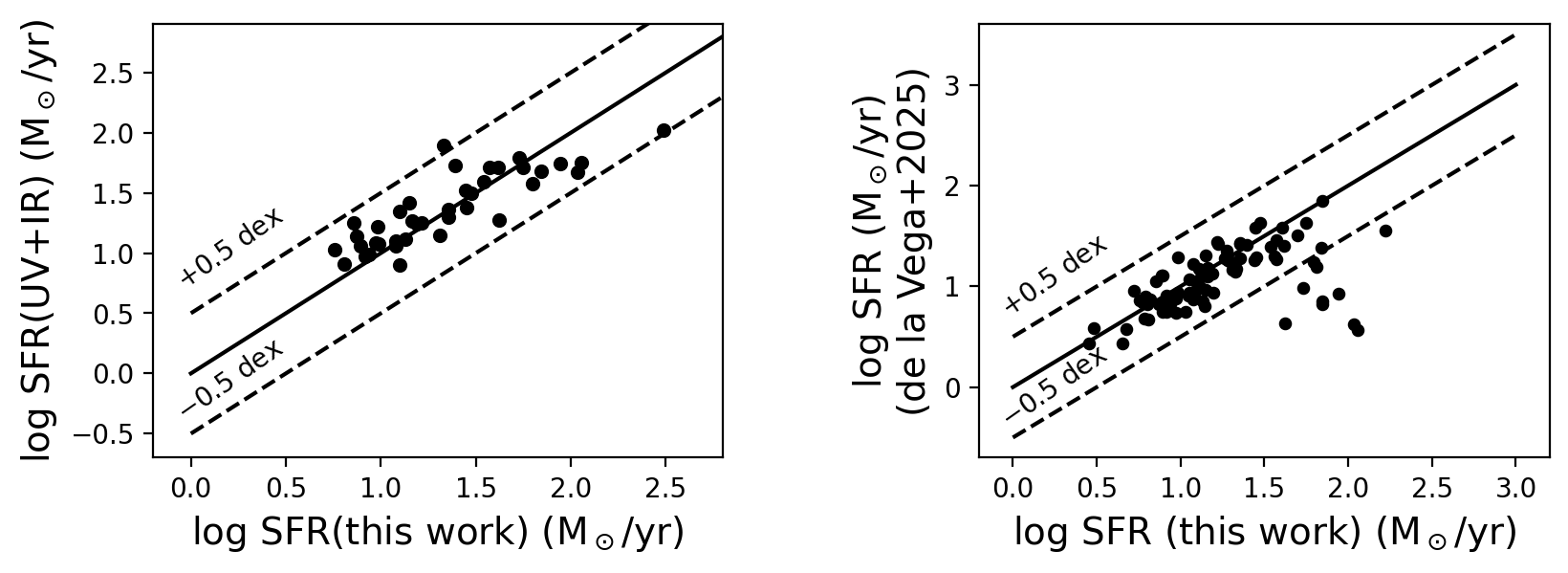}
\caption{In this paper we use SFRs measured from rest-frame NUV luminosities and corrected for dust.  Here we compare them with SFRs measured in two other ways and find them to be consistent: SFRs measured from rest-frame UV + IR luminosities (left) and SFRs measured from SED fitting to rest-frame UV-to-optical photometry (right; \citealt{delaVega2025}). For SFRs measured from UV+IR, only galaxies with reliable 24\,$\mu$m flux measurements, which are needed to infer the IR luminosities, are included and a total of 47 galaxies were removed. The solid lines in both panels mark the one-to-one relations and the dashed lines mark offsets of $\pm0.5$ dex. The majority of the galaxies lie within the two dashed lines in both panels and therefore have offsets smaller than 0.5 dex. \label{fig:sfrcalib}}
\end{figure}

In Figure \ref{fig:sfrcalib}, we compare the values of the SFR indicators adopted in this work with those inferred from the rest-frame UV and IR luminosities (left panel) and from SED fitting (right panel; \citealt{delaVega2025}). 
We find that the differences between the SFRs have an average offset smaller than 0.05 dex and a 1-$\sigma$ scatter smaller than 0.25 dex. 
In the left panel, For the SFRs measured from UV and IR luminosities, only those galaxies with reliable 24\,$\mu$m flux measurements, which are needed to infer the IR luminosities, are included. This removes 47 galaxies. 
For the SFRs measured from SED fitting, fluxes from the $u$-band to the \emph{HST} F160W band are used. For the SED fitter, we adopt the BayEsian Analysis of GaLaxy sEds tool (BEAGLE; \citealt{Chevallard2016}) and a set of physically-motivated priors described in \citet{delaVega2025}.

\section{Description of Literature Studies Used for Comparison}
\label{sec:literaturemeasurements}

\subsection{Studies of z$\sim$0 Starbursts}
\label{subsec:starburstmeasurements}

\begin{figure*}
	\centering
	\includegraphics[width = 7in]{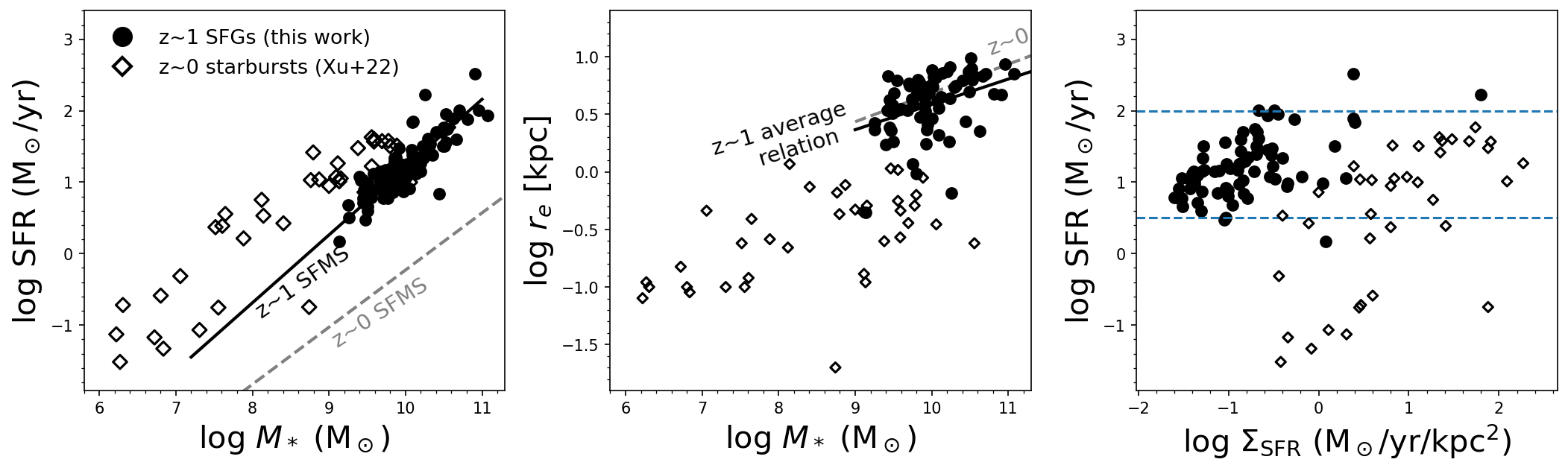}
	\caption{Our sample of $z\sim1$ SFGs (filled circles) and the $z\sim0$ starbursts (open diamonds; \citealt{Berg2022,Xu2022}) are compared on the SFR--$M_\star$ diagram, $r_e$--$M_\star$ diagram, and SFR--$\Sigma_\mathrm{SFR}$ diagram, from left to right.
    The $z\sim1$ SFGs in general have higher SFRs, larger sizes, and lower integrated SFR densities than the $z\sim0$ starbursts. \emph{Left}: 
    In the mass range where the two types of galaxies overlap, the SFGs have lower SFRs by $\sim$0.5 dex on average. Additionally, the starbursts lie significantly above the $z\sim0$ SFMS (grey dashed line; \citealt{Chang2015}) and even above the $z\sim1$ SFMS (solid line, from Figure~\ref{fig:halo7dsample}). \emph{Middle:} In the mass range where the two galaxy types overlap, the SFGs are larger by a factor of $\sim$1 dex. The average size--mass relations for the SFGs at $z\sim 1$ and $z\sim 0$ from \cite{VanderWel14} are also shown. \emph{Right:} In the SFR range where the two galaxy types overlap, the SFGs have on average 2.0 lower SFR densities than the starbursts. 
\label{fig:classy_sample_compare}}
\end{figure*}

We compare with a sample of 37 $z\sim0$ compact starburst galaxies from the Cosmic Origins Spectrograph (COS) Legacy Spectroscopic SurveY (CLASSY, \citealt{Berg2022}), which is compiled by \cite{Xu2022}. These galaxies are presented as open symbols in Figure \ref{fig:classy_sample_compare}, on the SFR--$M_\star$ diagram (left panel), $r_e$--$M_\star$ diagram (middle panel), and SFR--$\Sigma_\mathrm{SFR}$ diagram (right panel).

It is found from the figure that the starburst galaxies are exceptional objects at $z\sim 0$, because they lie above the $z\sim 0$ star-forming main sequence by $\sim2$ dex (left panel; \citealt{Berg2022}) and are substantially more compact than typical $z\sim 0$ SFGs by $\sim1$ dex (middle panel; \citealt{Berg2022}).

The starbursts are also compared with our sample of $z\sim 1$ SFGs (filled symbols) in the figure. In short, it is found that the starbursts generally have higher SFRs, smaller sizes, and higher SFR densities than the SFGs.

Specifically, in the left panel, the $z \sim 0$ starbursts span a larger range in stellar mass and SFR than our $z\sim 1$ SFGs.  Our galaxies overlap with the high mass end of the starbursts and extend to even higher masses.
For the masses where they overlap, the starbursts have slightly higher SFRs by about 0.5 dex. In the middle panel, the starbursts have typical sizes of around 0.3 kpc (-0.5 in log), much smaller than the SFGs which have typical sizes of 4 kpc (0.6 in log). As shown in the right panel, the SFR densities of the starbursts are substantially higher than the integrated SFR densities of the $z\sim 1$ SFGs ($\Sigma_\mathrm{SFR,\,avg}$). In the SFR range where the $z\sim0$ starbursts and our $z\sim 1$ SFGs overlap, $0.5<\log\,\mathrm{SFR}<2.0$, the former have on average 2.0 dex higher SFR densities than the latter.

Details of measurements of the physical properties and wind velocities of the starbursts are detailed below.

The stellar masses, SFRs, and sizes have already been measured by \cite{Berg2022} and \cite{Xu2022}, which we adopt in this work. As for the SFR density, we calculate it for each starburst by dividing the SFR by the area $2\pi r_e^2$, where $r_e$ is the galaxy half-light radius. 
As for the wind velocities, they were originally measured  by \citet{Xu2022} using multiple spectral lines. In this work, we choose instead to adopt the velocities measured from only the \ion{Si}{2} 1260\,\AA\ line, because it traces similar gas phases as the \ion{Mg}{2} 2796\,\AA\ line which we use for our $z\sim 1$ SFGs.   
The two spectral lines trace similar gas because the ionization potentials of Si and Mg are very similar (8.15~eV and 7.65~eV, respectively; \citealt{Zhu2015}).  
We calculate the wind velocities of the starbursts as the sum of $V_\mathrm{out}$ and 0.5$\,$FWHM$_\mathrm{out}$, where $V_\mathrm{out}$ and FWHM$_\mathrm{out}$ are the mean velocity and the full velocity width of the wind, respectively. These two quantities have been measured by \cite{Xu2022}, which we adopt in our calculation. The velocities calculated as such correspond to where the wind component reaches half of its maximum absorption depth; this approach is the same as the method used to calculate the wind velocities of our SFG sample.

\subsection{z$\sim$1 SFGs from Rubin et al.~(2014)}
\label{subsec:R14measurements}

We also compare with a sample of intermediate-redshift SFGs by \cite{Rubin2014}, which is similar to our sample. \cite{Rubin2014} measured wind velocities using the same absorption lines as in this work (\ion{Mg}{2}). To match the epoch of our study ($z \sim 1$), we include only the galaxies within $z$=0.7--1.5 from the ``Main Sample'' of \cite{Rubin2014}, defined as those with sufficient spectral S/N to detect winds. 

We compare the wind velocity as a function of stellar mass and SFR. We adopt the $\Delta v_\mathrm{max}$ calculated by \cite{Rubin2014} as wind velocities for the comparison, which were measured from the absorption wings of the \ion{Mg}{2} lines. However, we caution that this method of measuring velocity is not identical to that adopted by our work (\S \ref{subsec:analysis_vwindmeasure}): The $\Delta v_\mathrm{max}$ was calculated as the sum of the centroid velocity of the wind component and its standard deviation. As a result, 
the method by \cite{Rubin2014} traces the parts of the wind closer to the centroid velocity compared to ours, and thus will lead to smaller values which we find to be the case in Sections \ref{sec:result} and \ref{sec:result2}.

We do not compare the wind velocity as a function of the SFR density. This is because the SFR densities of the galaxies in \cite{Rubin2014} were measured in a different way from our work.  Specifically, in the calculation of SFR densities, \cite{Rubin2014} used the Source Extractor \citep{Bertin1996} parameters as galaxy sizes, whereas our work uses the half-light radii which are different from the Source Extractor parameters.

\section{Best-fit \ion{M\lowercase{g}}{2} Profile Models}
\label{appendix:mg2lineprofiles}

\begin{figure}
\centering
\includegraphics[width = 5.6 in]{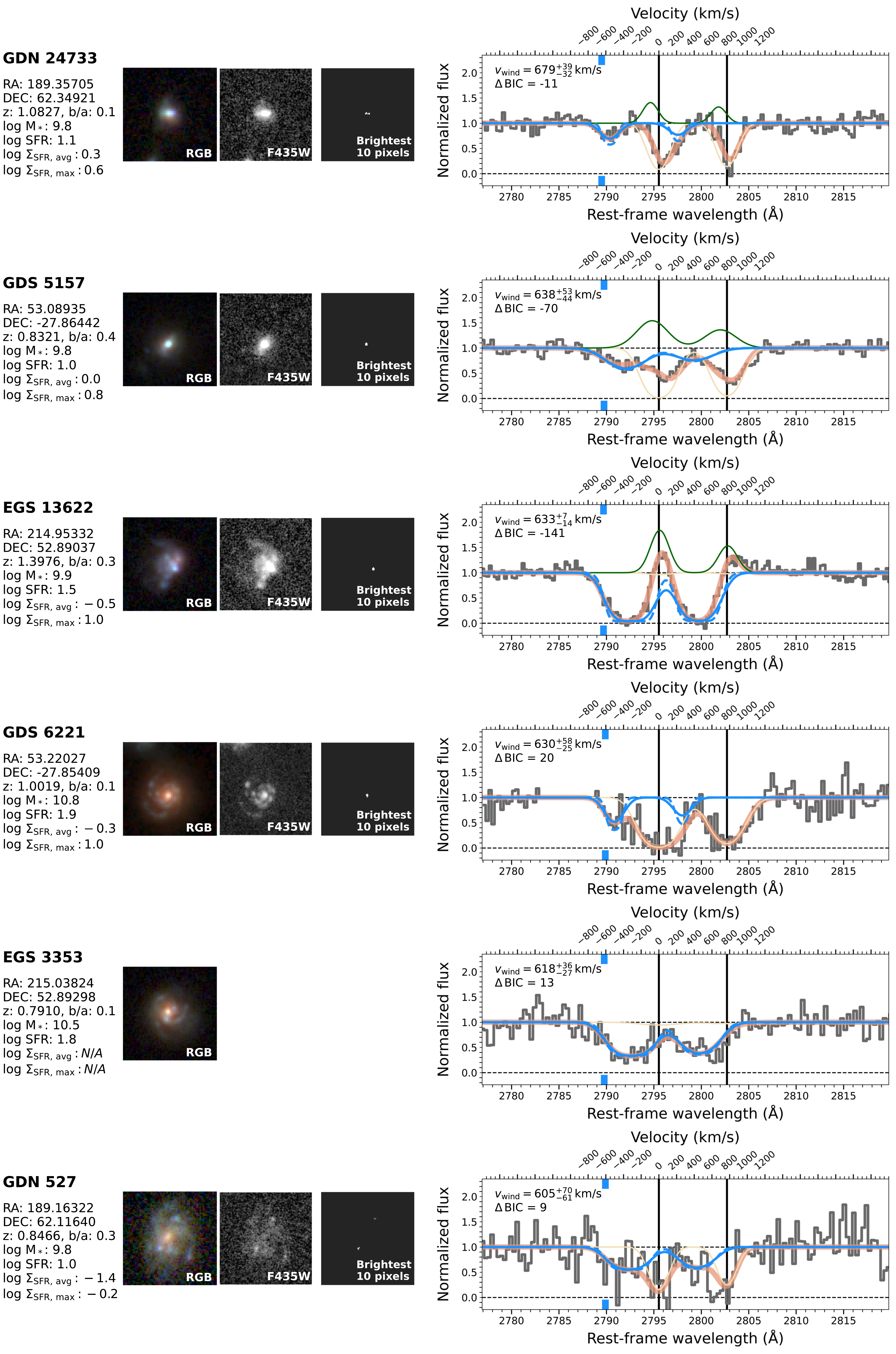}
\caption{Images and \ion{Mg}{2} line profile fits of the 50 SFGs at $z\sim 1$ with detected winds. For each galaxy, the galaxy ID and properties, RGB image, F435W image (if available), the top 10 brightest pixels of the F435W image (if available), and the \ion{Mg}{2} line profiles are shown from left to right. The images are all 3.\arcsec6 on a side, corresponding to 29 kpc at z=1. The RGB images are made using \emph{HST} images in the F160W, F814W, F606W wavebands as red, green, and blue channels, respectively. The spectrum plots adopt the same convention of lines and markers as described in the caption of Figure \ref{fig:linefittingexamples}. The measured wind velocity and $\Delta$BIC statistics are indicated in the spectrum plots. Galaxies are rank-ordered from high to low wind velocities. 
\label{fig:galaxyroster_windnoagn1}}

\end{figure}

\begin{figure}
\centering 
\includegraphics[width = 5.8 in]{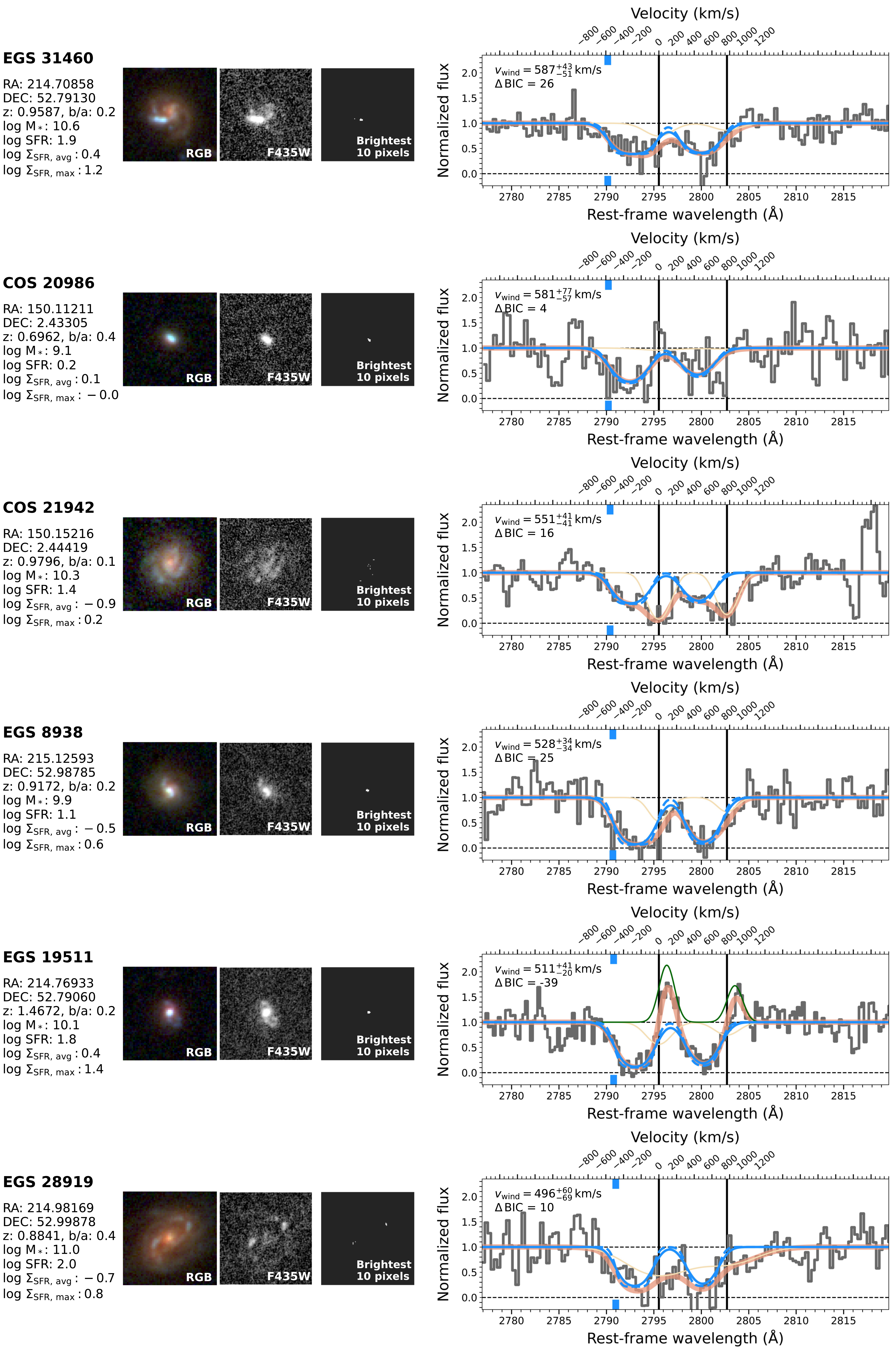}
\caption{Images and \ion{Mg}{2} line profiles of the SFGs in the $z\sim 1$ sample with detected winds (continued). \label{fig:galaxyroster_windnoagn2}}
\end{figure}

\begin{figure}
\centering 
\includegraphics[width = 5.8 in]{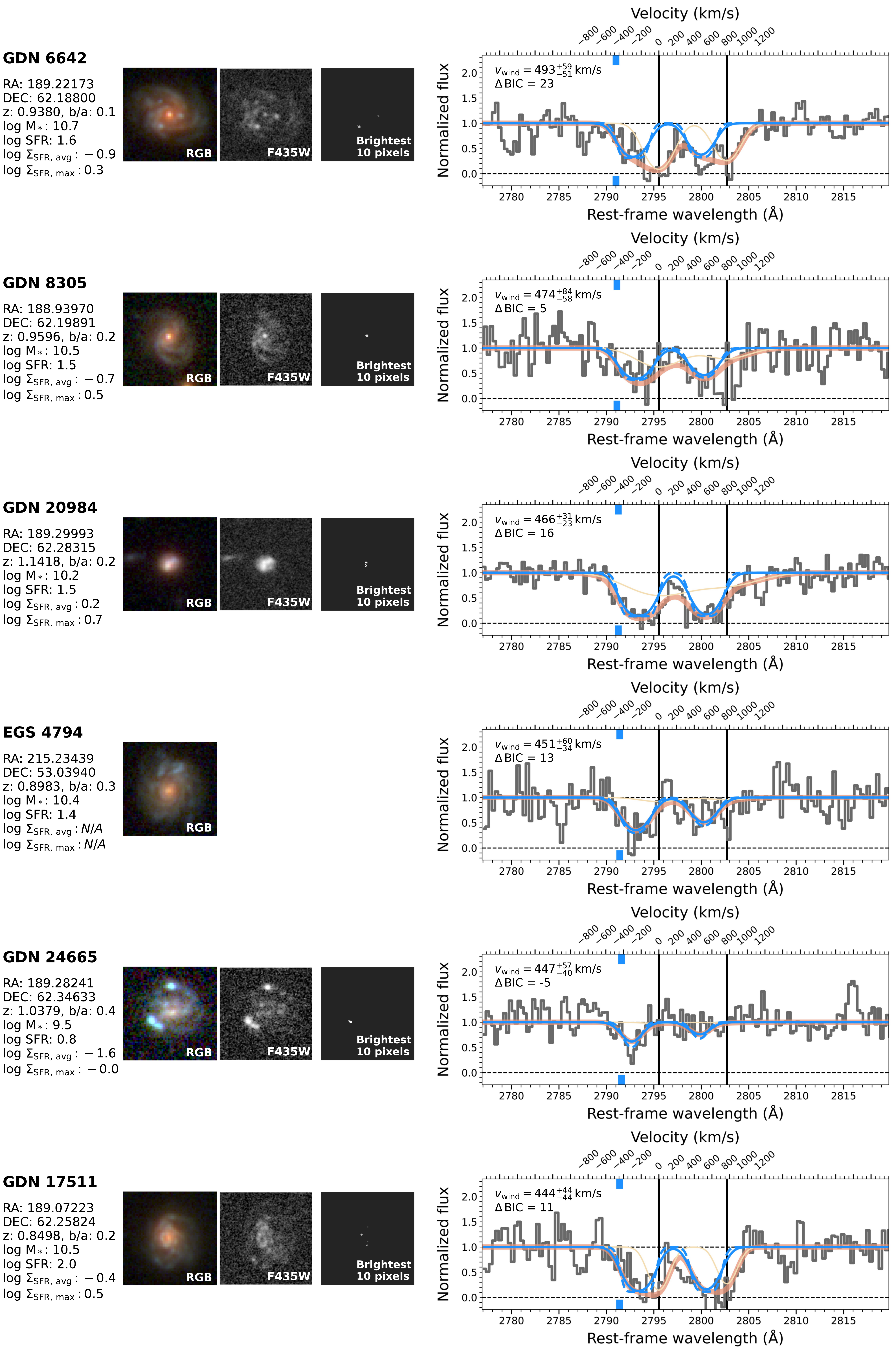}
\caption{Images and \ion{Mg}{2} line profiles of the SFGs in the $z\sim 1$ sample with detected winds (continued). \label{fig:galaxyroster_windnoagn3}}
\end{figure}

\begin{figure}
\centering 
\includegraphics[width = 5.8 in]{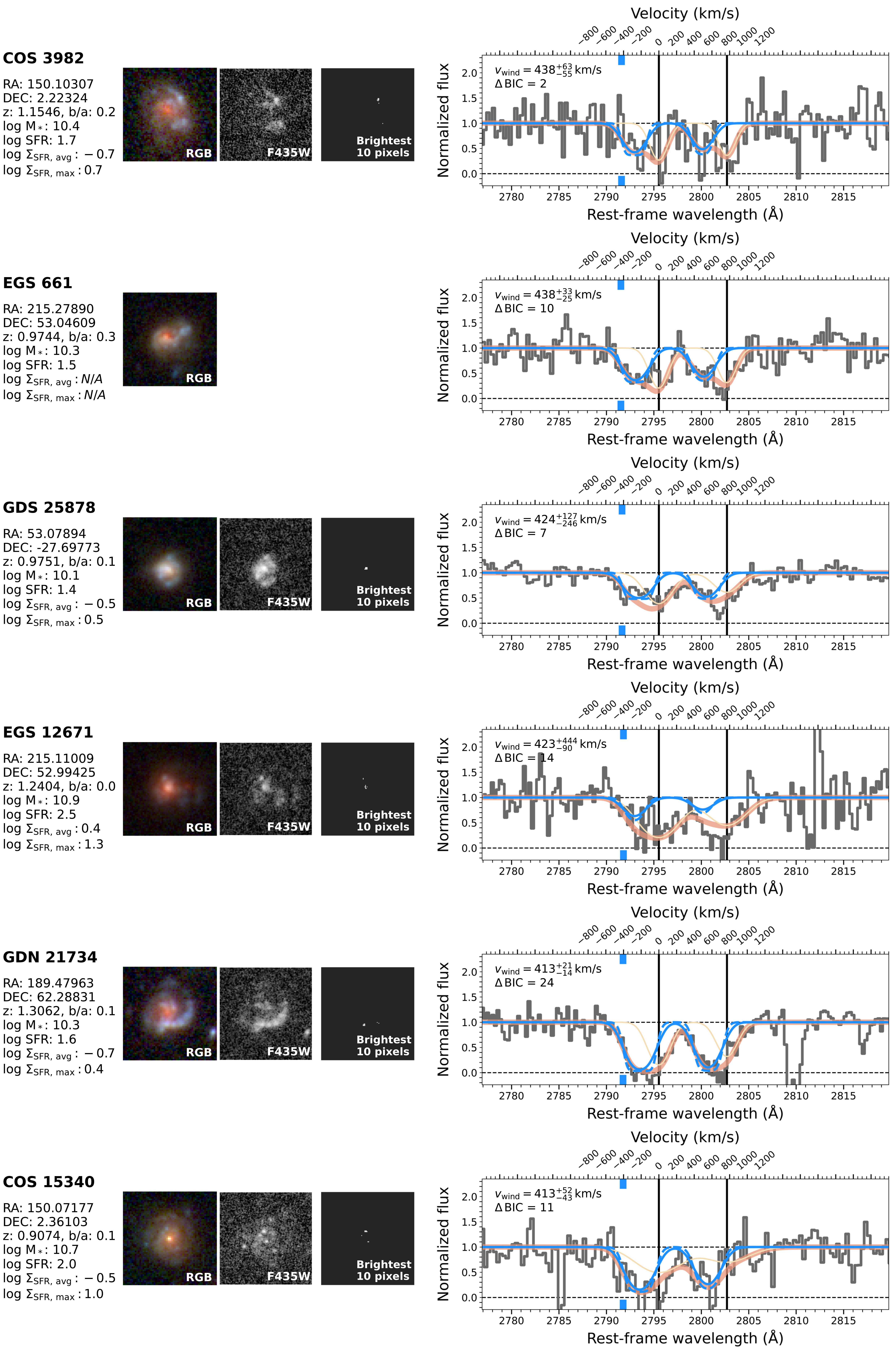}
\caption{Images and \ion{Mg}{2} line profiles of the SFGs in the $z\sim 1$ sample with detected winds (continued). \label{fig:galaxyroster_windnoagn4}}
\end{figure}

\begin{figure}
\centering 
\includegraphics[width = 5.8 in]{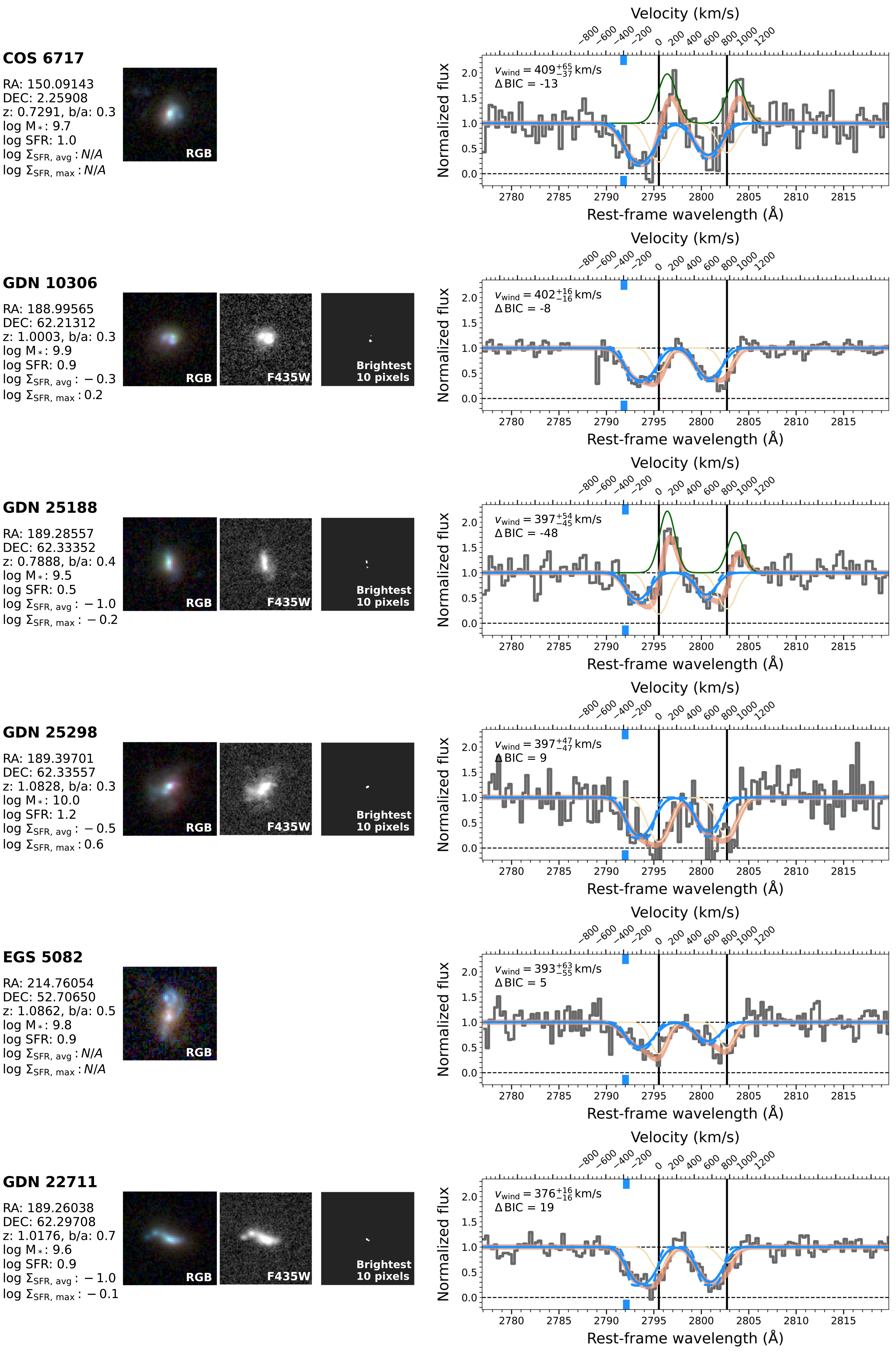}
\caption{Images and \ion{Mg}{2} line profiles of the SFGs in the $z\sim 1$ sample with detected winds (continued). \label{fig:galaxyroster_windnoagn5}}
\end{figure}

\begin{figure}
\centering 
\includegraphics[width = 5.8 in]{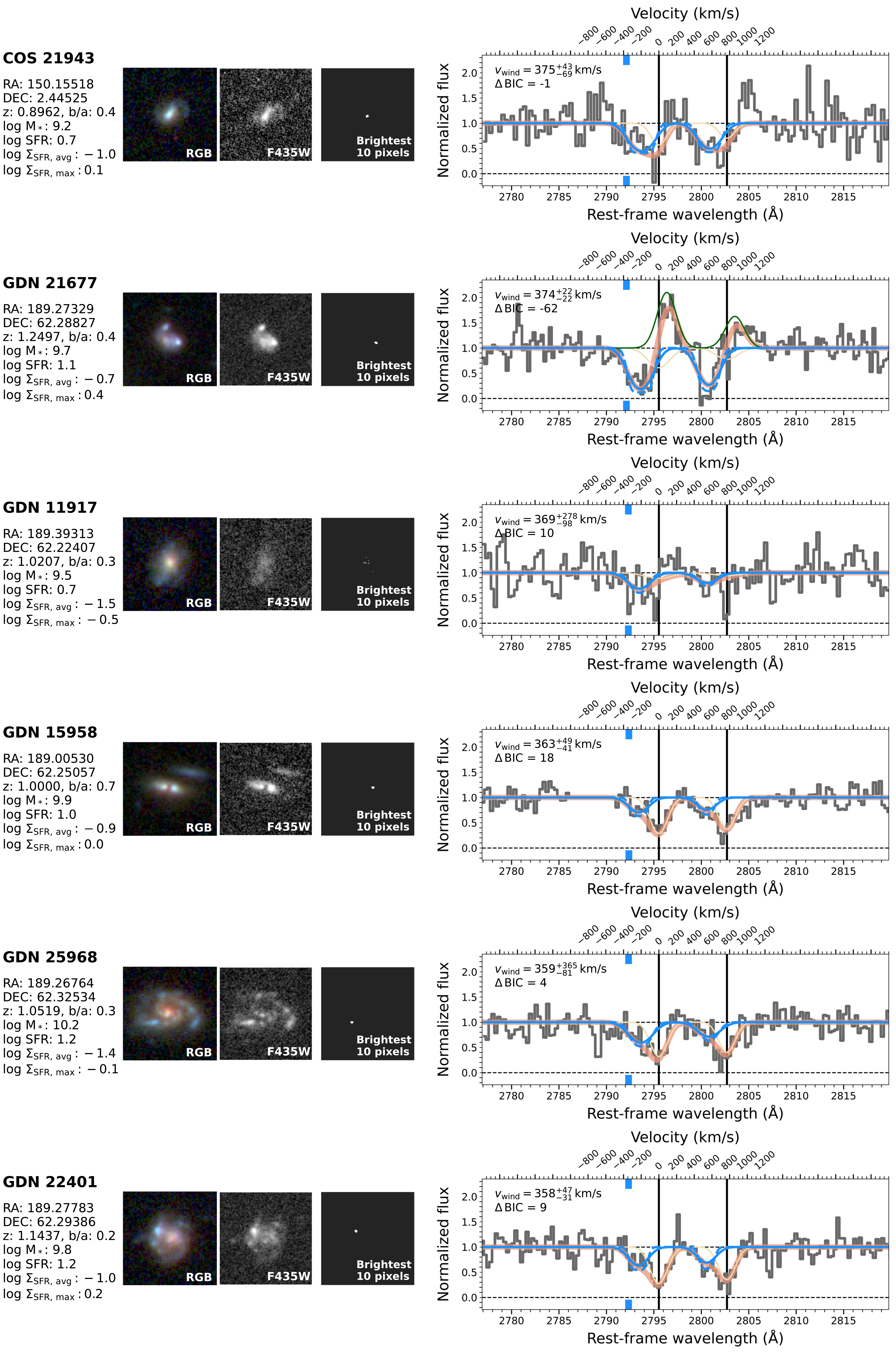}
\caption{Images and \ion{Mg}{2} line profiles of the SFGs in the $z\sim 1$ sample with detected winds (continued). \label{fig:galaxyroster_windnoagn6}}
\end{figure}

\begin{figure}
\centering 
\includegraphics[width = 5.8 in]{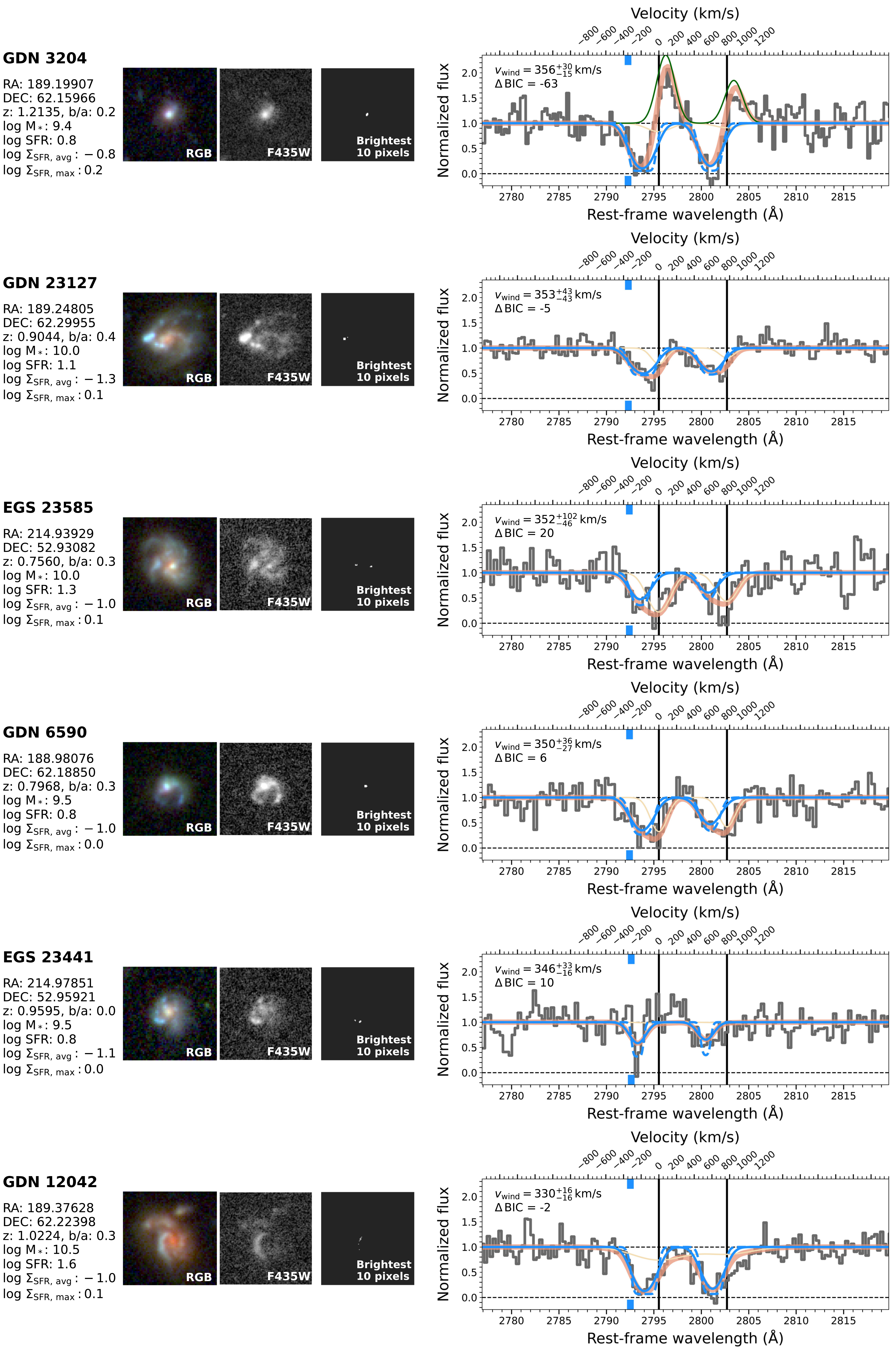}
\caption{Images and \ion{Mg}{2} line profiles of the SFGs in the $z\sim 1$ sample with detected winds (continued). \label{fig:galaxyroster_windnoagn7}}
\end{figure}

\begin{figure}
\centering 
\includegraphics[width = 5.8 in]{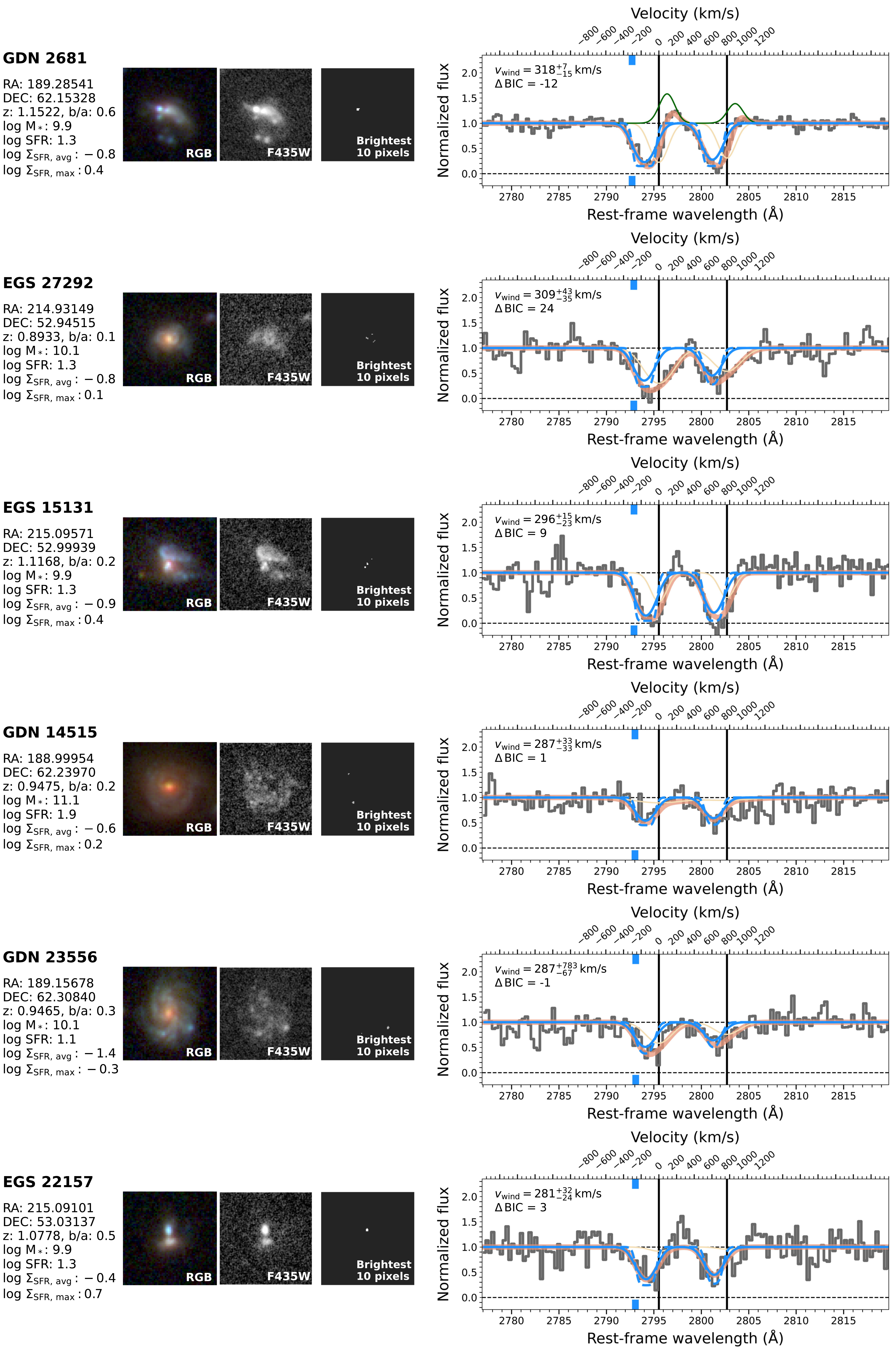}
\caption{Images and \ion{Mg}{2} line profiles of the SFGs in the $z\sim 1$ sample with detected winds (continued). \label{fig:galaxyroster_windnoagn8}}
\end{figure}

\begin{figure}
\centering 
\includegraphics[width = 5.8 in]{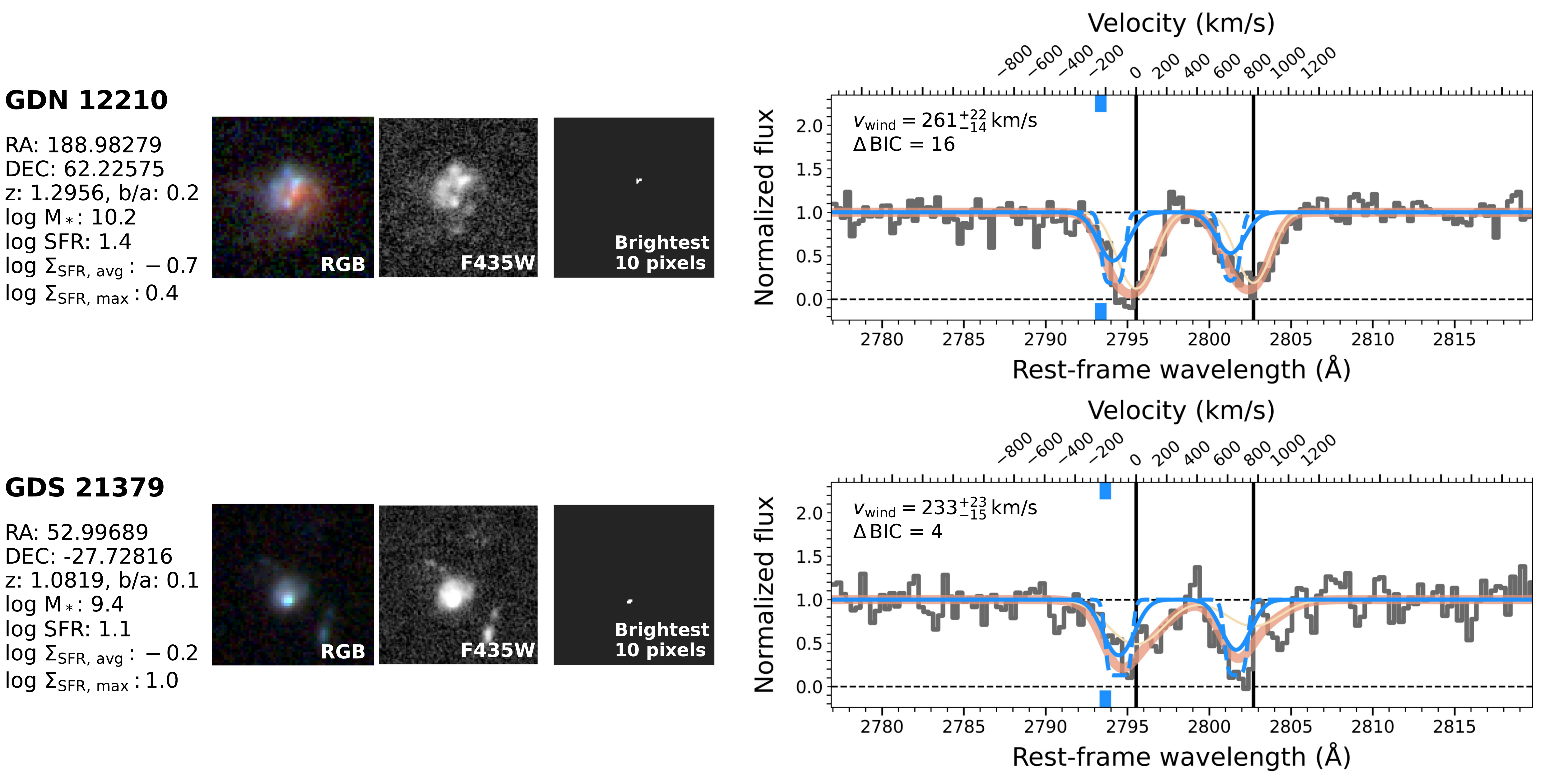}
\caption{Images and \ion{Mg}{2} line profiles of the SFGs in the $z\sim 1$ sample with detected winds (continued). \label{fig:galaxyroster_windnoagn9}}
\end{figure}

\begin{figure}
\centering 
\includegraphics[width = 5.8 in]{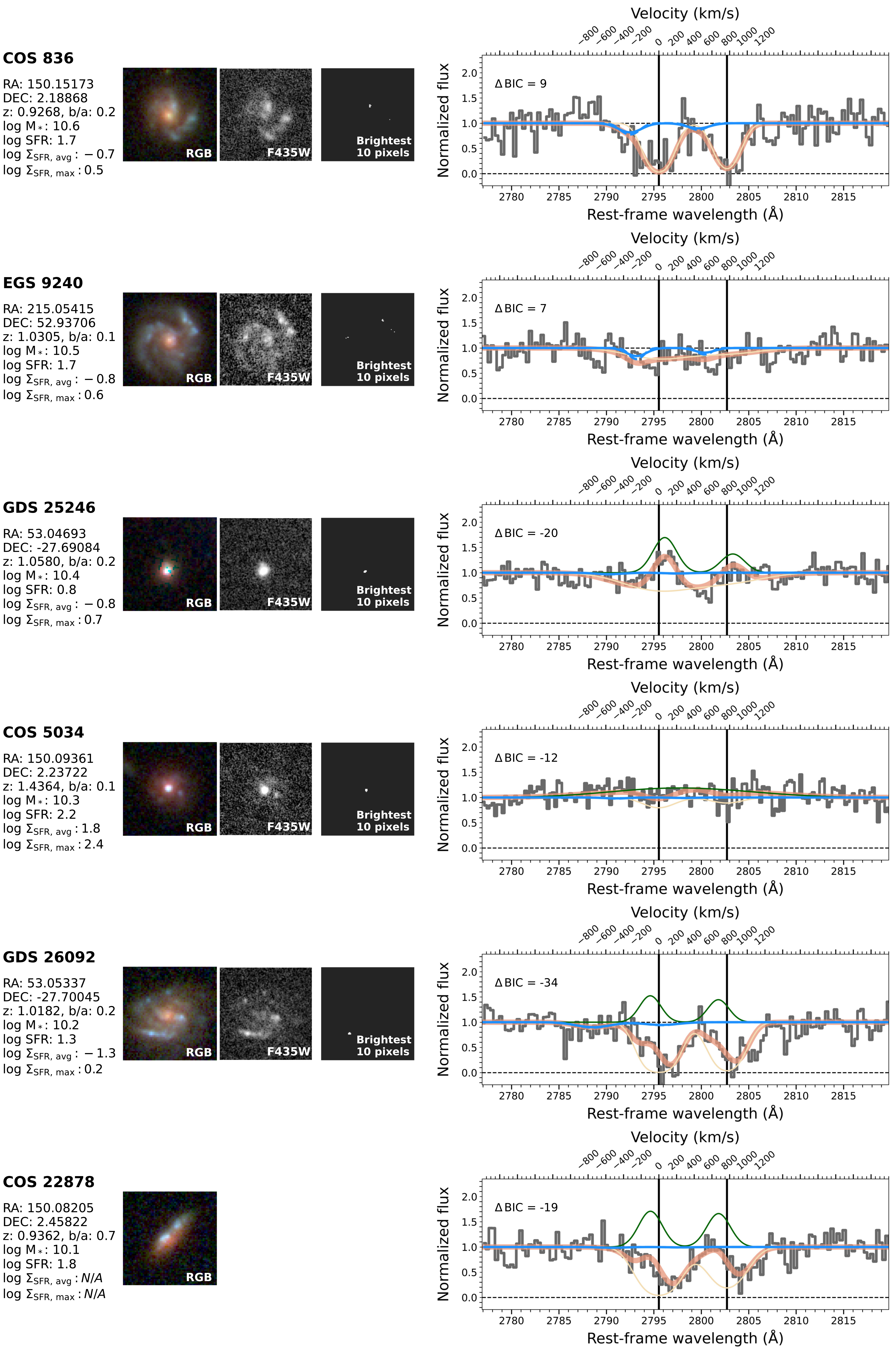}
\caption{Images and \ion{Mg}{2} line profiles of the SFGs in the $z\sim 1$ sample with no detected winds. Refer to the caption of Figure \ref{fig:galaxyroster_windnoagn1}. Galaxies are rank-ordered from high to low stellar masses. }
\label{fig:galaxyroster_nowindnoagn1}
\end{figure}

\begin{figure}
\centering 
\includegraphics[width = 5.8 in]{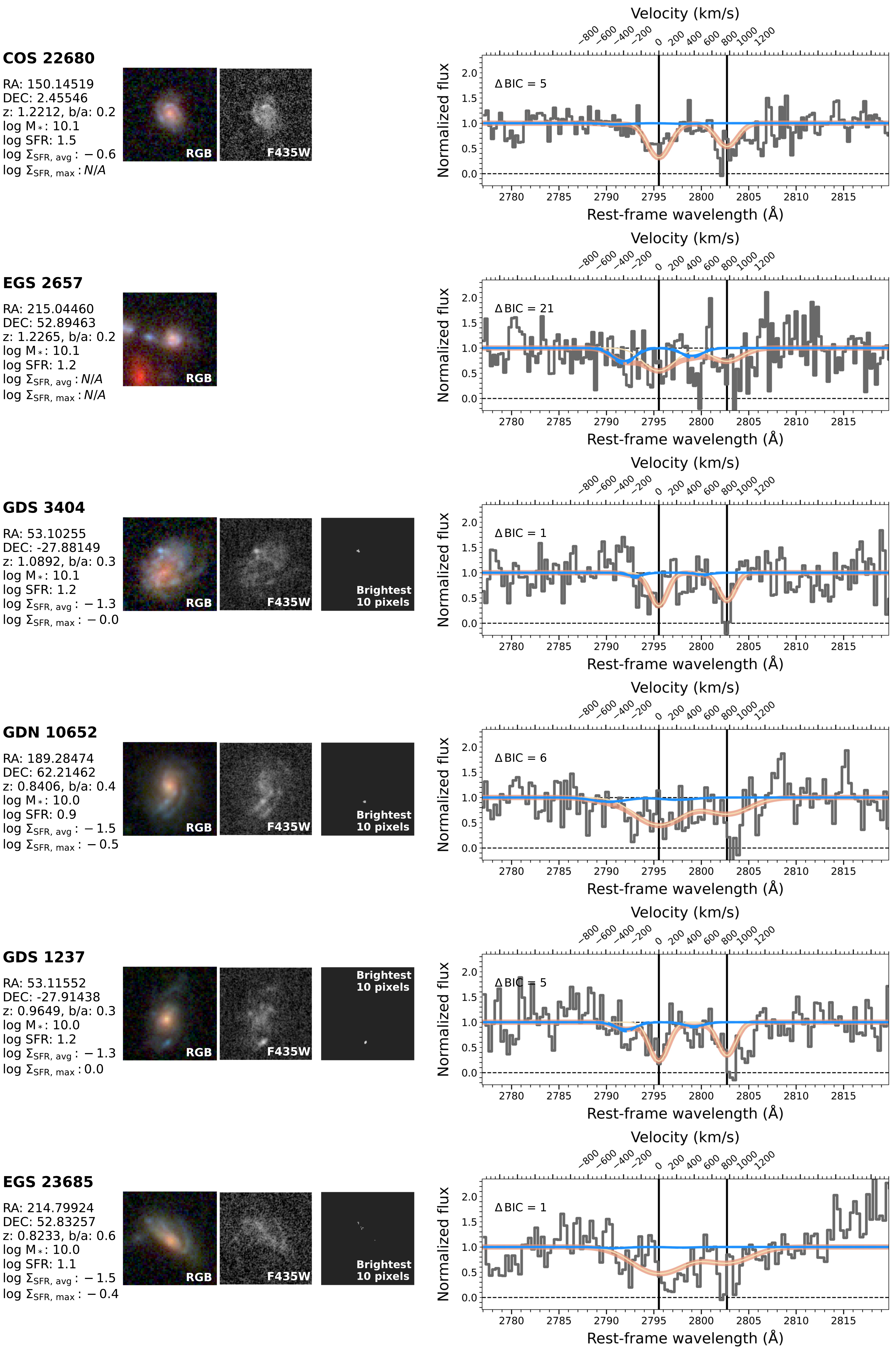}
\caption{Images and \ion{Mg}{2} line profiles of the SFGs in the $z\sim 1$ sample with no detected winds (continued). \label{fig:galaxyroster_nowindnoagn2}}
\end{figure}

\begin{figure}
\centering 
\includegraphics[width = 5.8 in]{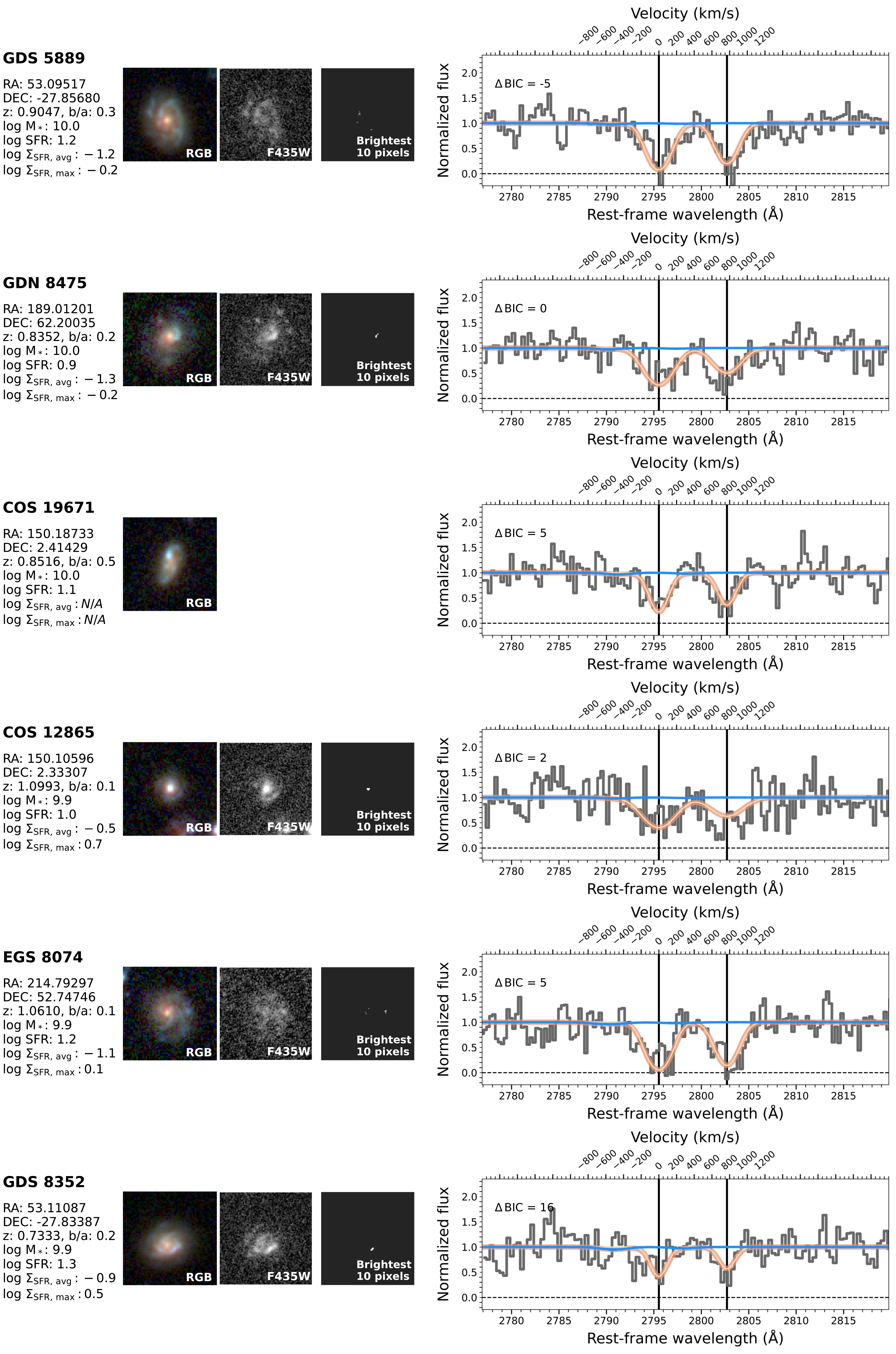}
\caption{Images and \ion{Mg}{2} line profiles of the SFGs in the $z\sim 1$ sample with no detected winds (continued). \label{fig:galaxyroster_nowindnoagn3}}
\end{figure}

\begin{figure}
\centering 
\includegraphics[width = 5.8 in]{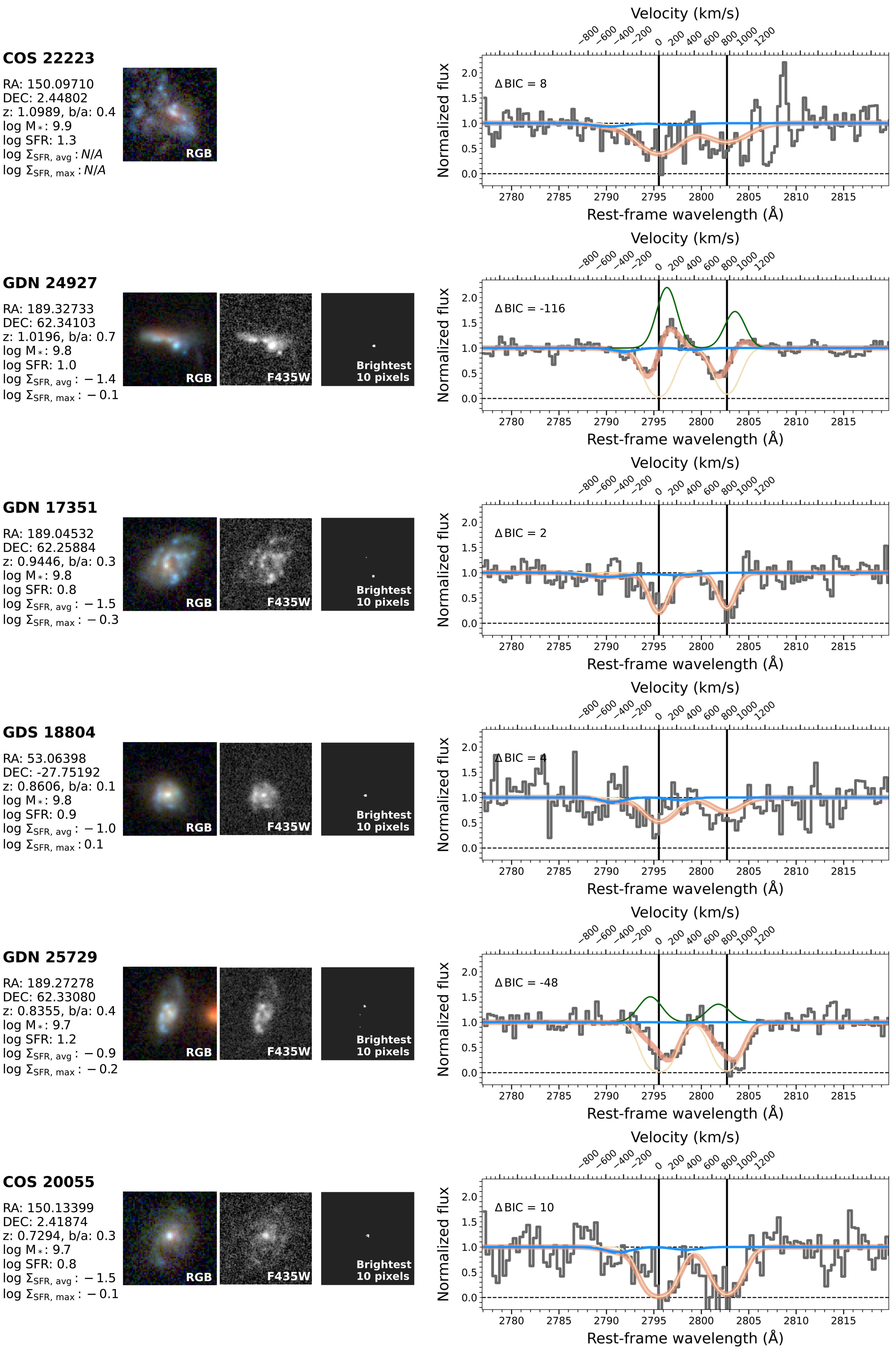}
\caption{Images and \ion{Mg}{2} line profiles of the SFGs in the $z\sim 1$ sample with no detected winds (continued). \label{fig:galaxyroster_nowindnoagn4}}
\end{figure}

\begin{figure}
\centering 
\includegraphics[width = 5.8 in]{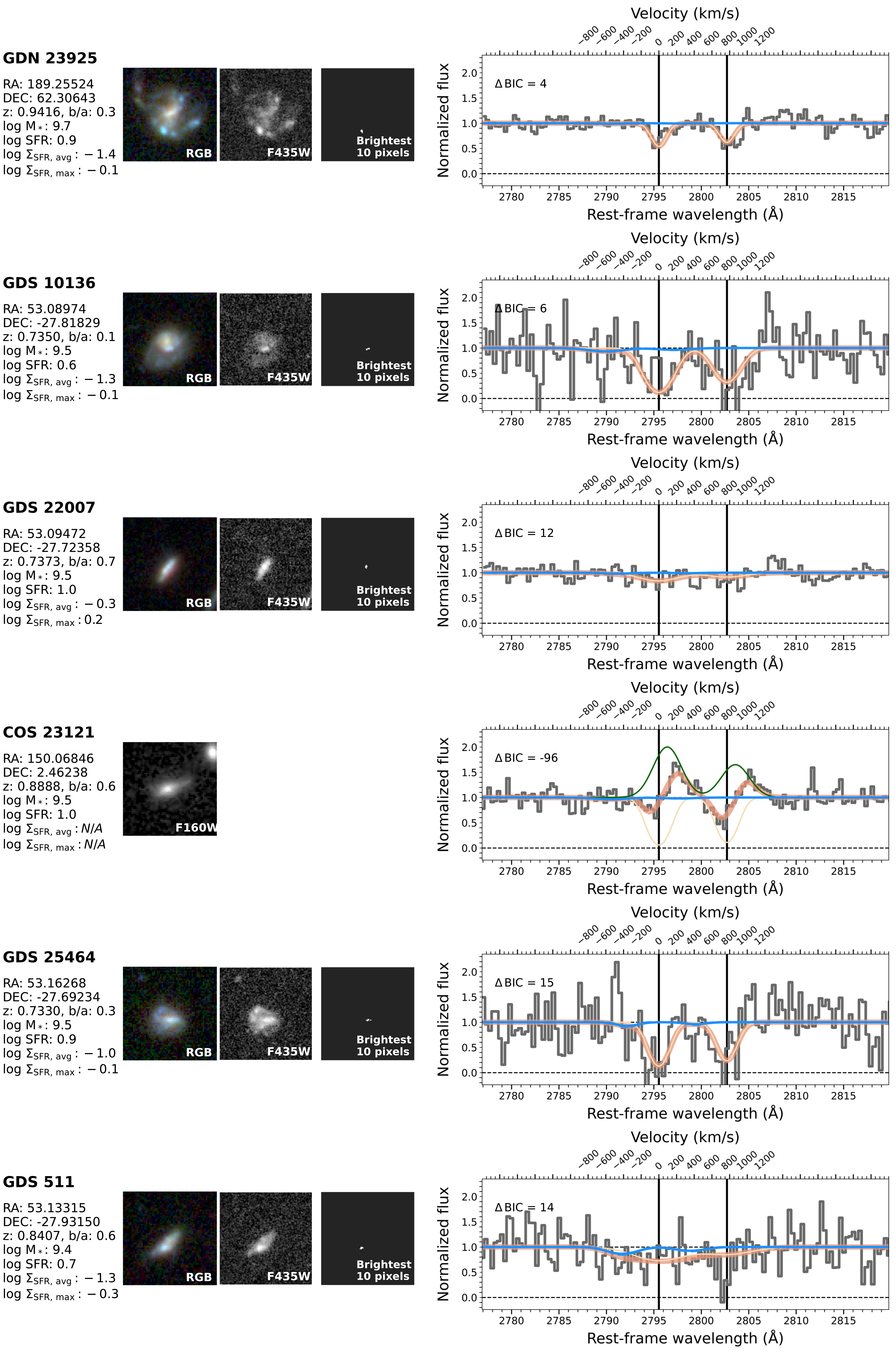}
\caption{Images and \ion{Mg}{2} line profiles of the SFGs in the $z\sim 1$ sample with no detected winds (continued). \label{fig:galaxyroster_nowindnoagn5}}
\end{figure}

\begin{figure}
\centering 
\includegraphics[width = 5.8 in]{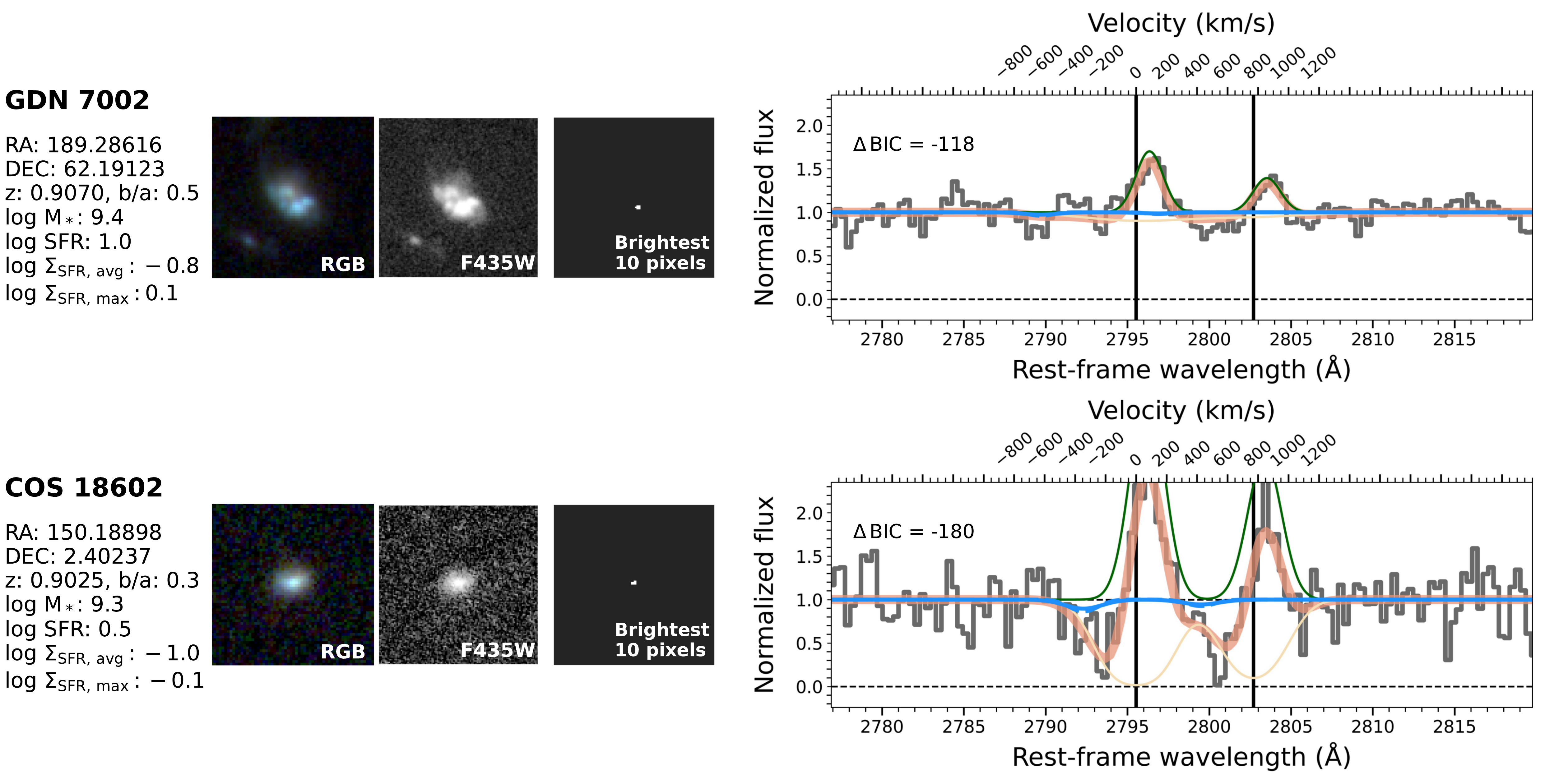}
\caption{Images and \ion{Mg}{2} line profiles of the SFGs in the $z\sim 1$ sample with no detected winds (continued). \label{fig:galaxyroster_nowindnoagn6}}
\end{figure}

\begin{figure}
\centering 
\includegraphics[width = 5.8 in]{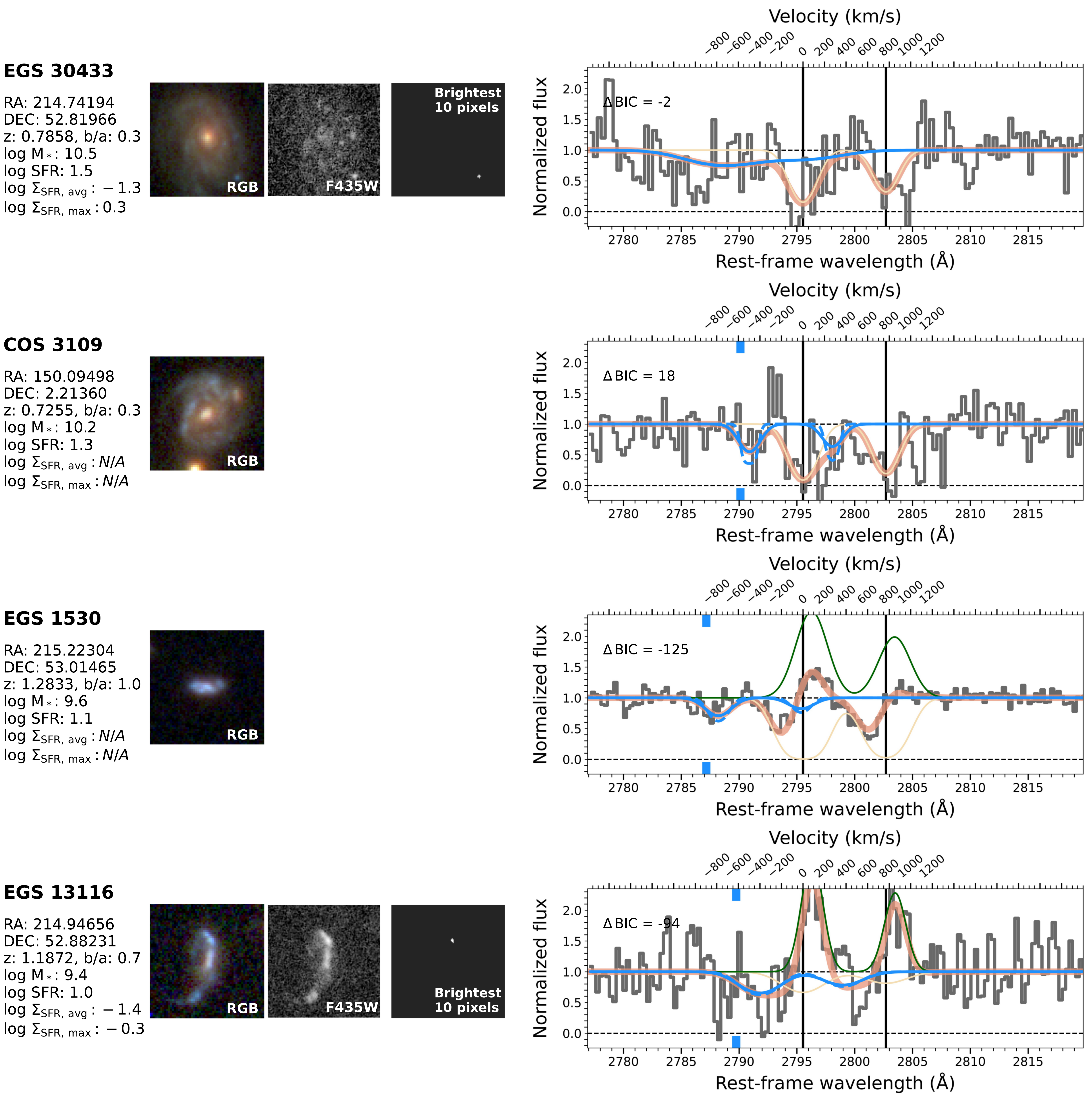}
\caption{Images and \ion{Mg}{2} line profiles of the SFGs in the $z\sim 1$ sample with potentially problematic \ion{Mg}{2} line fittings.  Four galaxies are flagged, which only account for less than $5\%$ of the total sample, and they are not included for wind velocity measurements. Refer to the caption of Figure \ref{fig:galaxyroster_windnoagn1}.   \label{fig:galaxyroster_flaggednoagn}}
\end{figure}

Best-fit \ion{Mg}{2} profile models of our sample of 86 $z\sim 1$ SFGs, along with their images, are presented in Figures \ref{fig:galaxyroster_windnoagn1}--\ref{fig:galaxyroster_nowindnoagn6}. Figures \ref{fig:galaxyroster_windnoagn1}--\ref{fig:galaxyroster_windnoagn9} present galaxies with detected winds, and Figures \ref{fig:galaxyroster_nowindnoagn1}--\ref{fig:galaxyroster_nowindnoagn6} present those without. Additionally, Figure \ref{fig:galaxyroster_flaggednoagn} presents four galaxies flagged due to potentially problematic line profile fits, and they are not included in our sample.

%
\FloatBarrier

\bibliography{sample631}{}

@ARTICLE{delaVega2025,
       author = {{de la Vega}, Alexander and {Kassin}, Susan A. and {Pacifici}, Camilla and {Charlot}, St{\'e}phane and {Curtis-Lake}, Emma and {Chevallard}, Jacopo and {Heckman}, Timothy M. and {Koekemoer}, Anton M. and {Wang}, Weichen},
        title = "{Improved SED-fitting Assumptions Result in Inside-out Quenching at z \raisebox{-0.5ex}\textasciitilde 0.5 and Quenching at All Radii Simultaneously at z \raisebox{-0.5ex}\textasciitilde 1}",
      journal = {\apj},
         year = 2025,
        month = feb,
       volume = {980},
       number = {2},
          eid = {168},
        pages = {168},
          doi = {10.3847/1538-4357/ada8a2},
archivePrefix = {arXiv},
       eprint = {2501.06297},
 primaryClass = {astro-ph.GA},
       adsurl = {https://ui.adsabs.harvard.edu/abs/2025ApJ...980..168D}
}

@ARTICLE{Guo2012,
       author = {{Guo}, Yicheng and {Giavalisco}, Mauro and {Ferguson}, Henry C. and {Cassata}, Paolo and {Koekemoer}, Anton M.},
        title = "{Multi-wavelength View of Kiloparsec-scale Clumps in Star-forming Galaxies at z \raisebox{-0.5ex}\textasciitilde 2}",
      journal = {\apj},
         year = 2012,
        month = oct,
       volume = {757},
       number = {2},
          eid = {120},
        pages = {120},
          doi = {10.1088/0004-637X/757/2/120},
}

@ARTICLE{Guo2015,
       author = {{Guo}, Yicheng and {Ferguson}, Henry C. and {Bell}, Eric F. and {Koo}, David C. and {Conselice}, Christopher J. and {Giavalisco}, Mauro and {Kassin}, Susan and {Lu}, Yu and {Lucas}, Ray and {Mandelker}, Nir and {McIntosh}, Daniel H. and {Primack}, Joel R. and {Ravindranath}, Swara and {Barro}, Guillermo and {Ceverino}, Daniel and {Dekel}, Avishai and {Faber}, Sandra M. and {Fang}, Jerome J. and {Koekemoer}, Anton M. and {Noeske}, Kai and {Rafelski}, Marc and {Straughn}, Amber},
        title = "{Clumpy Galaxies in CANDELS. I. The Definition of UV Clumps and the Fraction of Clumpy Galaxies at 0.5 < z < 3}",
      journal = {\apj},
         year = 2015,
        month = feb,
       volume = {800},
       number = {1},
          eid = {39},
        pages = {39},
          doi = {10.1088/0004-637X/800/1/39},
}

@article{Kornei2012,
	title = {The properties and prevalence of galactic outflows at z ∼ 1 in the {Extended} {Groth} {Strip}},
	volume = {758},
	issn = {15384357},
	doi = {10.1088/0004-637X/758/2/135},
	number = {2},
	journal = {\apj},
	author = {Kornei, Katherine A. and Shapley, Alice E. and Martin, Crystal L. and Coil, Alison L. and Lotz, Jennifer M. and Schiminovich, David and Bundy, Kevin and Noeske, Kai G.},
	year = {2012},
}

@article{Somerville2015,
	author = {Somerville, Rachel S and Dav{\'{e}}, Romeel},
	doi = {10.1146/annurev-astro-082812-140951},
	issn = {0066-4146},
	journal = {ARA{\&}A},
	pages = {51--113},
	title = {{Physical Models of Galaxy Formation in a Cosmological Framework}},
	volume = {53},
	year = {2015}
}

@ARTICLE{Wang2018,
       author = {{Wang}, Weichen and {Kassin}, Susan A. and {Pacifici}, Camilla and {Barro}, Guillermo and {de la Vega}, Alexander and {Simons}, Raymond C. and {Faber}, S.~M. and {Salmon}, Brett and {Ferguson}, Henry C. and {P{\'e}rez-Gonz{\'a}lez}, Pablo G. and {Snyder}, Gregory F. and {Gordon}, Karl D. and {Chen}, Zhu and {Kodra}, Dritan},
        title = "{Galaxy Inclination and the IRX-{\ensuremath{\beta}} Relation: Effects on UV Star Formation Rate Measurements at Intermediate to High Redshifts}",
      journal = {\apj},
         year = 2018,
        month = dec,
       volume = {869},
       number = {2},
          eid = {161},
        pages = {161},
          doi = {10.3847/1538-4357/aaef79},
       adsurl = {https://ui.adsabs.harvard.edu/abs/2018ApJ...869..161W}
}

@article{Wang2017,
	author = {Wang, Weichen and Faber, S. M. and Liu, F. S. and Guo, Yicheng and Pacifici, Camilla and Koo, David C. and Kassin, Susan A. and Mao, Shude and Fang, Jerome J. and Chen, Zhu and Koekemoer, Anton M. and Kocevski, Dale D. and Ashby, M. L. N.},
	doi = {10.1093/mnras/stx1148},
	issn = {0035-8711},
	journal = {\mnras},
	month = {aug},
	number = {4},
	pages = {4063--4082},
	title = {{UVI colour gradients of 0.4{\textless}z{\textless}1.4 star-forming main sequence galaxies in CANDELS: dust extinction and star formation profiles}},
	volume = {469},
	year = {2017}
}

@ARTICLE{Bordoloi2011,
	author = {{Bordoloi}, R. and {Lilly}, S.~J. and {Knobel}, C. and {Bolzonella}, M. and
	{Kampczyk}, P. and {Carollo}, C.~M. and {Iovino}, A. and {Zucca}, E. and
	{Contini}, T. and {Kneib}, J. -P. and {Le Fevre}, O. and
	{Mainieri}, V. and {Renzini}, A. and {Scodeggio}, M. and
	{Zamorani}, G. and {Balestra}, I. and {Bardelli}, S. and
	{Bongiorno}, A. and {Caputi}, K. and {Cucciati}, O. and
	{de la Torre}, S. and {de Ravel}, L. and {Garilli}, B. and
	{Kova{\v{c}}}, K. and {Lamareille}, F. and {Le Borgne}, J. -F. and
	{Le Brun}, V. and {Maier}, C. and {Mignoli}, M. and {Pello}, R. and
	{Peng}, Y. and {Perez Montero}, E. and {Presotto}, V. and
	{Scarlata}, C. and {Silverman}, J. and {Tanaka}, M. and {Tasca}, L. and
	{Tresse}, L. and {Vergani}, D. and {Barnes}, L. and {Cappi}, A. and
	{Cimatti}, A. and {Coppa}, G. and {Diener}, C. and {Franzetti}, P. and
	{Koekemoer}, A. and {L{\'o}pez-Sanjuan}, C. and {McCracken}, H.~J. and
	{Moresco}, M. and {Nair}, P. and {Oesch}, P. and {Pozzetti}, L. and
	{Welikala}, N.},
	title = "{The Radial and Azimuthal Profiles of Mg II Absorption around 0.5 \&lt; z \&lt; 0.9 zCOSMOS Galaxies of Different Colors, Masses, and Environments}",
	journal = {ApJ},
	year = 2011,
	month = dec,
	volume = {743},
	number = {1},
	eid = {10},
	pages = {10},
	doi = {10.1088/0004-637X/743/1/10},
	archivePrefix = {arXiv},
	eprint = {1106.0616},
	primaryClass = {astro-ph.CO},
}

@ARTICLE{Peck1972,
       author = {{Peck}, Edson R. and {Reeder}, Kaye},
        title = "{Dispersion of Air}",
      journal = {Journal of the Optical Society of America},
         year = 1972,
        month = aug,
       volume = {62},
       number = {8},
        pages = {958},
          doi = {10.1364/JOSA.62.000958},
       adsurl = {https://ui.adsabs.harvard.edu/abs/1972JOSA...62..958P}
}

@ARTICLE{Heckman1990,
	author = {{Heckman}, Timothy M. and {Armus}, Lee and {Miley}, George K.},
	title = "{On the Nature and Implications of Starburst-driven Galactic Superwinds}",
	journal = {ApJS},
	year = 1990,
	month = dec,
	volume = {74},
	pages = {833},
	doi = {10.1086/191522},
}

@ARTICLE{Thompson2024,
       author = {{Thompson}, Todd A. and {Heckman}, Timothy M.},
        title = "{Theory and Observation of Winds from Star-Forming Galaxies}",
      journal = {\araa},
         year = 2024,
        month = sep,
       volume = {62},
       number = {1},
        pages = {529-591},
          doi = {10.1146/annurev-astro-041224-011924},
archivePrefix = {arXiv},
       eprint = {2406.08561},
 primaryClass = {astro-ph.GA},
       adsurl = {https://ui.adsabs.harvard.edu/abs/2024ARA&A..62..529T}
}

@article{Noeske2007,
	author = {Noeske, K. G. and Weiner, B. J. and Faber, S. M. and Papovich, C. and Koo, D. C. and Somerville, R. S. and Bundy, K. and Conselice, C. J. and Newman, J. a. and Schiminovich, D. and {Le Floc'h}, E. and Coil, a. L. and Rieke, G. H. and Lotz, J. M. and Primack, J. R. and Barmby, P. and Cooper, M. C. and Davis, M. and Ellis, R. S. and Fazio, G. G. and Guhathakurta, P. and Huang, J. and Kassin, S. a. and Martin, D. C. and Phillips, a. C. and Rich, R. M. and Small, T. a. and Willmer, C. N. a. and Wilson, G.},
	journal = {ApJL},
	number = {1},
	pages = {L43--L46},
	volume = {660},
	year = {2007}
}

@article{Noeske2007a,
	author = {Noeske, K G and Faber, S M and Weiner, B J and Koo, D C and Primack, J R and Dekel, a and Papovich, C and Conselice, C J and {Le Floc'h}, E and Rieke, G H and Coil, a L and Lotz, J M and Somerville, R S and Bundy, K},
	journal = {ApJL},
	number = {1},
	pages = {L47--L50},
	volume = {660},
	year = {2007}
}

@article{Whitaker2014,
	author = {Whitaker, Katherine E. and Franx, Marijn and Leja, Joel and {Van Dokkum}, Pieter G. and Henry, Alaina and Skelton, Rosalind E. and Fumagalli, Mattia and Momcheva, Ivelina G. and Brammer, Gabriel B. and Labb{\'{e}}, Ivo and Nelson, Erica J. and Rigby, Jane R.},
	journal = {ApJ},
	number = {2},
	pages = {104},
	title = {{Constraining the low-mass slope of the star formation sequence at 0.5 {\textless} z {\textless} 2.5}},
	volume = {795},
	year = {2014}
}

@article{Madau2014,
	author = {Madau, Piero and Dickinson, Mark},
	journal = {ARA{\&}A},
	number = {1},
	pages = {415--488},
	title = {{Cosmic Star-Formation History}},
	volume = {52},
	year = {2014}
}

@ARTICLE{Chevalier1985,
       author = {{Chevalier}, R.~A. and {Clegg}, A.~W.},
        title = "{Wind from a starburst galaxy nucleus}",
      journal = {\nat},
         year = 1985,
        month = sep,
       volume = {317},
       number = {6032},
        pages = {44-45},
          doi = {10.1038/317044a0},
       adsurl = {https://ui.adsabs.harvard.edu/abs/1985Natur.317...44C}
}

@ARTICLE{Bordoloi2014,
	author = {{Bordoloi}, R. and {Lilly}, S.~J. and {Hardmeier}, E. and {Contini}, T. and
	{Kneib}, J. -P. and {Le Fevre}, O. and {Mainieri}, V. and
	{Renzini}, A. and {Scodeggio}, M. and {Zamorani}, G. and
	{Bardelli}, S. and {Bolzonella}, M. and {Bongiorno}, A. and
	{Caputi}, K. and {Carollo}, C.~M. and {Cucciati}, O. and
	{de la Torre}, S. and {de Ravel}, L. and {Garilli}, B. and
	{Iovino}, A. and {Kampczyk}, P. and {Kova{\v{c}}}, K. and {Knobel}, C. and
	{Lamareille}, F. and {Le Borgne}, J. -F. and {Le Brun}, V. and
	{Maier}, C. and {Mignoli}, M. and {Oesch}, P. and {Pello}, R. and
	{Peng}, Y. and {Perez Montero}, E. and {Presotto}, V. and
	{Silverman}, J. and {Tanaka}, M. and {Tasca}, L. and {Tresse}, L. and
	{Vergani}, D. and {Zucca}, E. and {Cappi}, A. and {Cimatti}, A. and
	{Coppa}, G. and {Franzetti}, P. and {Koekemoer}, A. and {Moresco}, M. and
	{Nair}, P. and {Pozzetti}, L.},
	title = "{The Dependence of Galactic Outflows on the Properties and Orientation of zCOSMOS Galaxies at z \raisebox{-0.5ex}\textasciitilde 1}",
	journal = {ApJ},
	year = 2014,
	month = oct,
	volume = {794},
	number = {2},
	eid = {130},
	pages = {130},
	doi = {10.1088/0004-637X/794/2/130},
	archivePrefix = {arXiv},
	eprint = {1307.6553},
	primaryClass = {astro-ph.CO},
}

@article{sugahara_evolution_2017,
title = {Evolution of {Galactic} {Outflows} at \$z{\textbackslash}sim 0{\textbackslash}mbox\{--\}2\$ {Revealed} with {SDSS}, {DEEP2}, and {Keck} {Spectra}},
volume = {850},
doi = {10.3847/1538-4357/aa956d},
number = {1},
journal = {ApJ},
author = {Sugahara, Yuma and Ouchi, Masami and Lin, Lihwai and Martin, Crystal L. and Ono, Yoshiaki and Harikane, Yuichi and Shibuya, Takatoshi and Yan, Renbin},
year = {2017},
pages = {51},
}

@ARTICLE{Berg2022,
       author = {{Berg}, Danielle A. and {James}, Bethan L. and {King}, Teagan and {McDonald}, Meaghan and {Chen}, Zuyi and {Chisholm}, John and {Heckman}, Timothy and {Martin}, Crystal L. and {Stark}, Dan P. and {Aloisi}, Alessandra and {Amor{\'\i}n}, Ricardo O. and {Arellano-C{\'o}rdova}, Karla Z. and {Bayliss}, Matthew and {Bordoloi}, Rongmon and {Brinchmann}, Jarle and {Charlot}, St{\'e}phane and {Chevallard}, Jacopo and {Clark}, Ilyse and {Erb}, Dawn K. and {Feltre}, Anna and {Gronke}, Max and {Hayes}, Matthew and {Henry}, Alaina and {Hernandez}, Svea and {Jaskot}, Anne and {Jones}, Tucker and {Kewley}, Lisa J. and {Kumari}, Nimisha and {Leitherer}, Claus and {Llerena}, Mario and {Maseda}, Michael and {Mingozzi}, Matilde and {Nanayakkara}, Themiya and {Ouchi}, Masami and {Plat}, Adele and {Pogge}, Richard W. and {Ravindranath}, Swara and {Rigby}, Jane R. and {Sanders}, Ryan and {Scarlata}, Claudia and {Senchyna}, Peter and {Skillman}, Evan D. and {Steidel}, Charles C. and {Strom}, Allison L. and {Sugahara}, Yuma and {Wilkins}, Stephen M. and {Wofford}, Aida and {Xu}, Xinfeng and {Classy Team}},
        title = "{The COS Legacy Archive Spectroscopy Survey (CLASSY) Treasury Atlas}",
      journal = {\apjs},
         year = 2022,
        month = aug,
       volume = {261},
       number = {2},
          eid = {31},
        pages = {31},
          doi = {10.3847/1538-4365/ac6c03},
archivePrefix = {arXiv},
       eprint = {2203.07357},
 primaryClass = {astro-ph.GA},
       adsurl = {https://ui.adsabs.harvard.edu/abs/2022ApJS..261...31B}
}

@ARTICLE{Xu2022,
       author = {{Xu}, Xinfeng and {Heckman}, Timothy and {Henry}, Alaina and {Berg}, Danielle A. and {Chisholm}, John and {James}, Bethan L. and {Martin}, Crystal L. and {Stark}, Daniel P. and {Aloisi}, Alessandra and {Amor{\'\i}n}, Ricardo O. and {Arellano-C{\'o}rdova}, Karla Z. and {Bordoloi}, Rongmon and {Charlot}, St{\'e}phane and {Chen}, Zuyi and {Hayes}, Matthew and {Mingozzi}, Matilde and {Sugahara}, Yuma and {Kewley}, Lisa J. and {Ouchi}, Masami and {Scarlata}, Claudia and {Steidel}, Charles C.},
        title = "{CLASSY III. The Properties of Starburst-driven Warm Ionized Outflows}",
      journal = {\apj},
         year = 2022,
        month = jul,
       volume = {933},
       number = {2},
          eid = {222},
        pages = {222},
          doi = {10.3847/1538-4357/ac6d56},
archivePrefix = {arXiv},
       eprint = {2204.09181},
 primaryClass = {astro-ph.GA},
       adsurl = {https://ui.adsabs.harvard.edu/abs/2022ApJ...933..222X}
}

@article{Heckman2015,
author = {{Heckman}, Timothy M. and Alexandroff, Rachel M. and Borthakur, Sanchayeeta and Overzier, Roderik and Leitherer, Claus},
doi = {10.1088/0004-637X/809/2/147},
issn = {1538-4357},
journal = {ApJ},
month = {aug},
number = {2},
pages = {147},
title = {{THE SYSTEMATIC PROPERTIES OF THE WARM PHASE OF STARBURST-DRIVEN GALACTIC WINDS}},
volume = {809},
year = {2015}
}

@article{Heckman2016,
author = {Heckman, Timothy M. and Borthakur, Sanchayeeta},
doi = {10.3847/0004-637X/822/1/9},
issn = {1538-4357},
journal = {ApJ},
month = {apr},
number = {1},
pages = {9},
title = {{THE IMPLICATIONS OF EXTREME OUTFLOWS FROM EXTREME STARBURSTS}},
volume = {822},
year = {2016}
}

@article{Brammer2008,
	author = {Brammer, Gabriel B. and van Dokkum, Pieter G. and Coppi, Paolo},
	journal = {\apj},
	number = {2},
	pages = {1503--1513},
	title = {{EAZY: A Fast, Public Photometric Redshift Code}},
	volume = {686},
	year = {2008}
}

@ARTICLE{Grogin2011,
	author = {{Grogin}, Norman A. and {Kocevski}, Dale D. and {Faber}, S.~M. and
	{Ferguson}, Henry C. and {Koekemoer}, Anton M. and {Riess}, Adam
	G. and {Acquaviva}, Viviana and {Alexander}, David M. and
	{Almaini}, Omar and {Ashby}, Matthew L.~N. and {Barden}, Marco
	and {Bell}, Eric F. and {Bournaud}, Fr{\'e}d{\'e}ric and
	{Brown}, Thomas M. and {Caputi}, Karina I. and {Casertano},
	Stefano and {Cassata}, Paolo and {Castellano}, Marco and
	{Challis}, Peter and {Chary}, Ranga-Ram and {Cheung}, Edmond and
	{Cirasuolo}, Michele and {Conselice}, Christopher J. and {Roshan
	Cooray}, Asantha and {Croton}, Darren J. and {Daddi}, Emanuele
	and {Dahlen}, Tomas and {Dav{\'e}}, Romeel and {de Mello},
	Du{\'\i}lia F. and {Dekel}, Avishai and {Dickinson}, Mark and
	{Dolch}, Timothy and {Donley}, Jennifer L. and {Dunlop}, James
	S. and {Dutton}, Aaron A. and {Elbaz}, David and {Fazio},
	Giovanni G. and {Filippenko}, Alexei V. and {Finkelstein},
	Steven L. and {Fontana}, Adriano and {Gardner}, Jonathan P. and
	{Garnavich}, Peter M. and {Gawiser}, Eric and {Giavalisco},
	Mauro and {Grazian}, Andrea and {Guo}, Yicheng and {Hathi},
	Nimish P. and {H{\"a}ussler}, Boris and {Hopkins}, Philip F. and
	{Huang}, Jia-Sheng and {Huang}, Kuang-Han and {Jha}, Saurabh W.
	and {Kartaltepe}, Jeyhan S. and {Kirshner}, Robert P. and {Koo},
	David C. and {Lai}, Kamson and {Lee}, Kyoung-Soo and {Li},
	Weidong and {Lotz}, Jennifer M. and {Lucas}, Ray A. and {Madau},
	Piero and {McCarthy}, Patrick J. and {McGrath}, Elizabeth J. and
	{McIntosh}, Daniel H. and {McLure}, Ross J. and {Mobasher},
	Bahram and {Moustakas}, Leonidas A. and {Mozena}, Mark and
	{Nandra}, Kirpal and {Newman}, Jeffrey A. and {Niemi}, Sami-
	Matias and {Noeske}, Kai G. and {Papovich}, Casey J. and
	{Pentericci}, Laura and {Pope}, Alexandra and {Primack}, Joel R.
	and {Rajan}, Abhijith and {Ravindranath}, Swara and {Reddy},
	Naveen A. and {Renzini}, Alvio and {Rix}, Hans-Walter and
	{Robaina}, Aday R. and {Rodney}, Steven A. and {Rosario}, David
	J. and {Rosati}, Piero and {Salimbeni}, Sara and {Scarlata},
	Claudia and {Siana}, Brian and {Simard}, Luc and {Smidt}, Joseph
	and {Somerville}, Rachel S. and {Spinrad}, Hyron and {Straughn},
	Amber N. and {Strolger}, Louis-Gregory and {Telford}, Olivia and
	{Teplitz}, Harry I. and {Trump}, Jonathan R. and {van der Wel},
	Arjen and {Villforth}, Carolin and {Wechsler}, Risa H. and
	{Weiner}, Benjamin J. and {Wiklind}, Tommy and {Wild}, Vivienne
	and {Wilson}, Grant and {Wuyts}, Stijn and {Yan}, Hao-Jing and
	{Yun}, Min S.},
	title = "{CANDELS: The Cosmic Assembly Near-infrared Deep Extragalactic Legacy Survey}",
	journal = {\apjs},
	year = 2011,
	month = Dec,
	volume = {197},
	eid = {35},
	pages = {35},
}

@article{Koekemoer2011,
	author = {Koekemoer, Anton M. and Faber, S. M. and Ferguson, Henry C. and Grogin, Norman A. and Kocevski, Dale D. and Koo, David C. and Lai, Kamson and Lotz, Jennifer M. and Lucas, Ray A. and McGrath, Elizabeth J. and Ogaz, Sara and Rajan, Abhijith and Riess, Adam G. and Rodney, Steve A. and Strolger, Louis and Casertano, Stefano and Castellano, Marco and Dahlen, Tomas and Dickinson, Mark and Dolch, Timothy and Fontana, Adriano and Giavalisco, Mauro and Grazian, Andrea and Guo, Yicheng and Hathi, Nimish P. and Huang, Kuang Han and {Van Der Wel}, Arjen and Yan, Hao Jing and Acquaviva, Viviana and Alexander, David M. and Almaini, Omar and Ashby, Matthew L.N. and Barden, Marco and Bell, Eric F. and Bournaud, Fr{\'{e}}d{\'{e}}ric and Brown, Thomas M. and Caputi, Karina I. and Cassata, Paolo and Challis, Peter J. and Chary, Ranga Ram and Cheung, Edmond and Cirasuolo, Michele and Conselice, Christopher J. and Cooray, Asantha Roshan and Croton, Darren J. and Daddi, Emanuele and Dav{\'{e}}, Romeel and {De Mello}, Duilia F. and {De Ravel}, Loic and Dekel, Avishai and Donley, Jennifer L. and Dunlop, James S. and Dutton, Aaron A. and Elbaz, David and Fazio, Giovanni G. and Filippenko, Alexei V. and Finkelstein, Steven L. and Frazer, Chris and Gardner, Jonathan P. and Garnavich, Peter M. and Gawiser, Eric and Gruetzbauch, Ruth and Hartley, Will G. and H{\"{a}}ussler, Boris and Herrington, Jessica and Hopkins, Philip F. and Huang, Jia Sheng and Jha, Saurabh W. and Johnson, Andrew and Kartaltepe, Jeyhan S. and Khostovan, Ali A. and Kirshner, Robert P. and Lani, Caterina and Lee, Kyoung Soo and Li, Weidong and Madau, Piero and McCarthy, Patrick J. and McIntosh, Daniel H. and McLure, Ross J. and McPartland, Conor and Mobasher, Bahram and Moreira, Heidi and Mortlock, Alice and Moustakas, Leonidas A. and Mozena, Mark and Nandra, Kirpal and Newman, Jeffrey A. and Nielsen, Jennifer L. and Niemi, Sami and Noeske, Kai G. and Papovich, Casey J. and Pentericci, Laura and Pope, Alexandra and Primack, Joel R. and Ravindranath, Swara and Reddy, Naveen A. and Renzini, Alvio and Rix, Hans Walter and Robaina, Aday R. and Rosario, David J. and Rosati, Piero and Salimbeni, Sara and Scarlata, Claudia and Siana, Brian and Simard, Luc and Smidt, Joseph and Snyder, Diana and Somerville, Rachel S. and Spinrad, Hyron and Straughn, Amber N. and Telford, Olivia and Teplitz, Harry I. and Trump, Jonathan R. and Vargas, Carlos and Villforth, Carolin and Wagner, Cory R. and Wandro, Pat and Wechsler, Risa H. and Weiner, Benjamin J. and Wiklind, Tommy and Wild, Vivienne and Wilson, Grant and Wuyts, Stijn and Yun, Min S.},
	journal = {ApJS},
	number = {2},
	pages = {36},
	title = {{Candels: The cosmic assembly near-infrared deep extragalactic legacy survey - The hubble space telescope observations, imaging data products, and mosaics}},
	volume = {197},
	year = {2011}
}

@article{Bertin1996,
	author = {Bertin, E. and Arnouts, S.},
	doi = {10.1051/aas:1996164},
	issn = {0365-0138},
	journal = {A{\&}AS},
	month = {jun},
	number = {2},
	pages = {393--404},
	pmid = {6547660},
	title = {{SExtractor: Software for source extraction}},
	volume = {117},
	year = {1996}
}

@article{Donley2012,
	author = {Donley, J. L. and Koekemoer, A. M. and Brusa, M. and Capak, P. and Cardamone, C. N. and Civano, F. and Ilbert, O. and Impey, C. D. and Kartaltepe, J. S. and Miyaji, T. and Salvato, M. and Sanders, D. B. and Trump, J. R. and Zamorani, G.},
	journal = {ApJ},
	number = {2},
	pages = {142},
	title = {{Identifying luminous active galactic nuclei in deep surveys: Revised IRAC selection criteria}},
	volume = {748},
	year = {2012}
}

@ARTICLE{Rodriguez-Gomez2019,
     author = {{Rodriguez-Gomez}, Vicente and {Snyder}, Gregory F. and {Lotz}, Jennifer M. and {Nelson}, Dylan and {Pillepich}, Annalisa and {Springel}, Volker and {Genel}, Shy and {Weinberger}, Rainer and {Tacchella}, Sandro and {Pakmor}, R{\"u}diger and {Torrey}, Paul and {Marinacci}, Federico and {Vogelsberger}, Mark and {Hernquist}, Lars and {Thilker}, David A.},
     title = "{The optical morphologies of galaxies in the IllustrisTNG simulation: a comparison to Pan-STARRS observations}",
     journal = {\mnras},
         year = 2019,
        month = mar,
       volume = {483},
       number = {3},
        pages = {4140-4159},
          doi = {10.1093/mnras/sty3345},
}

@article{Yesuf2017,
	author = {Yesuf, Hassen M. and Koo, David C. and Faber, S. M. and Prochaska, J. Xavier and Guo, Yicheng and Liu, F. S. and Cunningham, Emily C. and Coil, Alison L. and Guhathakurta, Puragra},
	journal = {ApJ},
	month = {may},
	number = {2},
	pages = {83},
	title = {{No evidence for feedback: Unexceptional Low-ionization winds in Host galaxies of Low Luminosity Active Galactic Nuclei at Redshift z {\~{}}1}},
	volume = {841},
	year = {2017}
}

@MISC{Cooper2012,
	author = {{Cooper}, Michael C. and {Newman}, Jeffrey A. and {Davis}, Marc and
	{Finkbeiner}, Douglas P. and {Gerke}, Brian F.},
	title = "{spec2d: DEEP2 DEIMOS Spectral Pipeline}",
	year = 2012,
	month = mar,
	eid = {ascl:1203.003},
	pages = {ascl:1203.003},
	archivePrefix = {ascl},
	eprint = {1203.003},
	adsurl = {https://ui.adsabs.harvard.edu/abs/2012ascl.soft03003C}
}

@article{Xue2016,
   author = {Y. Q. Xue and B. Luo and W. N. Brandt and D. M. Alexander and F. E. Bauer and B. D. Lehmer and G. Yang},
   doi = {10.3847/0067-0049/224/2/15},
   issn = {1538-4365},
   issue = {2},
   journal = {\apjs},
   month = {5},
   pages = {15},
   title = {The 2 Ms Chandra Deep Field-North Survey and the 250 ks Extended Chandra Deep Field-South Survey: Improved Point-Source Catalogs},
   volume = {224},
   year = {2016},
}

@article{Xue2011,
   author = {Y. Q. Xue and B. Luo and W. N. Brandt and F. E. Bauer and B. D. Lehmer and P. S. Broos and D. P. Schneider and D. M. Alexander and M. Brusa and A. Comastri and A. C. Fabian and R. Gilli and G. Hasinger and A. E. Hornschemeier and A. Koekemoer and T. Liu and V. Mainieri and M. Paolillo and D. A. Rafferty and P. Rosati and O. Shemmer and J. D. Silverman and I. Smail and P. Tozzi and C. Vignali},
   doi = {10.1088/0067-0049/195/1/10},
   isbn = {0067-0049\r1538-4365},
   issn = {00670049},
   issue = {1},
   journal = {\apjs},
   month = {7},
   pages = {10},
   title = {The Chandra Deep Field-South survey: 4 Ms source catalogs},
   volume = {195},
   year = {2011},
}

@article{Nandra2015,
   author = {K. Nandra and E. S. Laird and J. A. Aird and M. Salvato and A. Georgakakis and G. Barro and P. G. Perez-Gonzalez and P. Barmby and R.-R. Chary and A. Coil and M. C. Cooper and M. Davis and M. Dickinson and S. M. Faber and G. G. Fazio and P. Guhathakurta and S. Gwyn and L.-T. Hsu and J.-S. Huang and R. J. Ivison and D. C. Koo and J. A. Newman and C. Rangel and T. Yamada and C. Willmer},
   doi = {10.1088/0067-0049/220/1/10},
   issn = {1538-4365},
   issue = {1},
   journal = {\apjs},
   month = {9},
   pages = {10},
   title = {AEGIS-X: DEEP <i>CHANDRA</i> IMAGING OF THE CENTRAL GROTH STRIP},
   volume = {220},
   year = {2015},
}

@article{Civano2016,
   author = {F. Civano and S. Marchesi and A. Comastri and M. C. Urry and M. Elvis and N. Cappelluti and S. Puccetti and M. Brusa and G. Zamorani and G. Hasinger and T. Aldcroft and D. M. Alexander and V. Allevato and H. Brunner and P. Capak and A. Finoguenov and F. Fiore and A. Fruscione and R. Gilli and K. Glotfelty and R. E. Griffiths and H. Hao and F. A. Harrison and K. Jahnke and J. Kartaltepe and A. Karim and S. M. LaMassa and G. Lanzuisi and T. Miyaji and P. Ranalli and M. Salvato and M. Sargent and N. J. Scoville and K. Schawinski and E. Schinnerer and J. Silverman and V. Smolcic and D. Stern and S. Toft and B. Trakhenbrot and E. Treister and C. Vignali},
   doi = {10.3847/0004-637X/819/1/62},
   issn = {1538-4357},
   issue = {1},
   journal = {\apj},
   month = {2},
   pages = {62},
   title = {THE <i>CHANDRA</i> COSMOS LEGACY SURVEY: OVERVIEW AND POINT SOURCE CATALOG},
   volume = {819},
   year = {2016},
}

@ARTICLE{astropy2018,
	author = {{Astropy Collaboration} and {Price-Whelan}, A.~M. and
	{Sip{\H{o}}cz}, B.~M. and {G{\"u}nther}, H.~M. and {Lim}, P.~L. and
	{Crawford}, S.~M. and {Conseil}, S. and {Shupe}, D.~L. and
	{Craig}, M.~W. and {Dencheva}, N. and {Ginsburg}, A. and {Vand
	erPlas}, J.~T. and {Bradley}, L.~D. and {P{\'e}rez-Su{\'a}rez}, D. and
	{de Val-Borro}, M. and {Aldcroft}, T.~L. and {Cruz}, K.~L. and
	{Robitaille}, T.~P. and {Tollerud}, E.~J. and {Ardelean}, C. and
	{Babej}, T. and {Bach}, Y.~P. and {Bachetti}, M. and {Bakanov}, A.~V. and
	{Bamford}, S.~P. and {Barentsen}, G. and {Barmby}, P. and
	{Baumbach}, A. and {Berry}, K.~L. and {Biscani}, F. and {Boquien}, M. and
	{Bostroem}, K.~A. and {Bouma}, L.~G. and {Brammer}, G.~B. and
	{Bray}, E.~M. and {Breytenbach}, H. and {Buddelmeijer}, H. and
	{Burke}, D.~J. and {Calderone}, G. and {Cano Rodr{\'\i}guez}, J.~L. and
	{Cara}, M. and {Cardoso}, J.~V.~M. and {Cheedella}, S. and {Copin}, Y. and
	{Corrales}, L. and {Crichton}, D. and {D'Avella}, D. and {Deil}, C. and
	{Depagne}, {\'E}. and {Dietrich}, J.~P. and {Donath}, A. and
	{Droettboom}, M. and {Earl}, N. and {Erben}, T. and {Fabbro}, S. and
	{Ferreira}, L.~A. and {Finethy}, T. and {Fox}, R.~T. and
	{Garrison}, L.~H. and {Gibbons}, S.~L.~J. and {Goldstein}, D.~A. and
	{Gommers}, R. and {Greco}, J.~P. and {Greenfield}, P. and
	{Groener}, A.~M. and {Grollier}, F. and {Hagen}, A. and {Hirst}, P. and
	{Homeier}, D. and {Horton}, A.~J. and {Hosseinzadeh}, G. and {Hu}, L. and
	{Hunkeler}, J.~S. and {Ivezi{\'c}}, {\v{Z}}. and {Jain}, A. and
	{Jenness}, T. and {Kanarek}, G. and {Kendrew}, S. and {Kern}, N.~S. and
	{Kerzendorf}, W.~E. and {Khvalko}, A. and {King}, J. and {Kirkby}, D. and
	{Kulkarni}, A.~M. and {Kumar}, A. and {Lee}, A. and {Lenz}, D. and
	{Littlefair}, S.~P. and {Ma}, Z. and {Macleod}, D.~M. and
	{Mastropietro}, M. and {McCully}, C. and {Montagnac}, S. and
	{Morris}, B.~M. and {Mueller}, M. and {Mumford}, S.~J. and {Muna}, D. and
	{Murphy}, N.~A. and {Nelson}, S. and {Nguyen}, G.~H. and
	{Ninan}, J.~P. and {N{\"o}the}, M. and {Ogaz}, S. and {Oh}, S. and
	{Parejko}, J.~K. and {Parley}, N. and {Pascual}, S. and {Patil}, R. and
	{Patil}, A.~A. and {Plunkett}, A.~L. and {Prochaska}, J.~X. and
	{Rastogi}, T. and {Reddy Janga}, V. and {Sabater}, J. and
	{Sakurikar}, P. and {Seifert}, M. and {Sherbert}, L.~E. and
	{Sherwood-Taylor}, H. and {Shih}, A.~Y. and {Sick}, J. and
	{Silbiger}, M.~T. and {Singanamalla}, S. and {Singer}, L.~P. and
	{Sladen}, P.~H. and {Sooley}, K.~A. and {Sornarajah}, S. and
	{Streicher}, O. and {Teuben}, P. and {Thomas}, S.~W. and
	{Tremblay}, G.~R. and {Turner}, J.~E.~H. and {Terr{\'o}n}, V. and
	{van Kerkwijk}, M.~H. and {de la Vega}, A. and {Watkins}, L.~L. and
	{Weaver}, B.~A. and {Whitmore}, J.~B. and {Woillez}, J. and
	{Zabalza}, V. and {Astropy Contributors}},
	title = "{The Astropy Project: Building an Open-science Project and Status of the v2.0 Core Package}",
	journal = {\aj},
	year = 2018,
	month = sep,
	volume = {156},
	number = {3},
	eid = {123},
	pages = {123},
	doi = {10.3847/1538-3881/aabc4f},
	archivePrefix = {arXiv},
	eprint = {1801.02634},
	primaryClass = {astro-ph.IM},
	adsurl = {https://ui.adsabs.harvard.edu/abs/2018AJ....156..123A}
}

@ARTICLE{astropy2013,
	author = {{Astropy Collaboration} and {Robitaille}, Thomas P. and
	{Tollerud}, Erik J. and {Greenfield}, Perry and {Droettboom}, Michael and
	{Bray}, Erik and {Aldcroft}, Tom and {Davis}, Matt and
	{Ginsburg}, Adam and {Price-Whelan}, Adrian M. and
	{Kerzendorf}, Wolfgang E. and {Conley}, Alexander and {Crighton}, Neil and
	{Barbary}, Kyle and {Muna}, Demitri and {Ferguson}, Henry and
	{Grollier}, Fr{\'e}d{\'e}ric and {Parikh}, Madhura M. and
	{Nair}, Prasanth H. and {Unther}, Hans M. and {Deil}, Christoph and
	{Woillez}, Julien and {Conseil}, Simon and {Kramer}, Roban and
	{Turner}, James E.~H. and {Singer}, Leo and {Fox}, Ryan and
	{Weaver}, Benjamin A. and {Zabalza}, Victor and {Edwards}, Zachary I. and
	{Azalee Bostroem}, K. and {Burke}, D.~J. and {Casey}, Andrew R. and
	{Crawford}, Steven M. and {Dencheva}, Nadia and {Ely}, Justin and
	{Jenness}, Tim and {Labrie}, Kathleen and {Lim}, Pey Lian and
	{Pierfederici}, Francesco and {Pontzen}, Andrew and {Ptak}, Andy and
	{Refsdal}, Brian and {Servillat}, Mathieu and {Streicher}, Ole},
	title = "{Astropy: A community Python package for astronomy}",
	journal = {\aap},
	year = 2013,
	month = oct,
	volume = {558},
	eid = {A33},
	pages = {A33},
	doi = {10.1051/0004-6361/201322068},
	archivePrefix = {arXiv},
	eprint = {1307.6212},
	primaryClass = {astro-ph.IM},
	adsurl = {https://ui.adsabs.harvard.edu/abs/2013A&A...558A..33A}
}

@article{Chabrier2003,
	author = {Chabrier, Gilles},
	journal = {\pasp},
	number = {809},
	pages = {763--795},
	title = {{Galactic Stellar and Substellar Initial Mass Function}},
	volume = {115},
	year = {2003}
}

@BOOK{Spitzer1978,
	author = {{Spitzer}, Lyman},
	title = "{Physical processes in the interstellar medium}",
	year = "1978",
	doi = {10.1002/9783527617722}
}

@article{Foreman-Mackey2019,
	author = {Foreman-Mackey, Daniel and Farr, Will and Sinha, Manodeep and Archibald, Anne and Hogg, David and Sanders, Jeremy and Zuntz, Joe and Williams, Peter and Nelson, Andrew and de Val-Borro, Miguel and Erhardt, Tobias and Pashchenko, Ilya and Pla, Oriol},
	doi = {10.21105/joss.01864},
	issn = {2475-9066},
	journal = {Journal of Open Source Software},
	number = {43},
	pages = {1864},
	title = {{emcee v3: A Python ensemble sampling toolkit for affine-invariant MCMC}},
	volume = {4},
	year = {2019}
}

@article{Foreman-Mackey2013,
	author = {Foreman-Mackey, Daniel and Hogg, David W. and Lang, Dustin and Goodman, Jonathan},
	doi = {10.1086/670067},
	issn = {00046280},
	journal = {PASP},
	month = {mar},
	number = {925},
	pages = {306--312},
	publisher = {IOP Publishing},
	title = {{emcee : The MCMC Hammer }},
	volume = {125},
	year = {2013}
}

@article{Pacifici2023,
       author = {{Pacifici}, Camilla and {Iyer}, Kartheik G. and {Mobasher}, Bahram and {da Cunha}, Elisabete and {Acquaviva}, Viviana and {Burgarella}, Denis and {Calistro Rivera}, Gabriela and {Carnall}, Adam C. and {Chang}, Yu-Yen and {Chartab}, Nima and {Cooke}, Kevin C. and {Fairhurst}, Ciaran and {Kartaltepe}, Jeyhan and {Leja}, Joel and {Ma{\l}ek}, Katarzyna and {Salmon}, Brett and {Torelli}, Marianna and {Vidal-Garc{\'\i}a}, Alba and {Boquien}, M{\'e}d{\'e}ric and {Brammer}, Gabriel G. and {Brown}, Michael J.~I. and {Capak}, Peter L. and {Chevallard}, Jacopo and {Circosta}, Chiara and {Croton}, Darren and {Davidzon}, Iary and {Dickinson}, Mark and {Duncan}, Kenneth J. and {Faber}, Sandra M. and {Ferguson}, Harry C. and {Fontana}, Adriano and {Guo}, Yicheng and {Haeussler}, Boris and {Hemmati}, Shoubaneh and {Jafariyazani}, Marziye and {Kassin}, Susan A. and {Larson}, Rebecca L. and {Lee}, Bomee and {Mantha}, Kameswara Bharadwaj and {Marchi}, Francesca and {Nayyeri}, Hooshang and {Newman}, Jeffrey A. and {Pandya}, Viraj and {Pforr}, Janine and {Reddy}, Naveen and {Sanders}, Ryan and {Shah}, Ekta and {Shahidi}, Abtin and {Stevans}, Matthew L. and {Triani}, Dian Puspita and {Tyler}, Krystal D. and {Vanderhoof}, Brittany N. and {de la Vega}, Alexander and {Wang}, Weichen and {Weston}, Madalyn E.},
        title = {The Art of Measuring Physical Parameters in Galaxies: A Critical Assessment of Spectral Energy Distribution Fitting Techniques},
      journal = {\apj},
         year = {2023},
        month = {feb},
       volume = {944},
       number = {2},
          eid = {141},
        pages = {141},
          doi = {10.3847/1538-4357/acacff}
}

@article{Zhu2015,
	author = {Zhu, Guangtun Ben and Comparat, Johan and Kneib, Jean Paul and Delubac, Timoth{\'{e}}e and Raichoor, Anand and Dawson, Kyle S. and Newman, Jeffrey and Yeche, Christophe and Zhou, Xu and Schneider, Donald P.},
	doi = {10.1088/0004-637X/815/1/48},
	journal = {ApJ},
	month = {dec},
	number = {1},
	pages = {48},
	title = {{NEAR-ULTRAVIOLET SPECTROSCOPY of STAR-FORMING GALAXIES from eBOSS: SIGNATURES of UBIQUITOUS GALACTIC-SCALE OUTFLOWS}},
	volume = {815},
	year = {2015}
}

@ARTICLE{Chevallard2016,
       author = {{Chevallard}, Jacopo and {Charlot}, St{\'e}phane},
        title = "{Modelling and interpreting spectral energy distributions of galaxies with BEAGLE}",
      journal = {\mnras},
         year = 2016,
        month = oct,
       volume = {462},
       number = {2},
        pages = {1415-1443},
        doi = {10.1093/mnras/stw1756}
        }

@ARTICLE{Rubin2010,
	author = {{Rubin}, Kate H.~R. and {Weiner}, Benjamin J. and {Koo}, David C. and
	{Martin}, Crystal L. and {Prochaska}, J. Xavier and {Coil}, Alison L. and
	{Newman}, Jeffrey A.},
	title = "{The Persistence of Cool Galactic Winds in High Stellar Mass Galaxies between z \raisebox{-0.5ex}\textasciitilde 1.4 and \raisebox{-0.5ex}\textasciitilde1}",
	journal = {ApJ},
	year = 2010,
	month = aug,
	volume = {719},
	number = {2},
	pages = {1503-1525},
	doi = {10.1088/0004-637X/719/2/1503},
	archivePrefix = {arXiv},
	eprint = {0912.2343},
	primaryClass = {astro-ph.CO},
}

@article{Rubin2014,
	author = {{Rubin}, Kate H.~R. and Prochaska, J. Xavier and Koo, David C. and Phillips, Andrew C. and Martin, Crystal L. and Winstrom, Lucas O.},
	doi = {10.1088/0004-637X/794/2/156},
	journal = {ApJ},
	month = {oct},
	number = {2},
	pages = {156},
	title = {{EVIDENCE FOR UBIQUITOUS COLLIMATED GALACTIC-SCALE OUTFLOWS ALONG THE STAR-FORMING SEQUENCE AT z ∼ 0.5}},
	volume = {794},
	year = {2014}
}

@ARTICLE{McLure2002,
       author = {{McLure}, R.~J. and {Jarvis}, M.~J.},
        title = "{Measuring the black hole masses of high-redshift quasars}",
      journal = {\mnras},
         year = 2002,
        month = nov,
       volume = {337},
       number = {1},
        pages = {109-116},
          doi = {10.1046/j.1365-8711.2002.05871.x},
}

@article{Luo2017,
   author = {B. Luo and W. N. Brandt and Y. Q. Xue and B. Lehmer and D. M. Alexander and F. E. Bauer and F. Vito and G. Yang and A. R. Basu-Zych and A. Comastri and R. Gilli and Q.-S. Gu and A. E. Hornschemeier and A. Koekemoer and T. Liu and V. Mainieri and M. Paolillo and P. Ranalli and P. Rosati and D. P. Schneider and O. Shemmer and I. Smail and M. Sun and P. Tozzi and C. Vignali and J.-X. Wang},
   doi = {10.3847/1538-4365/228/1/2},
   issn = {1538-4365},
   issue = {1},
   journal = {\apjs},
   month = {12},
   pages = {2},
   title = {THE <i>CHANDRA</i> DEEP FIELD-SOUTH SURVEY: 7 MS SOURCE CATALOGS},
   volume = {228},
   year = {2017},
}

@article{Newman2013,
	author = {Newman, Jeffrey A. and Cooper, Michael C. and Davis, Marc and Faber, S. M. and Coil, Alison L. and Guhathakurta, Puragra and Koo, David C. and Phillips, Andrew C. and Conroy, Charlie and Dutton, Aaron A. and Finkbeiner, Douglas P. and Gerke, Brian F. and Rosario, David J. and Weiner, Benjamin J. and Willmer, C. N. A. and Yan, Renbin and Harker, Justin J. and Kassin, Susan A. and Konidaris, N. P. and Lai, Kamson and Madgwick, Darren S. and Noeske, K. G. and Wirth, Gregory D. and Connolly, A. J. and Kaiser, N. and Kirby, Evan N. and Lemaux, Brian C. and Lin, Lihwai and Lotz, Jennifer M. and Luppino, G. A. and Marinoni, C. and Matthews, Daniel J. and Metevier, Anne and Schiavon, Ricardo P.},
	journal = {ApJS},
	number = {1},
	pages = {5},
	title = {{THE DEEP2 GALAXY REDSHIFT SURVEY: DESIGN, OBSERVATIONS, DATA REDUCTION, AND REDSHIFTS}},
	volume = {208},
	year = {2013}
}

@ARTICLE{Cunningham2019b,
	author = {{Cunningham}, Emily C. and {Deason}, Alis J. and {Sanderson}, Robyn E. and
	{Sohn}, Sangmo Tony and {Anderson}, Jay and {Guhathakurta}, Puragra and
	{Rockosi}, Constance M. and {van der Marel}, Roeland P. and
	{Loebman}, Sarah R. and {Wetzel}, Andrew},
	title = "{HALO7D II: The Halo Velocity Ellipsoid and Velocity Anisotropy with Distant Main-sequence Stars}",
	journal = {\apj},
	year = 2019,
	month = jul,
	volume = {879},
	number = {2},
	eid = {120},
	pages = {120},
	doi = {10.3847/1538-4357/ab24cd},
	archivePrefix = {arXiv},
	eprint = {1810.12201},
	primaryClass = {astro-ph.GA},
	adsurl = {https://ui.adsabs.harvard.edu/abs/2019ApJ...879..120C}
}

@ARTICLE{Cunningham2019a,
	author = {{Cunningham}, Emily C. and {Deason}, Alis J. and
	{Rockosi}, Constance M. and {Guhathakurta}, Puragra and
	{Jennings}, Zachary G. and {Kirby}, Evan N. and {Toloba}, Elisa and
	{Barro}, Guillermo},
	title = "{HALO7D I. The Line-of-sight Velocities of Distant Main-sequence Stars in the Milky Way Halo}",
	journal = {\apj},
	year = 2019,
	month = may,
	volume = {876},
	number = {2},
	eid = {124},
	pages = {124},
	doi = {10.3847/1538-4357/ab16cb},
	archivePrefix = {arXiv},
	eprint = {1809.04082},
	primaryClass = {astro-ph.GA},
	adsurl = {https://ui.adsabs.harvard.edu/abs/2019ApJ...876..124C}
}

@INPROCEEDINGS{Faber2003,
	author = {{Faber}, Sandra M. and {Phillips}, Andrew C. and {Kibrick}, Robert I.
	and {Alcott}, Barry and {Allen}, Steven L. and {Burrous}, Jim
	and {Cantrall}, T. and {Clarke}, De and {Coil}, Alison L. and
	{Cowley}, David J. and {Davis}, Marc and {Deich}, William T.~S.
	and {Dietsch}, Ken and {Gilmore}, David K. and {Harper}, Carol
	A. and {Hilyard}, David F. and {Lewis}, Jeffrey P. and
	{McVeigh}, Molly and {Newman}, Jeffrey and {Osborne}, Jack and
	{Schiavon}, Ricardo and {Stover}, Richard J. and {Tucker}, Dean
	and {Wallace}, Vernon and {Wei}, Mingzhi and {Wirth}, Gregory
	and {Wright}, Christopher A.},
	title = "{The DEIMOS spectrograph for the Keck II Telescope: integration and testing}",
	booktitle = {Instrument Design and Performance for Optical/Infrared Ground-based Telescopes},
	year = 2003,
	editor = {{Iye}, Masanori and {Moorwood}, Alan F.~M.},
	series = {SPIE Conference Series},
	volume = {4841},
	month = Mar,
	pages = {1657-1669},
}

@ARTICLE{Wang2020AAS,
       author = {{Wang}, Xin and {Teplitz}, Harry I. and {Smith}, Brent M. and {Windhorst}, Rogier A. and {Rafelski}, Marc and {Mehta}, Vihang and {Alavi}, Anahita and {Ji}, Zhiyuan and {Brammer}, Gabriel and {Colbert}, James and {Grogin}, Norman and {Hathi}, Nimish P. and {Koekemoer}, Anton M. and {Prichard}, Laura and {Scarlata}, Claudia and {Sunnquist}, Ben and {Arrabal Haro}, Pablo and {Conselice}, Christopher and {Gawiser}, Eric and {Guo}, Yicheng and {Hayes}, Matthew and {Jansen}, Rolf A. and {Lucas}, Ray A. and {O'Connell}, Robert and {Robertson}, Brant and {Rutkowski}, Michael and {Siana}, Brian and {Vanzella}, Eros and {Ashcraft}, Teresa and {Bagley}, Micaela and {Baronchelli}, Ivano and {Barro}, Guillermo and {Blanche}, Alex and {Broussard}, Adam and {Carleton}, Timothy and {Chartab}, Nima and {Cheng}, Yingjie and {Codoreanu}, Alex and {Cohen}, Seth and {Dai}, Y. Sophia and {Darvish}, Behnam and {Dav{\'e}}, Romeel and {Degroot}, Laura and {de Mello}, Duilia and {Dickinson}, Mark and {Emami}, Najmeh and {Ferguson}, Henry and {Ferreira}, Leonardo and {Finkelstein}, Keely and {Finkelstein}, Steven and {Gardner}, Jonathan P. and {Gburek}, Timothy and {Giavalisco}, Mauro and {Grazian}, Andrea and {Gronwall}, Caryl and {Hemmati}, Shoubaneh and {Howell}, Justin and {Iyer}, Kartheik and {Kaviraj}, Sugata and {Kurczynski}, Peter and {Lazar}, Ilin and {MacKenty}, John and {Mantha}, Kameswara Bharadwaj and {Martin}, Alec and {Martin}, Garreth and {McCabe}, Tyler and {Mobasher}, Bahram and {Nedkova}, Kalina and {Olsen}, Charlotte and {Otteson}, Lillian and {Ravindranath}, Swara and {Redshaw}, Caleb and {Sattari}, Zahra and {Soto}, Emmaris and {Yung}, L.~Y. Aaron and {Zabelle}, Bonnabelle and {UVCANDELS Team}},
        title = "{The Lyman Continuum Escape Fraction of Star-forming Galaxies at 2.4 {\ensuremath{\lesssim}} z {\ensuremath{\lesssim}} 3.0 from UVCANDELS}",
      journal = {\apj},
         year = 2025,
        month = feb,
       volume = {980},
       number = {1},
          eid = {74},
        pages = {74},
          doi = {10.3847/1538-4357/ada4ab},
archivePrefix = {arXiv},
       eprint = {2308.09064},
 primaryClass = {astro-ph.GA},
       adsurl = {https://ui.adsabs.harvard.edu/abs/2025ApJ...980...74W}
}

@article{Concas2022,
author = {Concas, Alice and Maiolino, Roberto and Curti, Mirko and Hayden-Pawson, Connor and Cirasuolo, Michele and Jones, Gareth C and Mercurio, Amata and Belfiore, Francesco and Cresci, Giovanni and Cullen, Fergus and Mannucci, Filippo and Marconi, Alessandro and Cappellari, Michele and Cicone, Claudia and Peng, Yingjie and Troncoso, Paulina},
doi = {10.1093/mnras/stac1026},
eprint = {2203.11958},
journal = {\mnras},
number = {2},
pages = {2535--2562},
title = {{Being KLEVER at cosmic noon: Ionized gas outflows are inconspicuous in low-mass star-forming galaxies but prominent in massive AGN hosts}},
volume = {513},
year = {2022}
}

@article{Chang2015,
   author = {Yu Yen Chang and Arjen Van Der Wel and Elisabete Da Cunha and Hans Walter Rix},
   doi = {10.1088/0067-0049/219/1/8},
   issn = {00670049},
   issue = {1},
   journal = {\apjs},
   pages = {8},
   title = {STELLAR MASSES and STAR FORMATION RATES for 1 M GALAXIES from SDSS+WISE},
   volume = {219},
   year = {2015},
}

@ARTICLE{Weiner2006,
       author = {{Weiner}, Benjamin J. and {Willmer}, Christopher N.~A. and {Faber}, S.~M. and {Harker}, Justin and {Kassin}, Susan A. and {Phillips}, Andrew C. and {Melbourne}, Jason and {Metevier}, A.~J. and {Vogt}, N.~P. and {Koo}, D.~C.},
      journal = {\apj},
         year = 2006,
        month = dec,
       volume = {653},
       number = {2},
        pages = {1049-1069},
          doi = {10.1086/508922},
archivePrefix = {arXiv},
       eprint = {astro-ph/0609091},
 primaryClass = {astro-ph},
}

@article{Nelson2019,
   author = {Dylan Nelson and Annalisa Pillepich and Volker Springel and Rüdiger Pakmor and Rainer Weinberger and Shy Genel and Paul Torrey and Mark Vogelsberger and Federico Marinacci and Lars Hernquist},
   doi = {10.1093/mnras/stz2306},
   issn = {0035-8711},
   issue = {3},
   journal = {\mnras},
   month = {12},
   pages = {3234-3261},
   title = {First results from the TNG50 simulation: galactic outflows driven by supernovae and black hole feedback},
   volume = {490},
   year = {2019},
}

@ARTICLE{Hogg2010,
       author = {{Hogg}, David W. and {Bovy}, Jo and {Lang}, Dustin},
      journal = {arXiv e-prints},
         year = 2010,
        month = aug,
          eid = {arXiv:1008.4686},
        pages = {arXiv:1008.4686},
archivePrefix = {arXiv},
       eprint = {1008.4686},
}

@article{Avery2022,
author = {Avery, Charlotte R and Wuyts, Stijn and {F{\"{o}}rster Schreiber}, Natascha M and Villforth, Carolin and Bertemes, Caroline and Hamer, Stephen L and Sharma, Raman and Toshikawa, Jun and Zhang, Junkai},
doi = {10.1093/mnras/stac190},
eprint = {2201.08079},
issn = {0035-8711},
journal = {\mnras},
month = {feb},
number = {3},
pages = {4223--4237},
title = {{Cool outflows in MaNGA: a systematic study and comparison to the warm phase}},
volume = {511},
year = {2022}
}

@article{Swinbank2019,
author = {Swinbank, A. M. and Harrison, C. M. and Tiley, A. L. and Johnson, H. L. and Smail, Ian and Stott, J. P. and Best, P. N. and Bower, R. G. and Bureau, M. and Bunker, A. and Cirasuolo, M. and Jarvis, M. and Magdis, G. E. and Sharples, R. M. and Sobral, D.},
doi = {10.1093/mnras/stz1275},
eprint = {1906.05311},
isbn = {1906.05311v1},
issn = {13652966},
journal = {\mnras},
month = {jun},
number = {1},
pages = {381--393},
title = {{The energetics of starburst-driven outflows at z ∼1 from KMOS}},
volume = {487},
year = {2019}
}

@article{Kornei2013,
	author = {{Kornei}, Katherine A. and Shapley, Alice E. and Martin, Crystal L. and Coil, Alison L. and Lotz, Jennifer M. and Weiner, Benjamin J.},
	doi = {10.1088/0004-637X/774/1/50},
	issn = {0004-637X},
	journal = {ApJ},
	month = {aug},
	number = {1},
	pages = {50},
	title = {{FINE-STRUCTURE Fe II* EMISSION AND RESONANT Mg II EMISSION IN z ∼ 1 STAR-FORMING GALAXIES}},
	volume = {774},
	year = {2013}
}

@article{Feltre2018,
	author = {Feltre, Anna and Bacon, Roland and Tresse, Laurence and Finley, Hayley and Carton, David and Blaizot, J{\'{e}}r{\'{e}}my and Bouch{\'{e}}, Nicolas and Garel, Thibault and Inami, Hanae and Boogaard, Leindert A. and Brinchmann, Jarle and Charlot, St{\'{e}}phane and Chevallard, Jacopo and Contini, Thierry and Michel-Dansac, Leo and Mahler, Guillaume and Marino, Raffaella A. and Maseda, Michael V. and Richard, Johan and Schmidt, Kasper B. and Verhamme, Anne},
	issn = {0004-6361},
	journal = {A{\&}A},
	month = {sep},
	pages = {A62},
	title = {{The MUSE Hubble Ultra Deep Field Survey}},
	volume = {617},
	year = {2018}
}

@ARTICLE{Erb2012,
	author = {{Erb}, Dawn K. and {Quider}, Anna M. and {Henry}, Alaina L. and
	{Martin}, Crystal L.},
	title = "{Galactic Outflows in Absorption and Emission: Near-ultraviolet Spectroscopy of Galaxies at 1 \&lt; z \&lt; 2}",
	journal = {ApJ},
	year = 2012,
	month = nov,
	volume = {759},
	number = {1},
	eid = {26},
	pages = {26},
	doi = {10.1088/0004-637X/759/1/26},
	archivePrefix = {arXiv},
	eprint = {1209.4903},
	primaryClass = {astro-ph.CO},
}

@ARTICLE{Schwarz1978,
       author = {{Schwarz}, Gideon},
        title = "{Estimating the Dimension of a Model}",
      journal = {Annals of Statistics},
         year = 1978,
        month = jul,
       volume = {6},
       number = {2},
        pages = {461-464},
       adsurl = {https://ui.adsabs.harvard.edu/abs/1978AnSta...6..461S}
}

@article{Prochaska2011,
	author = {Prochaska, J. Xavier and Kasen, Daniel and Rubin, Kate},
	journal = {\apj},
	month = {jun},
	number = {1},
	pages = {24},
	publisher = {IOP Publishing},
	title = {{Simple models of metal-line absorption and emission from cool gas outflows}},
	volume = {734},
	year = {2011}
}

@article{VanderWel2012,
	author = {van der Wel, A. and Bell, E. F. and H{\"{a}}ussler, B. and McGrath, E. J. and Chang, Yu-Yen and Guo, Yicheng and McIntosh, D. H. and Rix, H.-W. and Barden, M. and Cheung, E. and Faber, S. M. and Ferguson, H. C. and Galametz, A. and Grogin, N. A. and Hartley, W. and Kartaltepe, J. S. and Kocevski, D. D. and Koekemoer, A. M. and Lotz, J. and Mozena, M. and Peth, M. A. and Peng, Chien Y.},
	journal = {\apjs},
	month = {dec},
	number = {2},
	pages = {24},
	title = {{STRUCTURAL PARAMETERS OF GALAXIES IN CANDELS}},
	volume = {203},
	year = {2012}
}

@software{Bradley2020,
author       = {Larry Bradley and
                Brigitta Sip{\H o}cz and
                Thomas Robitaille and
                Erik Tollerud and
                Z\`e Vin{\'{\i}}cius and
                Christoph Deil and
                Kyle Barbary and
                Tom J Wilson and
                Ivo Busko and
                Hans Moritz G{\"u}nther and
                Mihai Cara and
                Simon Conseil and
                Azalee Bostroem and
                Michael Droettboom and
                E. M. Bray and
                Lars Andersen Bratholm and
                P. L. Lim and
                Geert Barentsen and
                Matt Craig and
                Sergio Pascual and
                Gabriel Perren and
                Johnny Greco and
                Axel Donath and
                Miguel de Val-Borro and
                Wolfgang Kerzendorf and
                Yoonsoo P. Bach and
                Benjamin Alan Weaver and
                Francesco D'Eugenio and
                Harrison Souchereau and
                Leonardo Ferreira},
title        = {astropy/photutils: 1.0.0},
month        = sep,
year         = 2020,
publisher    = {Zenodo},
version      = {1.0.0},
doi          = {10.5281/zenodo.4044744},
url          = {https://doi.org/10.5281/zenodo.4044744}
}

@article{Kennicutt2012,
	author = {Kennicutt, Robert C. and Evans, Neal J.},
	doi = {10.1146/annurev-astro-081811-125610},
	issn = {0066-4146},
	journal = {ARA{\&}A},
	month = {sep},
	number = {1},
	pages = {531--608},
	title = {{Star Formation in the Milky Way and Nearby Galaxies}},
	volume = {50},
	year = {2012}
}

@article{Mobasher2015,
   author = {Bahram Mobasher and Tomas Dahlen and Henry C. Ferguson and Viviana Acquaviva and Guillermo Barro and Steven L. Finkelstein and Adriano Fontana and Ruth Gruetzbauch and Seth Johnson and Yu Lu and Casey J. Papovich and Janine Pforr and Mara Salvato and Rachel S. Somerville and Tommy Wiklind and Stijn Wuyts and Matthew L. N. Ashby and Eric Bell and Christopher J. Conselice and Mark E. Dickinson and Sandra M. Faber and Giovanni Fazio and Kristian Finlator and Audrey Galametz and Eric Gawiser and Mauro Giavalisco and Andrea Grazian and Norman A. Grogin and Yicheng Guo and Nimish Hathi and Dale Kocevski and Anton M. Koekemoer and David C. Koo and Jeffrey A. Newman and Naveen Reddy and Paola Santini and Risa H. Wechsler},
   doi = {10.1088/0004-637X/808/1/101},
   issn = {1538-4357},
   issue = {1},
   journal = {ApJ},
   month = {7},
   pages = {101},
   title = {A CRITICAL ASSESSMENT OF STELLAR MASS MEASUREMENT METHODS},
   volume = {808},
   year = {2015},
}

@ARTICLE{Martin2012,
	author = {{Martin}, Crystal L. and {Shapley}, Alice E. and {Coil}, Alison L. and
	{Kornei}, Katherine A. and {Bundy}, Kevin and {Weiner}, Benjamin J. and
	{Noeske}, Kai G. and {Schiminovich}, David},
	title = "{Demographics and Physical Properties of Gas Outflows/Inflows at 0.4 \&lt; z \&lt; 1.4}",
	journal = {ApJ},
	year = 2012,
	month = dec,
	volume = {760},
	number = {2},
	eid = {127},
	pages = {127},
	doi = {10.1088/0004-637X/760/2/127},
	archivePrefix = {arXiv},
	eprint = {1206.5552},
	primaryClass = {astro-ph.CO},
}

@article{Weiner2009,
	author = {Weiner, Benjamin J and Coil, Alison L and Prochaska, Jason X and Newman, Jeffrey A and Cooper, Michael C and Bundy, Kevin and Conselice, Christopher J and Dutton, Aaron A and Faber, S M and Koo, David C and Lotz, Jennifer M and Rieke, G H and Rubin, K H R},
	journal = {\apj},
	pages = {187--211},
	title = {{UBIQUITOUS OUTFLOWS IN DEEP2 SPECTRA OF STAR-FORMING GALAXIES AT z = 1.4}},
	volume = {692},
	year = {2009}
}

@article{Santini2015,
	author = {Santini, P. and Ferguson, H. C. and Fontana, A. and Mobasher, B. and Barro, G. and Castellano, M. and Finkelstein, S. L. and Grazian, A. and Hsu, L. T. and Lee, B. and Lee, S.-K. and Pforr, J. and Salvato, M. and Wiklind, T. and Wuyts, S. and Almaini, O. and Cooper, M. C. and Galametz, A. and Weiner, B. and Amorin, R. and Boutsia, K. and Conselice, C. J. and Dahlen, T. and Dickinson, M. E. and Giavalisco, M. and Grogin, N. A. and Guo, Y. and Hathi, N. P. and Kocevski, D. and Koekemoer, A. M. and Kurczynski, P. and Merlin, E. and Mortlock, A. and Newman, J. A. and Paris, D. and Pentericci, L. and Simons, R. and Willner, S. P.},
	journal = {\apj},
	month = {mar},
	number = {2},
	pages = {97},
	title = {{STELLAR MASSES FROM THE CANDELS SURVEY: THE GOODS-SOUTH AND UDS FIELDS}},
	volume = {801},
	year = {2015}
}

@ARTICLE{Giavalisco2004,
       author = {{Giavalisco}, M. and {Ferguson}, H.~C. and {Koekemoer}, A.~M. and {Dickinson}, M. and {Alexander}, D.~M. and {Bauer}, F.~E. and {Bergeron}, J. and {Biagetti}, C. and {Brandt}, W.~N. and {Casertano}, S. and {Cesarsky}, C. and {Chatzichristou}, E. and {Conselice}, C. and {Cristiani}, S. and {Da Costa}, L. and {Dahlen}, T. and {de Mello}, D. and {Eisenhardt}, P. and {Erben}, T. and {Fall}, S.~M. and {Fassnacht}, C. and {Fosbury}, R. and {Fruchter}, A. and {Gardner}, J.~P. and {Grogin}, N. and {Hook}, R.~N. and {Hornschemeier}, A.~E. and {Idzi}, R. and {Jogee}, S. and {Kretchmer}, C. and {Laidler}, V. and {Lee}, K.~S. and {Livio}, M. and {Lucas}, R. and {Madau}, P. and {Mobasher}, B. and {Moustakas}, L.~A. and {Nonino}, M. and {Padovani}, P. and {Papovich}, C. and {Park}, Y. and {Ravindranath}, S. and {Renzini}, A. and {Richardson}, M. and {Riess}, A. and {Rosati}, P. and {Schirmer}, M. and {Schreier}, E. and {Somerville}, R.~S. and {Spinrad}, H. and {Stern}, D. and {Stiavelli}, M. and {Strolger}, L. and {Urry}, C.~M. and {Vandame}, B. and {Williams}, R. and {Wolf}, C.},
        title = "{The Great Observatories Origins Deep Survey: Initial Results from Optical and Near-Infrared Imaging}",
      journal = {\apjl},
         year = 2004,
        month = jan,
       volume = {600},
       number = {2},
        pages = {L93-L98},
          doi = {10.1086/379232},
archivePrefix = {arXiv},
       eprint = {astro-ph/0309105},
 primaryClass = {astro-ph},
       adsurl = {https://ui.adsabs.harvard.edu/abs/2004ApJ...600L..93G}
}

@article{Pacifici2012,
author = {Pacifici, Camilla and Charlot, St{\'{e}}phane and Blaizot, J{\'{e}}r{\'{e}}my and Brinchmann, Jarle},
doi = {10.1111/j.1365-2966.2012.20431.x},
journal = {MNRAS},
month = {apr},
number = {3},
pages = {2002--2024},
title = {{Relative merits of different types of rest-frame optical observations to constrain galaxy physical parameters}},
volume = {421},
year = {2012}
}

@article{Pacifici2015,
author = {Pacifici, Camilla and da Cunha, Elisabete and Charlot, St{\'{e}}phane and Rix, Hans-Walter and Fumagalli, Mattia and van der Wel, Arjen and Franx, Marijn and Maseda, Michael V. and van Dokkum, Pieter G. and Brammer, Gabriel B. and Momcheva, Ivelina and Skelton, Rosalind E. and Whitaker, Katherine and Leja, Joel and Lundgren, Britt and Kassin, Susan A. and Yi, Sukyoung K.},
doi = {10.1093/mnras/stu2447},
issn = {1365-2966},
journal = {MNRAS},
month = {feb},
number = {1},
pages = {786--805},
title = {{On the importance of using appropriate spectral models to derive physical properties of galaxies at 0.7 < z < 2.8}},
volume = {447},
year = {2015}
}

@ARTICLE{Oke1983,
       author = {{Oke}, J.~B. and {Gunn}, J.~E.},
        title = "{Secondary standard stars for absolute spectrophotometry.}",
      journal = {\apj},
         year = 1983,
        month = mar,
       volume = {266},
        pages = {713-717},
          doi = {10.1086/160817},
}

@ARTICLE{VanderWel14,
       author = {{van der Wel}, A. and {Franx}, M. and {van Dokkum}, P.~G. and {Skelton}, R.~E. and {Momcheva}, I.~G. and {Whitaker}, K.~E. and {Brammer}, G.~B. and {Bell}, E.~F. and {Rix}, H. -W. and {Wuyts}, S. and {Ferguson}, H.~C. and {Holden}, B.~P. and {Barro}, G. and {Koekemoer}, A.~M. and {Chang}, Yu-Yen and {McGrath}, E.~J. and {H{\"a}ussler}, B. and {Dekel}, A. and {Behroozi}, P. and {Fumagalli}, M. and {Leja}, J. and {Lundgren}, B.~F. and {Maseda}, M.~V. and {Nelson}, E.~J. and {Wake}, D.~A. and {Patel}, S.~G. and {Labb{\'e}}, I. and {Faber}, S.~M. and {Grogin}, N.~A. and {Kocevski}, D.~D.},
        title = "{3D-HST+CANDELS: The Evolution of the Galaxy Size-Mass Distribution since z = 3}",
      journal = {\apj},
         year = 2014,
        month = jun,
       volume = {788},
       number = {1},
          eid = {28},
        pages = {28},
          doi = {10.1088/0004-637X/788/1/28}
}

@article{Pacifici2016,
author = {Pacifici, Camilla and Kassin, Susan A. and Weiner, Benjamin J. and Holden, Bradford and Gardner, Jonathan P. and Faber, Sandra M. and Ferguson, Henry C. and Koo, David C. and Primack, Joel R. and Bell, Eric F. and Dekel, Avishai and Gawiser, Eric and Giavalisco, Mauro and Rafelski, Marc and Simons, Raymond C. and Barro, Guillermo and Croton, Darren J. and Dav{\'{e}}, Romeel and Fontana, Adriano and Grogin, Norman A. and Koekemoer, Anton M. and Lee, Seong-Kook and Salmon, Brett and Somerville, Rachel and Behroozi, Peter},
doi = {10.3847/0004-637x/832/1/79},
issn = {1538-4357},
journal = {ApJ},
month = {nov},
number = {1},
pages = {79},
title = {{the Evolution of Star Formation Histories of Quiescent Galaxies}},
volume = {832},
year = {2016}
}

@article{Barro2019,
author = {Barro, Guillermo and P{\'{e}}rez-Gonz{\'{a}}lez, Pablo G. and Cava, Antonio and Brammer, Gabriel and Pandya, Viraj and Moral, Carmen Eliche and Esquej, Pilar and Dom{\'{i}}nguez-S{\'{a}}nchez, Helena and Pampliega, Belen Alcalde and Guo, Yicheng and Koekemoer, Anton M. and Trump, Jonathan R. and Ashby, Matthew L. N. and Cardiel, Nicolas and Castellano, Marco and Conselice, Christopher J. and Dickinson, Mark E. and Dolch, Timothy and Donley, Jennifer L. and Briones, N{\'{e}}stor Espino and Faber, Sandra M. and Fazio, Giovanni G. and Ferguson, Henry and Finkelstein, Steve and Fontana, Adriano and Galametz, Audrey and Gardner, Jonathan P. and Gawiser, Eric and Giavalisco, Mauro and Grazian, Andrea and Grogin, Norman A. and Hathi, Nimish P. and Hemmati, Shoubaneh and Hern{\'{a}}n-Caballero, Antonio and Kocevski, Dale and Koo, David C. and Kodra, Dritan and Lee, Kyoung-Soo and Lin, Lihwai and Lucas, Ray A. and Mobasher, Bahram and McGrath, Elizabeth J. and Nandra, Kirpal and Nayyeri, Hooshang and Newman, Jeffrey A. and Pforr, Janine and Peth, Michael and Rafelski, Marc and Rodr{\'{i}}guez-Munoz, Lucia and Salvato, Mara and Stefanon, Mauro and van der Wel, Arjen and Willner, Steven P. and Wiklind, Tommy and Wuyts, Stijn},
doi = {10.3847/1538-4365/ab23f2},
issn = {1538-4365},
journal = {ApJS},
month = {jul},
number = {2},
pages = {22},
title = {{The CANDELS/SHARDS Multiwavelength Catalog in GOODS-N: Photometry, Photometric Redshifts, Stellar Masses, Emission-line Fluxes, and Star Formation Rates}},
volume = {243},
year = {2019}
}

@ARTICLE{Pharo2022,
       author = {{Pharo}, John and {Guo}, Yicheng and {Calvo}, Guillermo Barro and {Carleton}, Timothy and {Faber}, S.~M. and {Guhathakurta}, Puragra and {Kassin}, Susan A. and {Koo}, David C. and {Lonergan}, Jack and {Teppala}, Teja and {Wang}, Weichen and {Yesuf}, Hassen M. and {Bian}, Fuyan and {Dav{\'e}}, Romeel and {Forbes}, John C. and {Keres}, Dusan and {Perez-Gonzalez}, Pablo and {Martin}, Alec and {Puleo}, A.~J. and {Williams}, Lauryn and {Winningham}, Benjamin},
        title = "{The Dwarf Galaxy Population at z   0.7: A Catalog of Emission Lines and Redshifts from Deep Keck Observations}",
      journal = {\apjs},
         year = 2022,
        month = aug,
       volume = {261},
       number = {2},
          eid = {12},
        pages = {12},
          doi = {10.3847/1538-4365/ac6cdf},
archivePrefix = {arXiv},
       eprint = {2203.09588},
 primaryClass = {astro-ph.GA},
       adsurl = {https://ui.adsabs.harvard.edu/abs/2022ApJS..261...12P}
}

@article{Wang2022,
author = {Wang, Weichen and Kassin, Susan A. and Faber, S. M. and Koo, David C. and Cunningham, Emily C. and Yesuf, Hassen M. and Barro, Guillermo and Guhathakurta, Puragra and Weiner, Benjamin J. and de la Vega, Alexander and Guo, Yicheng and Heckman, Timothy M. and Pacifici, Camilla and Wang, Bingjie and Welker, Charlotte},
doi = {10.3847/1538-4357/ac6592},
issn = {0004-637X},
journal = {ApJ},
month = {may},
number = {2},
pages = {146},
title = {{The Baltimore Oriole's Nest: Cool Winds from the Inner and Outer Parts of a Star-forming Galaxy at z = 1.3}},
volume = {930},
year = {2022}
}

@article{Prusinski2021,
author = {Prusinski, Nikolaus Z and Erb, Dawn K and Martin, Crystal L},
doi = {10.3847/1538-3881/abe85b},
issn = {0004-6256},
journal = {AJ},
month = {may},
number = {5},
pages = {212},
title = {{Connecting Galactic Outflows and Star Formation: Inferences from H$\alpha$ Maps and Absorption-line Spectroscopy at 1 ≲ z ≲ 1.5}},
volume = {161},
year = {2021}
}

@ARTICLE{Liddle2007,
       author = {{Liddle}, Andrew R.},
        title = "{Information criteria for astrophysical model selection}",
      journal = {\mnras},
         year = 2007,
        month = may,
       volume = {377},
       number = {1},
        pages = {L74-L78},
          doi = {10.1111/j.1745-3933.2007.00306.x},
archivePrefix = {arXiv},
       eprint = {astro-ph/0701113},
 primaryClass = {astro-ph},
}

@article{Williams2009,
	author = {Williams, Rik J. and Quadri, Ryan F. and Franx, Marijn and van Dokkum, Pieter and Labb{\'{e}}, Ivo},
	journal = {ApJ},
	number = {2},
	pages = {1879--1895},
	volume = {691},
	year = {2009}
	}

@ARTICLE{Calabro2022,
       author = {{Calabr{\`o}}, A. and {Pentericci}, L. and {Talia}, M. and {Cresci}, G. and {Castellano}, M. and {Belfiori}, D. and {Mascia}, S. and {Zamorani}, G. and {Amor{\'\i}n}, R. and {Fynbo}, J.~P.~U. and {Ginolfi}, M. and {Guaita}, L. and {Hathi}, N.~P. and {Koekemoer}, A. and {Llerena}, M. and {Mannucci}, F. and {Santini}, P. and {Saxena}, A. and {Schaerer}, D.},
        title = "{Properties of the interstellar medium in star-forming galaxies at redshifts 2 {\ensuremath{\leq}} z {\ensuremath{\leq}} 5 from the VANDELS survey}",
      journal = {\aap},
         year = 2022,
        month = nov,
       volume = {667},
          eid = {A117},
        pages = {A117},
          doi = {10.1051/0004-6361/202244364},
archivePrefix = {arXiv},
       eprint = {2206.14918},
 primaryClass = {astro-ph.GA},
       adsurl = {https://ui.adsabs.harvard.edu/abs/2022A&A...667A.117C}
}

@ARTICLE{Kehoe2025,
       author = {{Kehoe}, Emily and {Shapley}, Alice E. and {Sanders}, Ryan L. and {Reddy}, Naveen A. and {Topping}, Michael W. and {Lam}, Natalie and {Clarke}, Leonardo and {Cullen}, Fergus and {Ellis}, Richard S. and {F{\"o}rster Schreiber}, N.~M. and {Jones}, Tucker and {Khostovan}, Ali Ahmad and {McLeod}, Derek J. and {McLure}, Ross J. and {Narayanan}, Desika and {Oesch}, Pascal and {Pahl}, Anthony J.},
        title = "{The AURORA Survey: Tracing Galactic Outflows at z {\ensuremath{\gtrsim}} 2.5 with JWST/NIRSpec Near-ultraviolet Absorption Lines}",
      journal = {\apj},
         year = 2025,
        month = dec,
       volume = {994},
       number = {2},
          eid = {170},
        pages = {170},
          doi = {10.3847/1538-4357/ae10b3},
archivePrefix = {arXiv},
       eprint = {2506.17381},
 primaryClass = {astro-ph.GA},
       adsurl = {https://ui.adsabs.harvard.edu/abs/2025ApJ...994..170K}
}

@ARTICLE{Lyu2025,
       author = {{Lyu}, Cheqiu and {Yu}, Haoran and {Wang}, Enci and {Wang}, Junxian and {Jia}, Cheng and {Song}, Jie and {Chen}, Yangyao and {Wang}, Jinyang and {Chen}, Zeyu and {Ma}, Chengyu and {Wang}, Yifan and {Kong}, Xu},
        title = "{First Statistical Detection of Mg II-traced Cool Gas Outflows with JWST toward Cosmic Dawn}",
      journal = {\apjl},
         year = 2026,
        month = mar,
       volume = {1000},
       number = {1},
          eid = {L3},
        pages = {L3},
          doi = {10.3847/2041-8213/ae48ee},
archivePrefix = {arXiv},
       eprint = {2512.05622},
 primaryClass = {astro-ph.GA},
       adsurl = {https://ui.adsabs.harvard.edu/abs/2026ApJ..1000L...3L}
}

@ARTICLE{Claeyssens2025,
       author = {{Claeyssens}, Ad{\'e}la{\"\i}de and {Adamo}, Angela and {Messa}, Matteo and {Dessauges-Zavadsky}, Miroslava and {Richard}, Johan and {Kramarenko}, Ivan and {Matthee}, Jorryt and {Naidu}, Rohan P.},
        title = "{Tracing star formation across cosmic time at tens of parsec-scales in the lensing cluster field Abell 2744}",
      journal = {\mnras},
         year = 2025,
        month = mar,
       volume = {537},
       number = {3},
        pages = {2535-2558},
          doi = {10.1093/mnras/staf058},
archivePrefix = {arXiv},
       eprint = {2410.10974},
 primaryClass = {astro-ph.GA},
       adsurl = {https://ui.adsabs.harvard.edu/abs/2025MNRAS.537.2535C}
}

@ARTICLE{Cameron2011,
       author = {{Cameron}, Ewan},
        title = "{On the Estimation of Confidence Intervals for Binomial Population Proportions in Astronomy: The Simplicity and Superiority of the Bayesian Approach}",
      journal = {\pasa},
         year = 2011,
        month = jun,
       volume = {28},
       number = {2},
        pages = {128-139},
          doi = {10.1071/AS10046},
archivePrefix = {arXiv},
       eprint = {1012.0566},
 primaryClass = {astro-ph.IM},
       adsurl = {https://ui.adsabs.harvard.edu/abs/2011PASA...28..128C}
}
\bibliographystyle{aasjournal}



\end{document}